\documentclass[12pt,preprint]{aastex}
\newcommand{\PREPRINT}{}
\newcommand{\COFigure}[3]{
\begin{figure}
\includegraphics[height=3.5cm]{#1-eps-converted-to.pdf}
\caption{Disk emission in CO J=#3 of #2.}
\label{fig:#1}
\end{figure}
}
\newcommand{\COFigSet}[3]{
\begin{figure}
\includegraphics[height=3.5cm]{#1-eps-converted-to.pdf}
\caption{Disk emission in CO J=#3 of #2.
\textbf{The complete figure set for all sources (13 images) is available in the online journal}
}
\label{fig:aj}
\end{figure}
}
\newcommand{\COpanel}[3]{\includegraphics[width=8.0cm]{#1-eps-converted-to.pdf}}

\newcommand{\COSet}[4]{
    \figsetgrpstart
    \figsetgrpnum{\ref{fig:aj}.#4}
    \figsetgrptitle{CO J=3-2 towards #2}
\figsetplot{#1-eps-converted-to.pdf}
    \figsetgrpend
}
\newcommand{\teff}{\hbox{$T_\mathrm{eff}$}}
\begin{document}

\def \lsun          {\hbox{L$_{\odot}$}}
\def \msun          {\hbox{M$_\odot$}}

\title{Dynamical Masses of Low Mass Stars in the Taurus and Ophiuchus Star Forming Regions}

\author{M. Simon\altaffilmark{1,8}, S. Guilloteau\altaffilmark{2},
E. Di Folco\altaffilmark{2}, A. Dutrey\altaffilmark{2}, N. Grosso\altaffilmark{3},
V. Pi\'etu\altaffilmark{4}. E. Chapillon\altaffilmark{2,4}, L. Prato\altaffilmark{5},
G.H. Schaefer\altaffilmark{6}, E. Rice\altaffilmark{7,8}, Y. Boehler\altaffilmark{9}}

\altaffiltext{1}{Dept. of Physics and Astronomy, Stony Brook University, Stony Brook, NY
11794-3800, USA; michal.simon@stonybrook.edu}
\altaffiltext{2}{Laboratoire d'astrophysique de Bordeaux, Univ. Bordeaux, CNRS, B18N, allée Geoffroy Saint-Hilaire, 33615 Pessac, France}
\altaffiltext{3}{Universit\'e de Strasbourg, CNRS, Observatoire Astronomique de Strasbourg, 
UMR 7550, 67000 Strasbourg, France}
\altaffiltext{4}{IRAM, 300 rue de la piscine, F-38406 Saint Martin d'H\`eres, France}
\altaffiltext{5}{Lowell Observatory, 1400 West Mars Hill Road, Flagstaff, AZ 86001, USA}
\altaffiltext{6}{The CHARA Array of Georgia State University, Mount Wilson Observatory,
Mount Wilson, CA 91023, USA}
\altaffiltext{7}{Dept. of Engineering Science and Physics, College of Staten Island,
Staten Island, NY 10314, USA}
\altaffiltext{8}{Dept. of Astrophysics, American Museum of Natural History, New York, NY 10024, USA}
\altaffiltext{9}{Physics and Astronomy Dept., Rice University, Houston, TX 77005-1827, USA}

\begin{abstract}

We report new dynamical masses  for 5 pre-main sequence (PMS)
stars in the L1495 region of the Taurus star-forming region (SFR) and 6
in the L1688 region of the Ophiuchus SFR.  Since these regions have VLBA
parallaxes these are absolute measurements of the stars' masses
and are independent of their effective temperatures and luminosities. 
Seven of the stars have masses $<0.6$~\msun ~thus providing data in a mass range
with little data, and of these, 6  are measured to precision $< 5 \%$.
We find  8 stars with masses in the range 0.09 to 1.1 \msun~
that agree well with the current generation of PMS evolutionary models.
The ages of the stars we measured in the Taurus SFR are in the range 1-3 MY,
and $<1$~MY for those in L1688.  We also measured the dynamical masses
of 14 stars in the ALMA archival data for Akeson~\&~Jensen's Cycle 0
project on binaries in the Taurus SFR.  We find that the masses of 7 of the
targets are so large that they cannot be reconciled with reported values 
of their luminosity and effective temperature.  We suggest that 
these targets are themselves binaries or triples.

\end{abstract}


\section{Introduction}

Astronomers deduce the masses and ages of stars by their positions on
Hertzsprung-Russell diagrams (HRDs) relative to models of stellar evolution
and their isochrones.  This procedure is regarded as reliable for stars
on the main sequence or approaching it.  The reasons
for this confidence are that the theoretical models for these stars are
in good agreement, that the models are calibrated by many accurate and precise
 measurements of stellar masses, and that the properties of the stars
are well-understood.
The situation is more problematic for pre-main sequence (PMS) stars.
For stars of young age,  age  $<\sim 10$~MY, and  low mass,
 $<\sim 1$~\msun, there has been considerable scatter among the
models of their evolution, precise measurements of their masses
are sparse, and measurements of properties such as the luminosity and
effective temperature are often complicated  by activity associated with
PMS stars. Among the reasons why this gap in our understanding needs to be
filled is that we now know that it includes the era of planet formation among
sun-like stars.

In this paper we present new dynamical masses  of single, low mass young
 stars measured by the rotation of their circumstellar disks 
e.g. \citet{Guilloteau+etal_2014}. We focused our ALMA Cycle 2 program 
on young stars
expected to have masses smaller than 0.5 \msun~ on the basis of their
spectral types.  The measured masses are absolute because
our targets are at known distances.  We measured 5 new masses in the
L1495 region of Taurus and 6 in the L1688 region of Ophiuchus, all
with high precision.

We also used  archival data from \citet{Akeson+Jensen_2014}'s ALMA
program  designed to study the disks of the components of young
binaries in the Taurus SFR.  We were able to measure  14 new dynamical
masses, most more massive  than the stars in our ALMA program.
Seven of these new measurements probably represent the detection of new
multiples that are angularly unresolved at present.

For simplicity we designate the two data sets we use in
this paper as follows.  ALMA Cycle 2 refers to data obtained for
our project on single stars in  Taurus (L1495) and Ophiuchus (L1688)
(ALMA Project 2013.2.00163.S). ALMA Cycle 0 designates 
\citet{Akeson+Jensen_2014}'s project (AJ14) directed at binaries 
distributed over the  Taurus SFR (ALMA Project 2011.0.00150.S).

\section{Targets}

\subsection{ALMA Cycle 2: Taurus and Ophiuchus}
\label{sub:ours}

We sought  to observe single stars with circumstellar disks expected
on the basis of their spectral types to have masses $<\sim 0.5$~\msun~
and to lie at known distances.  We proceeded as follows:

\parindent = 0.0in

1){\it Star Forming Regions (SFRs):}  We chose to observe PMS stars in the
Taurus and Ophiuchus  SFRs because lying at average distances 140 and 120 pc,
respectively, they are relatively nearby \citep{{Kenyon+etal_2008},{Wilking+etal_2008}}.

Also, extensive recent studies of their members are available: 
\citet[][And13]{Andrews+etal_2013} and 
\citet[][He14]{Herczeg+Hillenbrand_2014} for Taurus;
\citet[][R10a]{Ricci+etal_2010}, \citet{McClure+etal_2010}, and  
\citet{Najita+etal_2015} for Ophiuchus.

2) {\it Distances, Environments, and Ages:} PMS stars in the Taurus and
Ophiuchus SFRs are mostly located in small dark clouds \citep{Lynds_1962}
which are also identifiable in CO \citep[][see for example
Fig. 6 in Guilloteau et al. 2014]{Dame+etal_2001}.  Several of
these stellar groups have measured distances more precise than the averages
to the Tauurus and Ophiuchus SFRs.
Three stars in  L1495 in Taurus have VLBA parallaxes  placing its
average distance at $131.4\pm 2.4$ pc \citep{Torres+etal_2012}. Two stars
in L1688 in Ophiuchus have VLBA parallaxes \citep{Loinard+etal_2008}.
We use their average, $119.4\pm 4.6$ pc for the distance to L1688.
We drew our targets from the L1495 and L1688 regions.  The environments
of these regions are quite different.  The stellar and molecular gas
density in L1688 in Ophiuchus is far greater \citep{Wilking+etal_2008}
than in the L1495 in Taurus \citep{Kenyon+etal_2008}.  We were interested
to learn whether observations of the two regions
would provide information on environmental effects on the measurement
of stellar masses by disk rotation.

\begin{table}
\caption{Stellar Properties in Taurus L1495 and Ophiuchus L1688}
\begin{tabular}{rl|rrrrrr}
\hline
\hline
ID & Name    & SpTy & log \teff\  & log $L/\lsun$ & SpTy & log \teff\  & log$L/\lsun$ \\
&&\multicolumn{3}{c}{~~~~~Andrews et al. (2013)}&\multicolumn{3}{c}{Herczeg \& Hillenbrand (2014)}\\
\hline
 &&\multicolumn{6}{c}{Stars in L1495}\\
&&\multicolumn{6}{c}{{\bf Bold} identifies spectral types differing by more than one subclass (see text)}\\
1& FN Tau & {\bf M5}   & $3.495\pm0.020$&$-0.140\pm0.097$&{\bf M3.5}&$3.516\pm0.013$ &$-0.20\pm0.20$ \\
2& MHO 1  & M2.5       & $3.543\pm0.018$&$0.172\pm0.883$ & N/A  &                &                 \\
3& CIDA 1 & {\bf M5.5} & $3.485\pm0.020$&$-0.959\pm0.089$&{\bf M3.5}&$3.516\pm0.013$ &$-0.72\pm0.20$   \\
4& CY Tau   & M1.5     & $3.560\pm0.017$&$-0.456\pm0.090$&M2.5&$3.542\pm0.006$&$-0.58\pm0.20$\\
5& FP Tau   & {\bf M4} & $3.514\pm0.019$&$-0.549\pm0.048$&{\bf M2.6}&$3.540\pm0.006$&$-0.78\pm0.20$\\
6& CX Tau   & M2.5     & $3.543\pm0.015$&$-0.489\pm0.065$&M2.5&$3.542\pm0.006$&$-0.58\pm0.20$ \\
7& V410 X-Ray 1   & M4 & $3.514\pm0.019$&$-0.409\pm0.066$& M3.7&$3.507\pm0.013$&$-1.55\pm0.20$ \\
8& IP Tau   & M0       & $ 3.586\pm0.024$&$-0.389\pm0.126$&M0.6&$3.583\pm0.006$&$-0.47\pm0.20$ \\
9& FM Tau   & {\bf M0} & $3.586\pm0.024$ &$-0.45\pm0.12$ &{\bf M4.5}&$3.489\pm0.012$&$-1.15\pm0.20$\\
\hline
 ID & Name     &SpTy & log \teff\ & log $L/\lsun$&SpTy & log \teff\ & log $L/\lsun$ \\
 &&\multicolumn{3}{c}{~~Ricci et al. (2010b)}&\multicolumn{3}{c}{Najita et al. (2015), McClure et al. (2010)}\\
 \hline
 &&\multicolumn{6}{c}{Stars in L1688}\\
 1&    GSS 26  &K8   &$3.599\pm0.010$  &$0.14\pm0.13$&K7&$3.609\pm0.023$&$0.91\pm0.60$ \\
 2&    GSS 39  &M0   &$3.586\pm0.021$&$-0.11\pm0.13$ &M0&$3.586\pm0.024$&$0.11\pm0.24$\\
 3&    YLW16C  &M1  &$3.566\pm0.020$&$0.045\pm0.13$  &M1&$3.569\pm0.017$&$0.14\pm0.25$ \\
 4&    ROXs 25 &{\bf M2}  &$3.550\pm0.018$&$0.45\pm0.13$&{\bf K7}&$3.609\pm0.023$&$0.49\pm0.09$\\
 5&    YLW 58  &M4  &$3.505\pm 0.020$&$-0.64\pm 0.13$ &M4.5&$3.495\pm0.019$&$-0.70\pm0.04$\\
 6&Flying Saucer&  &$3.544\pm0.040$&$-1.05\pm0.13$   & \multicolumn{3}{c}{See text for Flying Saucer Parameters}\\

 7&    WL 18   &N/A   &               &              &K6.5&$3.615\pm0.016$&$-0.57\pm0.13$ \\
 8&    WL 14   &N/A   &               &              &M4&$3.514\pm0.019$&$-0.85\pm0.15$   \\
 9&    GY 284  &N/A  &               &            &M3.25&$3.525\pm0.018$&$-0.92\pm0.05$\\
 \hline
 \end{tabular}
 \label{tab:1}
 \end{table}

\parindent=0.5in

Table \ref{tab:1} lists the stars observed in the Taurus L1495 region.  All are Class II
YSOs. Columns 1 and 2 provide an ID and name. Cols. 3, 4, and 5 list the spectral type,
effective temperature, and luminosity as listed by And13 and cols. 6, 7, and 8 provide the
same parameters as provided by He14. We did not propagate the distance uncertainty into
the uncertainty of the luminosity. And13's spectral types are drawn from the
literature.  We used their listing of the corresponding {\teff}s  and  uncertainties.
He14 derived stellar parameters from an
analysis of a large homogeneous set of spectra in the visible.  Col. 6 lists He14's
independent SpTy  assignments and col. 7 gives the corresponding {\teff}s using their
conversion table and their recommended uncertainty of $\pm 0.3$ sub-type.  We highlighted
in bold the cases for which the And13 and He14 spectral type designations differ by more
than one subclass. Taking together this sample and Akeson and Jensen's  (see \S 2.2)
there is no clear pattern that one set of spectral types is  hotter or cooler than the
other in the spectral type range considered.  And13 derived luminosities and their
uncertainties from spectral energy distributions (SEDs). Since And13's  values are
calculated at the 140 pc average distance to the Taurus SFR we scaled them to the 131 pc
distance of L1495. He14 derived luminosities using their observed spectra referred to the
BT/Settl models at $log~g=4.0$~provided by \citet{Allard+etal_2012}. We list He14's 
luminosities with
their recommended uncertainty $\pm 0.2$ dex. He14 evaluated these luminosities at 131 pc.
The last star in the table, FM Tau, was not a part of our ALMA program; it was
observed in CO J=2-1 by Y. Boehler et al. in ALMA project 2013.2.00426.S using the same
antenna configuration as our Taurus observations on Jul. 19 and Aug 8, 2015, and in CO
J=3-2 on Jul 24, 2015 with about 0.3$''$ resolution. 
CX Tau was also observed in this project.

Table \ref{tab:1} also provides the same parameters for selected stars in 
the Ophiuchus SFR.  All are in L1688. Except for the Flying Saucer 
\citep[2MASS J16281370-2431391][]{Grosso+etal_2003}, the stellar parameters are from
\citet[][R10a]{Ricci+etal_2010} and \citet{Najita+etal_2015}. We converted  the SpTy's to
{\teff}s using the look-up table from \citet{Pecaut+Mamajek_2013}, applied an uncertainty
of $\pm 1$ sub-type. R10a evaluated the luminosities from the stars' near IR SEDs after
dereddening the near IR colors and correcting for extinction.  Their values are given for
a distance of 130 pc; accordingly we scaled the luminosities to the 119 pc distance adopted
here for the L1688 region and  applied an uncertainty of $\pm 0.13$ dex as suggested in
\citet{Ricci+etal_2010_a}.  Luminosities listed by \citet{Najita+etal_2015} are based on
the analysis of \citet{McLure+etal_2010} of Spitzer Observatory observations.  The
luminosities are presented at the distance of L1688 and do not include a propagated
distance uncertainty.   The uncertainties depend on
extinction to the star; we used values kindly provided by McClure (priv. comm.). We
converted the spectral types to {\teff}s using Pecaut and Mamajek's look-up 
table again. \citet{McClure+etal_2010}
identify all the stars as Class  II PMS stars except for GSS 26 and GY 284 listed as flat
spectrum sources (FS) and YLW16C (= GY 262) for which a designation is not available
from the SED slope between 2 and 25 $\mu$m (but for which an envelope is excluded
from the SED slope between 5 and 12 $\mu$m).
The Flying Saucer is an edge-on disk dominated by scattered light; its parameters
are from  \citet{Grosso+etal_2003} and \citet{Pontoppidan+etal_2007}. Both groups
 assumed a \citet{Kurucz_1979} stellar atmosphere model at \teff\ = 3500K to derive
the luminosity of the central star from SED modeling.  From an extinction map
around the Flying Saucer, \citet{Grosso+etal_2003} derived $A_v$=2.1 mag and 
L$_*$/\lsun =0.14 at a distance of 140 pc corresponding to L$_*$/\lsun =0.10
at 119 pc. \citet{Pontoppidan+etal_2007} assumed $A_v$=0.50 mag, added an accretion
luminosity $L_{acc}/\lsun = 0.6$ and L$_*$/\lsun =0.084 at 125 pc which scales to
$L_*/\lsun =0.076$~at 119 pc. In Table \ref{tab:1} we quote 
the \teff\ with an uncertainty of $\pm 200$K which corresponds to
approximately to one spectral class sub-type at $\sim 3500$K, and the luminosity as an
average of the two values $0.089\pm 0.025$ with the uncertainty equal to the range. The
derived stellar luminosity depends on the assumed  foreground  extinction.  The 
uncertainty of the Flying Saucer's luminosity may be larger than the value we adopt.
\vskip 0.5cm

\subsection{ALMA Cycle 0: Taurus}
\label{sub:aj}

AJ14  obtained ALMA Cycle 0 observations of 17 young binaries in the Taurus SFR.
We list their targets  in Table \ref{tab:2} in the same format as Table \ref{tab:1}.  
The components of the binaries in their program are mostly
of earlier spectral type than in ours.  GK Tau, HO Tau, and DS Tau, are probably single
because AJ14 found that their presumed companions seem to be  chance associations. Most
of the stars in their sample are distributed over the Taurus SFR; we consider 140 pc as
their distance. However, we can assign a more precise value for 2 stars, GK Tau and HO
Tau. GK Tau's position (l=174.2, b=-15.7) places it near L1495: we assume its distance
is 131 pc. HO Tau lies near the star HP Tau/G2 which has a VLBA parallax
measurement $161.2\pm0.09$pc \citep{Torres+etal_2009}. These stars seem to be
associated with L1529 region.  We assume both lie at 161 pc.  


\begin{table}
\caption{Stellar Properties Akeson\&Jensen Archival Data}
\begin{tabular}{rl|rrrrrr}
\hline
\hline
ID & Name &SpTy & log \teff\ & log $L/\lsun$&SpTy &log \teff\ & log$L/\lsun$ \\
&&\multicolumn{3}{c}{Andrews et al. (2013)}&\multicolumn{3}{c}{Herczeg \& Hillenbrand (2014)}\\
\hline
&&\multicolumn{6}{c}{{\bf Bold} identifies spectral types differing by more than one subclass (see text)}\\
1 & FV Tau A &{\bf K5}&$3.639\pm0.024$&$0.369\pm0.143$&{\bf M0.0}&$3.591\pm0.006$&$-0.48\pm0.20$ \\
2 & FX Tau A&{\bf M1} &$3.569\pm0.017$&$-0.285\pm0.215$&{\bf M2.2}&$3.545\pm0.008$& $-0.29\pm0.20$  \\
3 & HBC 411 B&    M4.5&$3.505\pm0.039$&$-0.699\pm0.130$&     M4.3&$3.501\pm0.004$&$-0.80\pm0.20$\\
4 & CIDA-9 A&{\bf K8}&$3.601\pm0.023$&$-1.010\pm0.255$&{\bf M1.8}&$3.560\pm0.005$&$-0.88\pm0.20$\\
5A & HK Tau A&M0.5  &$3.577\pm0.020$&$-0.353\pm0.126$&M1.5&$3.564\pm0.007$&$-0.52\pm0.20$\\
5B & HK Tau B&M2    &$3.552\pm0.020$&$-1.571\pm0.194$& N/A  &         &       \\
6  & IT Tau B&{\bf M4}&$3.515\pm0.019$&$-0.684\pm0.165$&{\bf K6.0}&$3.615\pm0.003$&$-0.01\pm0.20$\\
7A & DK Tau A&K8     &$3.601\pm0.023$&$0.119\pm0.189$&K8.5&$3.599\pm0.003$&$-0.27\pm0.20$\\
7B & DK Tau B&M1     &$3.569\pm0.020$&$-0.498\pm0.158$&M1.7&$3.560\pm0.006$&$-0.76\pm0.20$\\
8 & GK Tau  &K7      &$3.609\pm0.023$&$0.129\pm0.234$ &K6.5&$3.609\pm0.003$&$-0.03\pm0.20$\\
9 & HN Tau A &{\bf K5}&$3.639\pm0.024$&$-0.376\pm0.365$&{\bf K3}&$3.657\pm0.006$&$-0.77\pm0.20$\\
10 & V710 Tau A&{\bf M0.5}&$3.577\pm0.020$&$-0.241\pm0.124$&{\bf M3.3}&$3.524\pm0.009$&$-0.43\pm0.20$\\
11 & HO Tau  &{\bf M0.5}&$3.577\pm0.020$&$-0.886\pm0.087$&{\bf M3.2}&$3.527\pm0.008$&$-0.85\pm0.20$\\
12 & DS Tau  &{\bf K5  }   &$3.639\pm0.024$&$-0.119\pm0.162$&{\bf M0.4}&$3.581\pm0.007$&$-0.72\pm0.20$\\
\hline
\end{tabular}
\label{tab:2}
\end{table}

\section{Observations, Data Reduction, and Analysis }

\subsection{Observations}
\label{sub:obs}

\paragraph{ALMA 2013.2.00163.S} Our ALMA observations were made with a frequency
setup covering the CO J=2-1 and H$_2$CO 3$_{13}$-2$_{12}$
 transitions, as well as all the stronger hyperfine components of the
CN N=2-1 line. The Ophiuchus targets
were observed on May 23, 2015 with a baseline configuration that
provided a typical angular resolution
of $0.6''\times 0.4''$ at PA $90^\circ$.  The typical noise is $\sim 7$
mJy/beam (about 0.6\,K in brightness) at 0.2 km\,s$^{-1}$ spectral resolution.
The Taurus targets were observed on Sept. 19, 2015 with the same
frequency setup but a baseline configuration that provided a higher
angular resolution
$ 0.25'' \times 0.21''$  at PA $180^\circ$  and hence a much higher
brightness noise, typically about 3\,K. We calibrated the data using the
ALMA pipeline in the CASA software package (Version 4.2.2) and
applied doppler correction using the ``cvel'' task to transform the
data in the LSRK frame.  Data were then exported through UVFITS format
to the GILDAS package for imaging and data analysis. For all sources
except WL 14 and GY 284 (which have no or weak signal), phase-only self-calibration solutions were
derived from the continuum and applied to the spectral line tables.
All data were smoothed to 0.2 km\,s$^{-1}$ spectral resolution.

\paragraph{ALMA 2011.0.00150.S}

AJ14's spectral line observations were made in the CO J=2-1 and 3-2 lines.  We processed
data for both lines as above using CASA 3.4 software and exported it to GILDAS for
imaging and analysis. The angular resolution of the J=2-1 data is about
$0.85\times0.50''$ and the spectral resolution is 0.85 km\,s$^{-1}$. The typical
brightness sensitivity is 0.5 K. The J=2-1 data has both lower angular resolution
$1.0\times0.8''$ and spectral resolution 1.26 km\,s$^{-1}$, but the brightness
sensitivity is around 0.15\,K. In both cases, phase-only self-calibration was derived
from the continuum data and applied to the spectral line data. 

\paragraph{ALMA 2013.2.00426.S}
In addition to our survey data, two sources observed in ALMA 2013.2.00426.S (PI
Y.Boehler) in CO J=2-1 and CO J=3-2 were suitable for our purpose: CX Tau and FM Tau. The
CO J=2-1 observations were done with similar velocity and angular resolutions and
sensitivity as in 2012.2.00163.S, while the CO J=3-2 data provided about
$0.3\times0.17''$ resolution, with about 3\,K brightness sensitivity at 0.21 km\,s$^{-1}$
spectral resolution. 

\begin{figure}
\includegraphics[height=8.5cm]{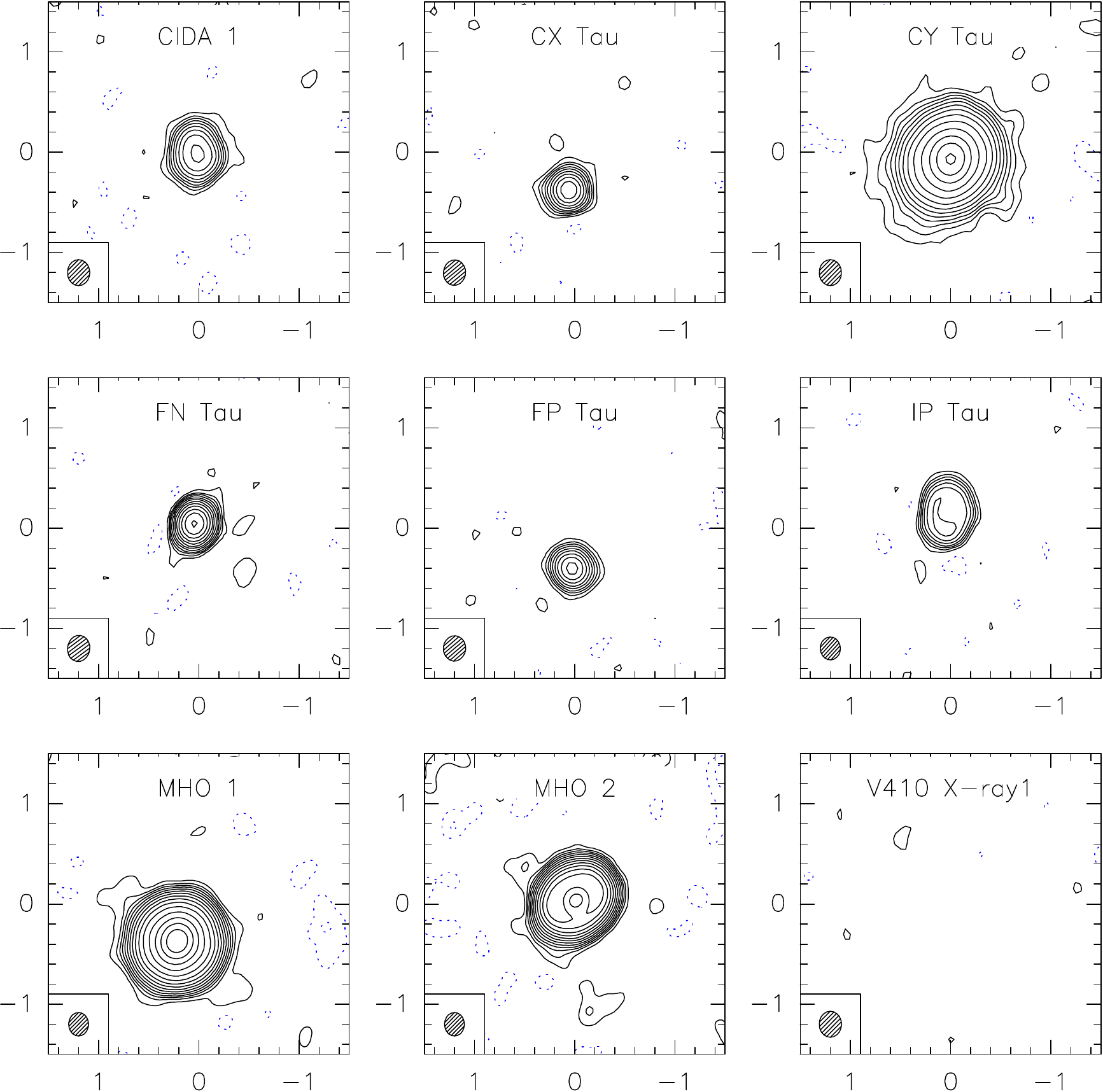}
\caption{Continuum images at 228 GHz of the Taurus sources. Contour levels 
are -3,-2,-1,1,2,3 times
the contour step (0.25 mJy/beam), and then increase exponentially by a factor $\sqrt{2}$. }
\label{cont:taurus}
\end{figure}

\begin{figure}
\includegraphics[height=8.5cm]{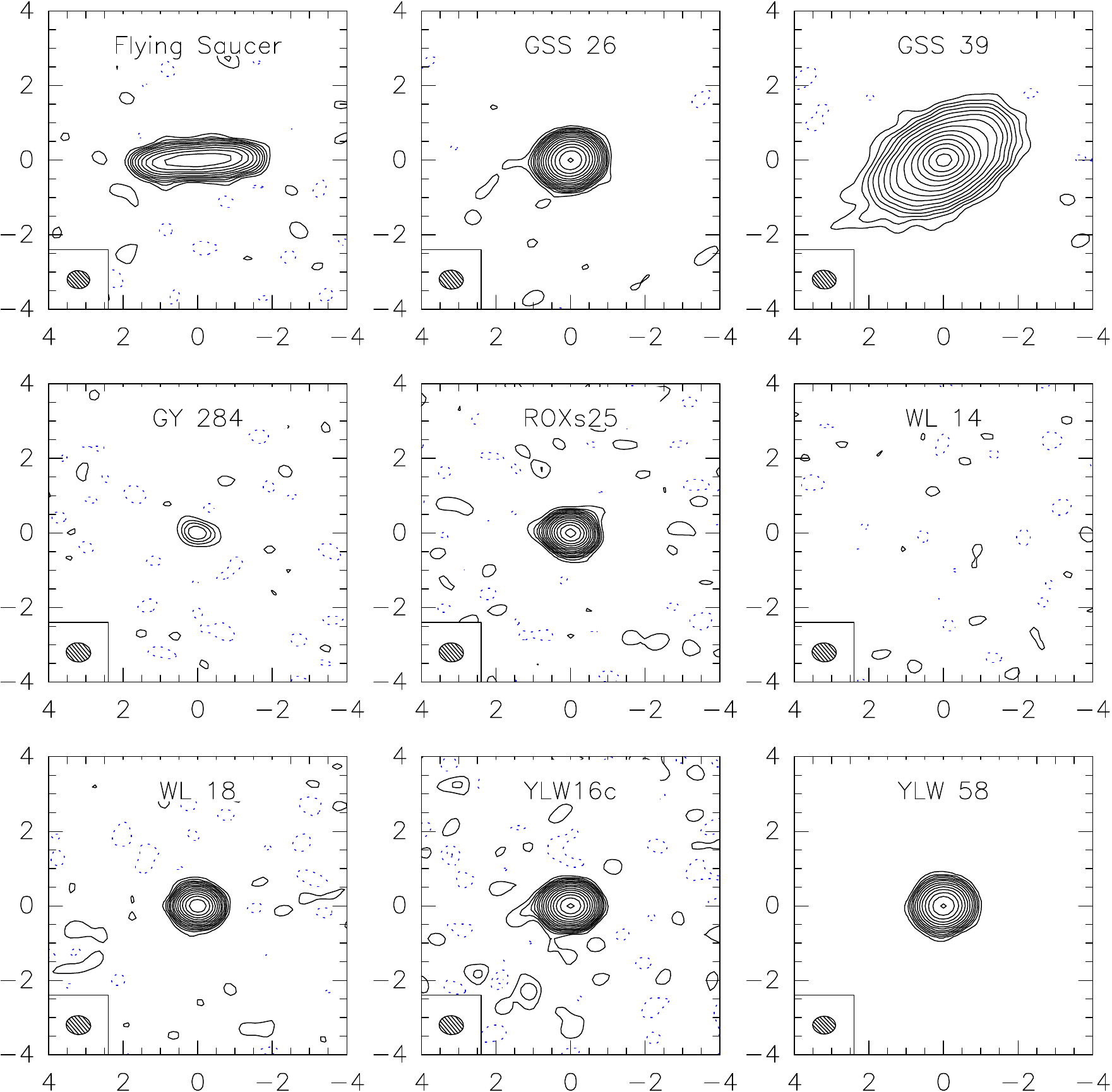}
\caption{As Fig. \ref{cont:taurus} for $\rho$ Oph sources. Contour step is 0.45 mJy/beam
for GSS 26, GSS 39 and YLW 58, 0.15 mJy/beam for the other sources (approximately 2$\sigma$).}
\label{cont:oph}
\end{figure}

 Continuum images for the observed stars are shown in 
Fig.\ref{cont:taurus}-\ref{cont:oph}.  We do not report 
values of the continuum flux for two reasons.
Quoting reliable uncertainties would imply a proper flux scale
calibration which is not reliable in our data set.  The 
absolute flux has no impact on our analysis which relies only
on the morphology of the emission.


\subsection{Analysis}
\label{sub:analysis}

\begin{table}
\caption{Measured Parameters in Taurus L1495 and Ophiuchus L1688}
\begin{tabular}{rrrrrrrrr}
\hline
ID & Name &Dist& Lines& V$_{LSR}$&$i_{line}$&$i_{cont}$&$R_{out}$& M$_{*}$ \\
   &      & pc &      & km\,s$^{-1}$     &  deg     &  deg     &  AU  & \msun \\
\hline
&&\multicolumn{7}{c}{ Stars in L1495} \\
4& CY Tau &131 & CO, CN & $7.27\pm0.03$ & $30\pm2$ & $32\pm1$ & $290\pm10$ & $0.31\pm0.02$  \\
5& FP Tau &131 & CO, CN & $8.32\pm0.06$ & $66\pm2$ & $66\pm4$ & $95\pm5$  & $0.37\pm0.02$ \\
6& CX Tau &131 & CO, CN & $8.40\pm0.03$ & $61\pm1$ & $60\pm5$ & $160\pm20$ & $0.37\pm0.02$ \\
8& IP Tau &131 & CO, CN & $6.30\pm0.02$ & $34\pm1$ & $35\pm2$ & $95\pm20$ & $0.95\pm0.05$ \\
9& FM Tau &131 & CO     & $5.98\pm0.08$ & $52\pm2$ & $55\pm2$ & $50\pm2$ & $0.36\pm0.01$\\
\hline
&&\multicolumn{7}{c}{ Stars in L1688} \\
1 &    GSS 26 &119&CO& $2.75\pm0.01$&$40\pm1$ &$39\pm1$&$325\pm5$&$1.51\pm0.02$ \\
2 & GSS 39&119&CO, H$_{2}$CO&$2.00\pm0.01$ & $54.3\pm0.2$ &$57\pm2$&$600\pm1$&$0.47\pm0.01$ \\
3 & YLW16C&119&H$_{2}$CO& $5.43\pm0.17$&$17\pm1$ &$17\pm4$&$32\pm2$ &$1.80\pm0.10$ \\
4 & ROXs 25&119&CO, CN&$4.94\pm0.05$&$39\pm4$ &$44\pm4$&$60\pm3$&$1.10\pm0.07$ \\
5 & YLW 58 &119&H$_{2}$CO& $3.76\pm0.04$&$30\pm5$ &$30\pm1$&$130\pm3$&$0.09\pm0.01$ \\
6 &Flying Saucer&120&CS& $3.555\pm0.003$&$85.4\pm0.5$&$89.2\pm0.4$&$200\pm7$&$ 0.58\pm0.01$\\
\hline
\end{tabular}
\label{tab:3}
\end{table}

\begin{figure}
\includegraphics[height=3.5cm]{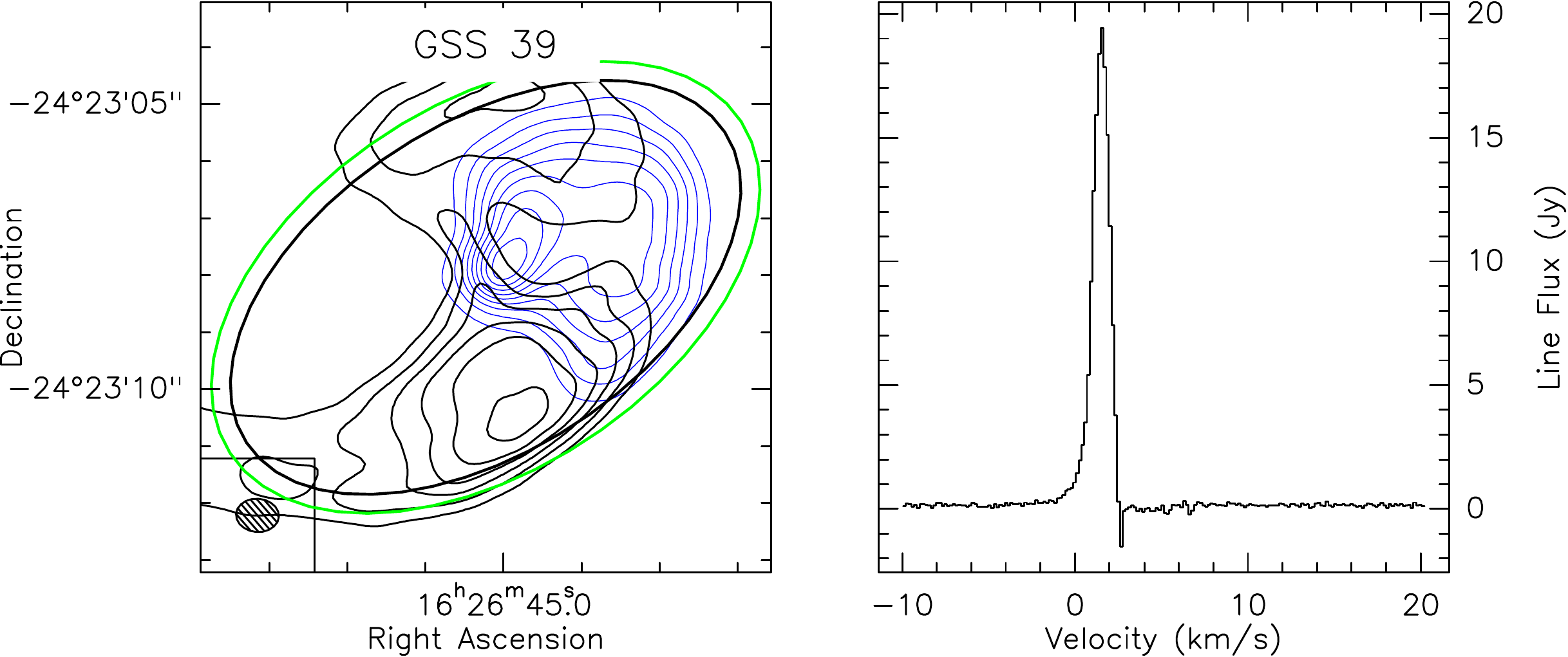}\hspace{0.5cm}
\includegraphics[height=3.5cm]{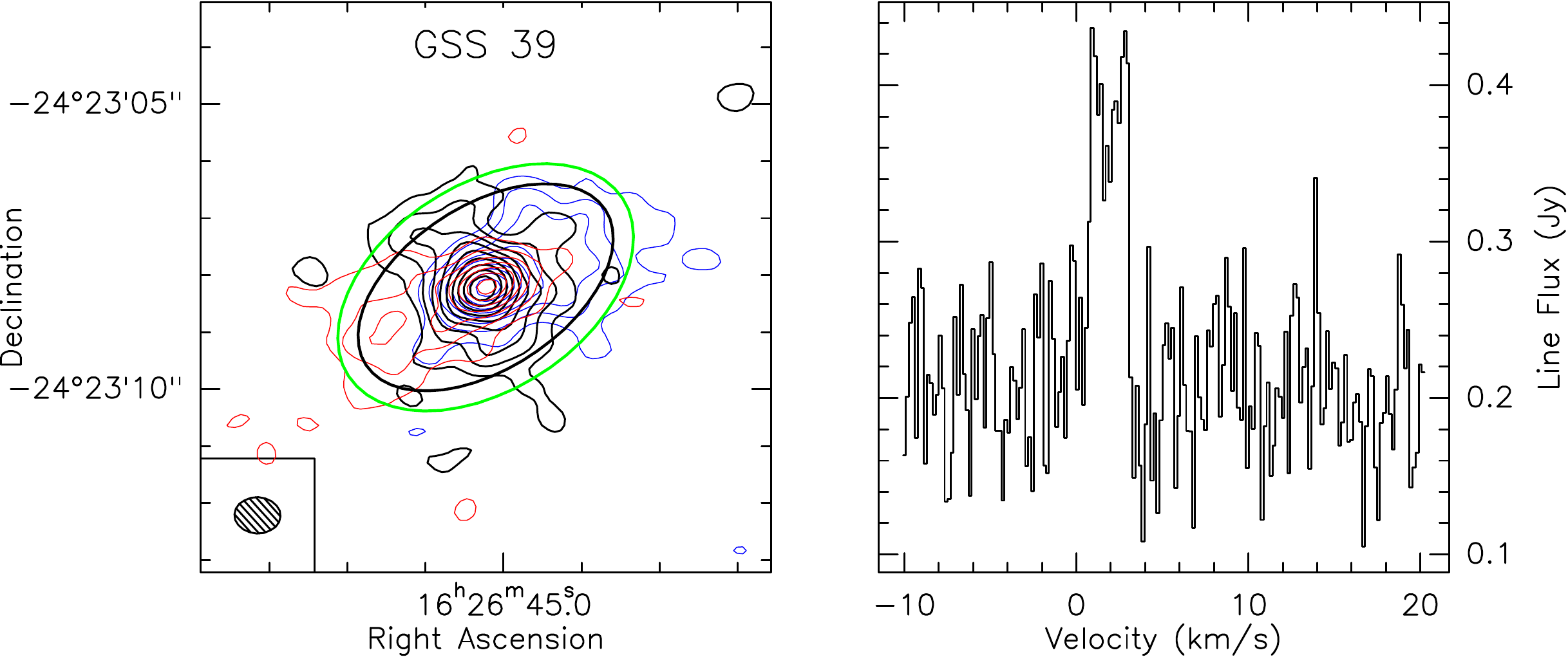}
\includegraphics[height=3.5cm]{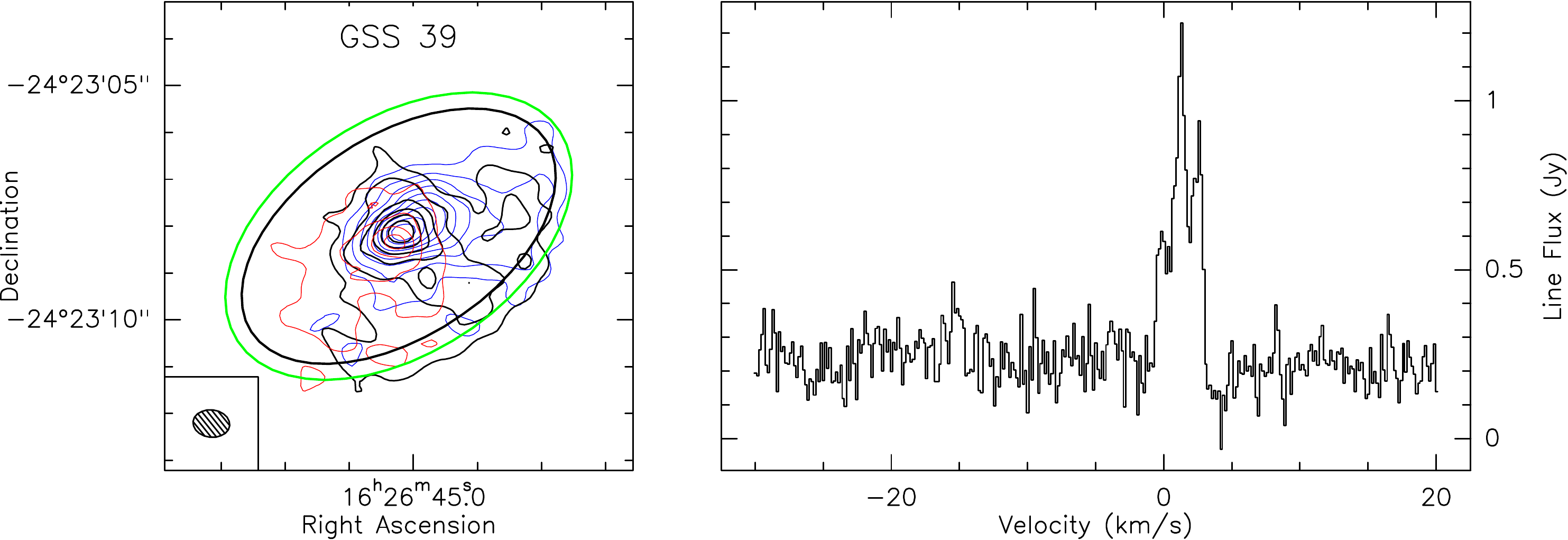}
\caption{Molecules towards GSS 39: CO (top left), H$_2$CO (top right), CN (bottom).
In the maps, red (resp. blue) contours indicate red-shifted (blue-shifted) emission, black
contours emission near the systemic velocity. The black ellipse is the location
of the disk outer
 radius, and the green ellipse the region used to derive
the integrated spectra. The CN lines have hyperfine structure.
\ifdefined\PREPRINT
  See Appendix for other sources.
\else
  \textbf{The complete figure set for all sources (9 images) is available in the online journal}
\fi
.}
\label{fig:gss39}
\end{figure}

We analyzed the data for disk parameters using the
{\it DiskFit} tool \citep{Pietu+etal_2007} exactly as in our earlier work 
\citep{Guilloteau+etal_2014} and  give details in Appendix A.
Table \ref{tab:3}  presents the results of ALMA Cycle 2 observations of the
stars in Taurus and Ophiuchus.  In our L1495 sample, ALMA did not
detect a signal at V410 X-Ray 1, confusion at MHO 1-2, and
confusion and weak signal at CIDA 1 prevented  reliable measurements.
We discuss the confusion problem further in \S 5.4.
No line emission was detected at FN Tau despite its strong continuum so
we were unable to measure its mass.
In L1688 we were unable to measure masses of WL 18, WL 14, and
GY 284 because the line emission of their disks proved too weak for detection
with the observational set-up we used.  Spectral line images of the 
detected disks are given in  
\ifdefined\PREPRINT
  Figs. \ref{fig:gss39} and \ref{fig:gss_26}-\ref{fig:mho_2}. 
\else
  \textbf{Figure Set \ref{fig:gss39}}.
\fi 

In Table \ref{tab:3} Columns 1
and 2 give the sample number  and name.  We assumed that the entire sample is
associated with L1495 at its distance of 131 pc (Col. 3). We did not propagate
uncertainties in the distance. Col. 4 lists
the lines analyzed. Col. 5 gives the
velocity of the line center with respect to the local standard of rest.
Cols. 6 and 7 give the disk inclination as measured for the line, $i_{line}$,
and continuum emission, $i_{cont}$ ($i=0$ corresponds to face-on).
Because the continuum emission arises from dust in the disk, its
emission region is more compact than that of the  line emission.
Hence we were not able to measure $i_{cont}$ reliably for all disks.
Because an interferometer provides spatial measurements on an angular scale,
radial distances in the disk scale as distance D to the star
and because the disk is seen at an inclination along the line-of-sight,
the velocity measured is the radial velocity.  Therefore the measured
mass scales as $M_* \propto D/sin^2 i$ 
\citep[see for example Table 3 in][]{Dutrey+etal_2003}. $R_\mathrm{out}$ (col. 8) is the
outer radius of the line emission  and $M_*$~ is the measured mass (col. 9).
For stars with precise distances, uncertainties of the inclination
dominate the mass uncertainty.  When only the average distance is
known, we use the distance-independent parameter $L/M^2$~to present the
results on a modified Hertzsprung-Russell Diagram.

\parindent=0.5in
Table \ref{tab:4} gives the  observational results for the Akeson-Jensen sample.
Its format is identical to Table \ref{tab:3} except it also
lists the binary separations (col. 3) as given
by AJ14;  ``S'' designates the stars they identified as probably single.
A and B  denote the primary and secondary.
We were able to measure masses of both components for
HK Tau, as did \citet{Jensen+Akeson_2014} also, and for DK Tau.

Table \ref{tab:4} provides disk inclinations for several stars which AJ14 describe as
point sources.  AJ14 analyzed the data in the image plane, 
while we derived disk sizes and orientations by fitting a 
disk model in the UV plane, fitting both frequencies simultaneously.  
We list the 
masses at 140 pc except for GK Tau and HO Tau as discussed in \S 2.2.  
Star names in bold have small values of $L/M^2$; see \S 4.2 and 5.3 for 
details. 
\ifdefined\PREPRINT
  Fig. \ref{fig:aj} (in Appendix) 
\else
  %
  \textbf{Figure Set \ref{fig:aj} (in Appendix)}
\fi
shows the disk emission in CO 3-2 towards the observed sources.

\begin{table}
\caption{Measured Parameters in Akeson\& Jensen Sample }
\begin{tabular}{rrrrrrrrrr}
\hline
ID & Name &sep.&Dist& Lines& V$_{LSR}$&$i_{line}$&$i_{cont}$&$R_{out}$& M$_{*}$ \\
   &      &arc sec& pc &      & km\,s$^{-1}$     &  deg     &  deg     &  AU  & \msun \\
\hline
\multicolumn{10}{c}{{\bf Bold} identifies stars with small $L/M^2$ values, see \S 4.3}\\
1 &{\bf FV Tau A}&0.70&140&CO&  $7.30\pm0.30$&$81\pm2$&$79\pm7$&$57\pm2$& $2.30\pm0.17$  \\
2 &{\bf FX Tau A}&0.68&140&CO& $6.70\pm0.08$&$40\pm4$&$20\pm20$&$40\pm10$& $1.70\pm0.18$   \\
3 &{\bf HBC 411 B}&2.8&140&CO& $5.50\pm0.10$&$25\pm1$&        &$33\pm2$& $2.05\pm0.20$   \\
4 &{\bf CIDA-9 A} &2.2&140&CO& $6.48\pm0.01$&$33\pm2$&$34\pm6$&$192\pm2$& $1.08\pm0.20$   \\
5A & HK Tau A&2.3 &140&CO& $5.98\pm0.04$&$51\pm2$&$54\pm7$&$90\pm10$& $0.58\pm0.05$   \\
5B &{\bf HK Tau B}&&140&CO& $6.43\pm0.05$&$81\pm2$&        &$120\pm10$& $1.00\pm0.03$   \\
6  & IT Tau B&2.4&140&CO& $6.40\pm0.30$&$66\pm12$&        &$50\pm3$& $0.50\pm0.08$   \\
7A & DK Tau A&3.4&140&CO& $5.57\pm0.20$&$20\pm5$&$65\pm8$&$38\pm4$& $0.60\pm0.14$   \\
7B &{\bf DK Tau B}&&140&CO& $6.30\pm0.30$&$55\pm10$&        &$60\pm12$& $1.30\pm0.30$   \\
8  & GK Tau &S&131&CO& $6.26\pm0.05$&$71\pm5$&        &$82\pm2$& $0.79\pm0.07$   \\
9  &{\bf HN Tau A}&3.0&140&CO& $5.65\pm0.13$&$75\pm4$&$50\pm15$&$52\pm10$& $1.57\pm0.15$\\
10 & V710 Tau A&3.2&140&CO& $6.64\pm0.04$&$46\pm5$&$47\pm4$&$82\pm6$& $0.66\pm0.06$   \\
11 & HO Tau & S&161&CO& $5.65\pm0.15$&$64\pm3$&$35\pm20$&$62\pm5$& $0.37\pm0.03$   \\
12 & DS Tau &S &140&CO& $5.68\pm0.02$&$71\pm1$&$69\pm6$&$164\pm2$& $0.73\pm0.02$   \\
\hline
\end{tabular}
\label{tab:4}
\end{table}

\section{Derived Masses in Taurus and Ophiuchus}

\subsection{Masses for ALMA Cycle 2 Stars in L1495
and Other Stars with Known Distances}

\label{sub:taurus:dist}

\begin{figure}
\figurenum{4}
\centering
\includegraphics[width=7cm]{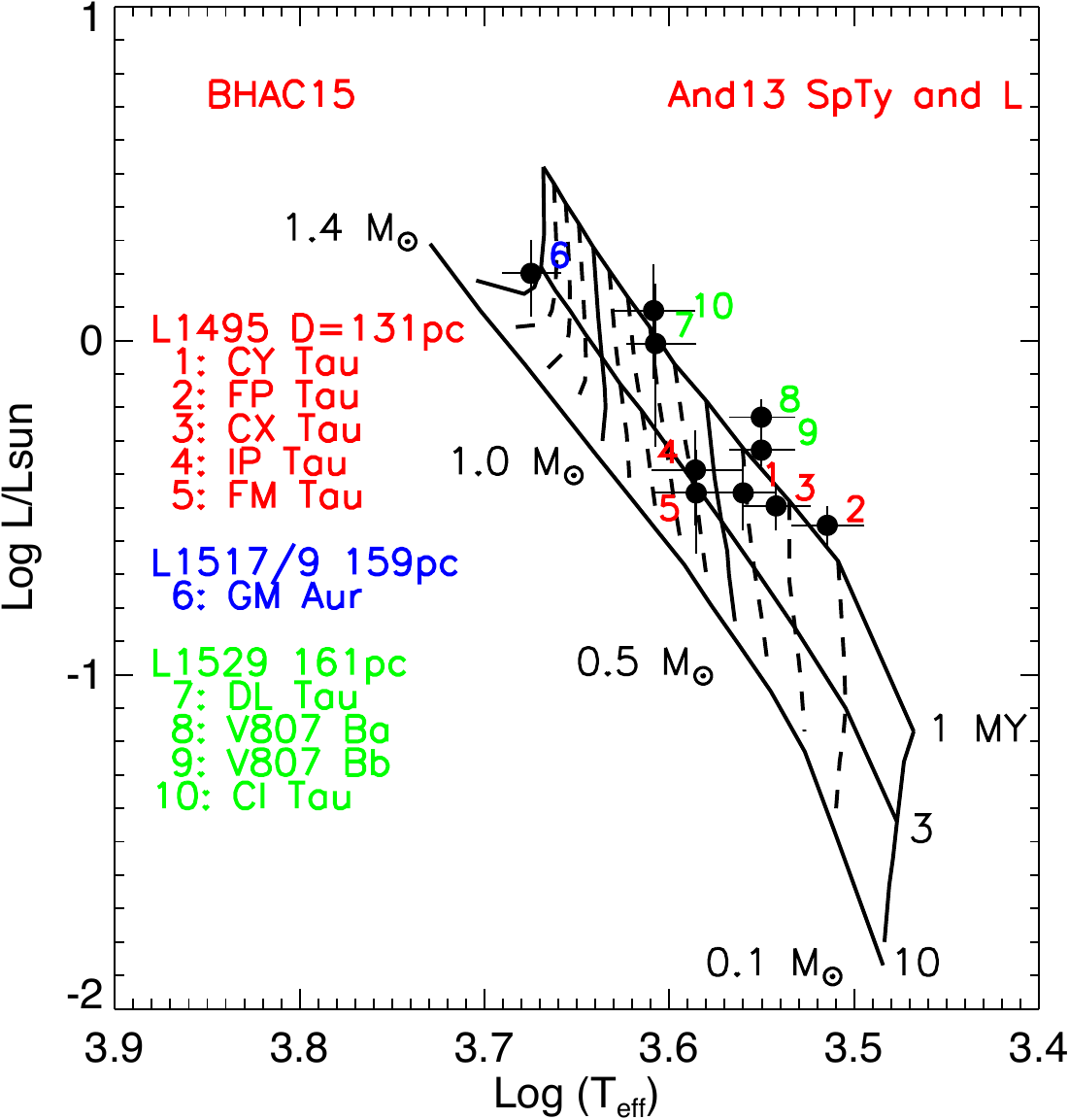}
\includegraphics[width=7cm]{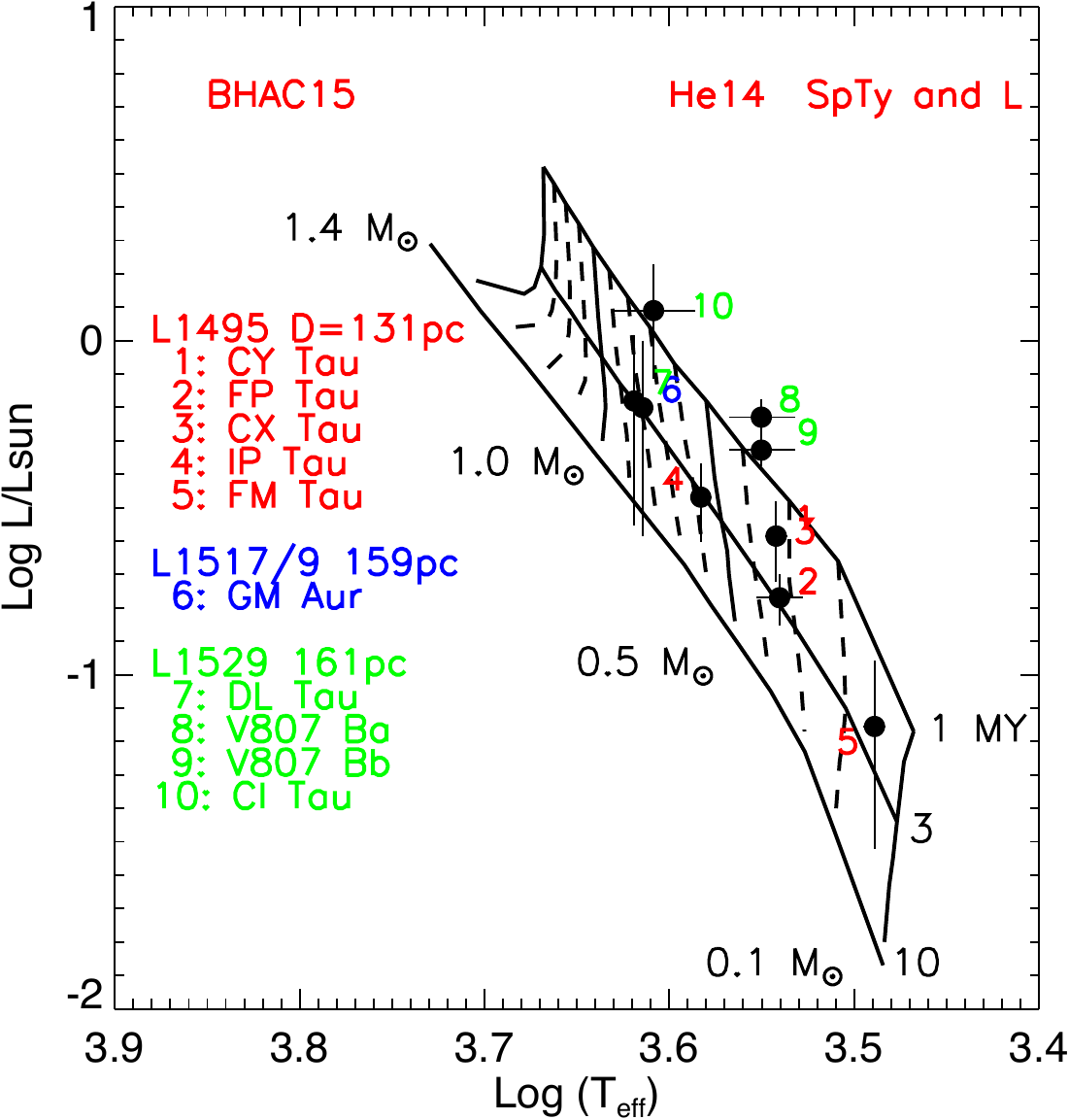}
\caption{An HRD for single PMS stars in the L1495 region in the Taurus SFR 
with absolute masses measured dynamically in our ALMA Cycle 2 observations.  
We include data for stars in L1517 and L1529 in Taurus from Guilloteau et
al. 2014 and Schaefer et al. 2012.  Left: luminosities and effective 
temperatures given by Andrews et al. (2013, And13), and right from Herczeg 
and Hillenbrand (2014, He14).  In the right panel, the positions of stars 
CY Tau (1) and CX Tau (3) are identical. }
\label{fig:4}
\end{figure}

Stars with measured mass to an internal precision smaller than $5\%$
and accurate and precise distances may be considered to have good values
of their absolute mass, an intrinsic property of the star.
This is the case for the stars in Table \ref{tab:3}  and we may analyze them 
on a conventional Hertzsprung-Russell diagram (HRD). Fig.~\ref{fig:4} places the stars 
on HRDs using the ($L$,{\teff}) values of And13 and He14.
The figures also include several other stars in Taurus with masses measured to
internal precision better than 5\% and with reliable distances that we have
published previously:  the single stars GM Aur, DL Tau,  and CI Tau
\citep{Guilloteau+etal_2014} and the binary 
V807 Tau Ba, Bb~\citep{Schaefer+etal_2012}. \citet{Schaefer+etal_2012} 
used a distance 140 pc for V807 Tau; here we use the 161 pc
distance of L1529. The figures use evolutionary models calculated by 
\citet[][BHAC15]{Baraffe+etal_2015}. 

\begin{figure}
\epsscale{0.8}
\figurenum{5}
\centering
\includegraphics[width=2.77in]{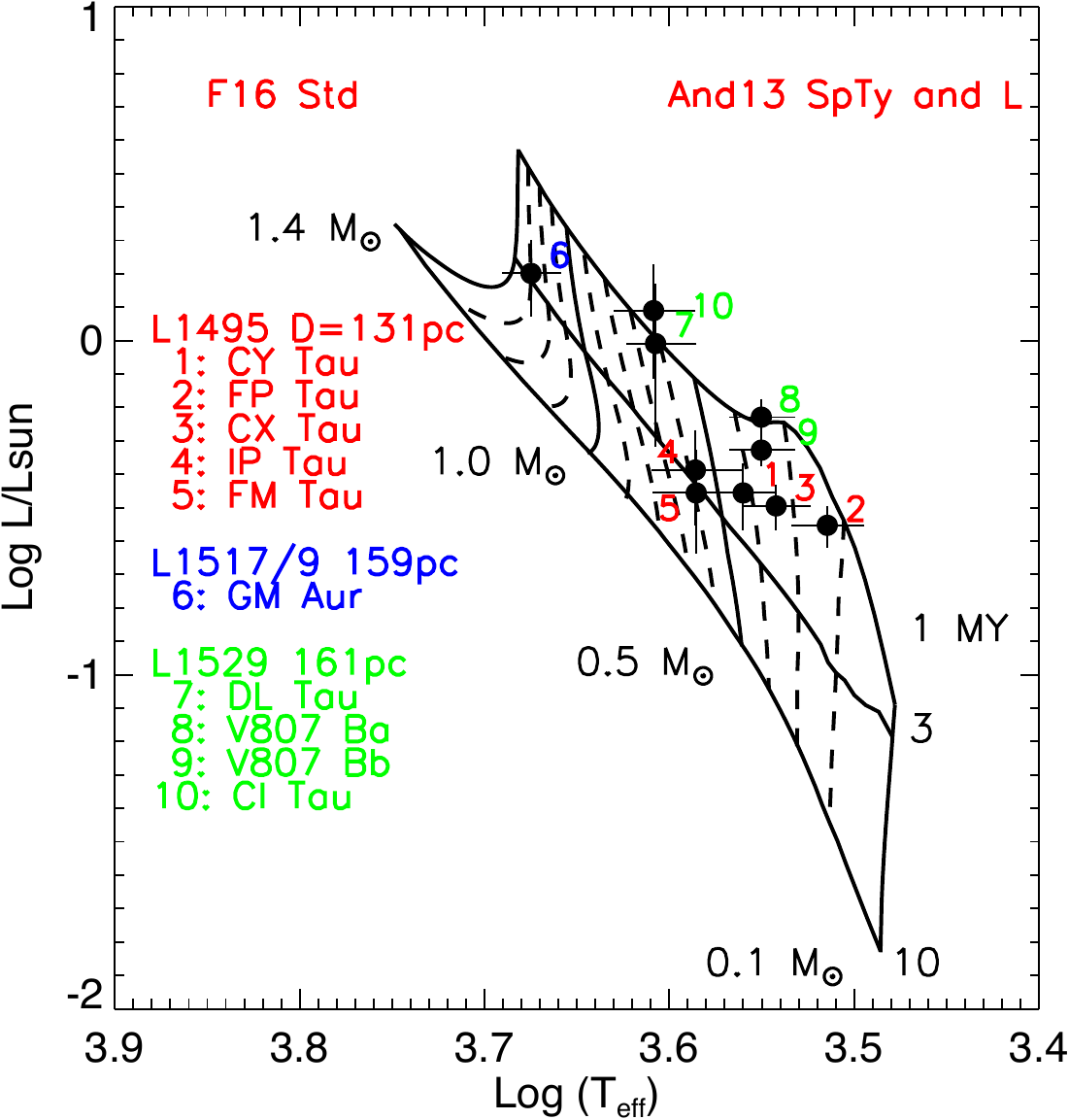}
\caption{Identical to Fig.~\ref{fig:4} except that the HRD uses 
the non-magnetic evolutionary models calculated by Feiden (2016, F16) 
and only the luminosities and effective  temperatures given by And13. }
\label{fig:5}
\end{figure}

 More recently, \citet[][F16]{Feiden_2016} published models of PMS evolution
that include models with and without internal magnetic fields.
It is of interest to apply F16's non-magnetic models  to the stars with 
well-determined absolute masses.  We do this in Fig.~\ref{fig:5} for the same 
stars as in Fig.~\ref{fig:4} 
and use And13's stellar parameters. Fig.~\ref{fig:5} is  nearly identical to Fig.~\ref{fig:4} 
(left). The only significant difference is seen in the 1 MY isochrone: 
stars with mass in the range  $\sim 0.2 - 0.5 \msun$ appear more 
luminous than in the BHAC15
models.  This may be the effect of different treatments of deuterium burning.
The effect is to make stars CY Tau, FP Tau, CX Tau, V807 Tau Ba and Bb appear
slightly older than the $\sim 1$MY indicated in Fig.~\ref{fig:4}.   Given the similarity
of the F16 and BHAC15 models we will continue to apply the BHAC15 models to
our mass measurements.

\begin{figure}
\figurenum{6}
\centering
\includegraphics[width=7cm]{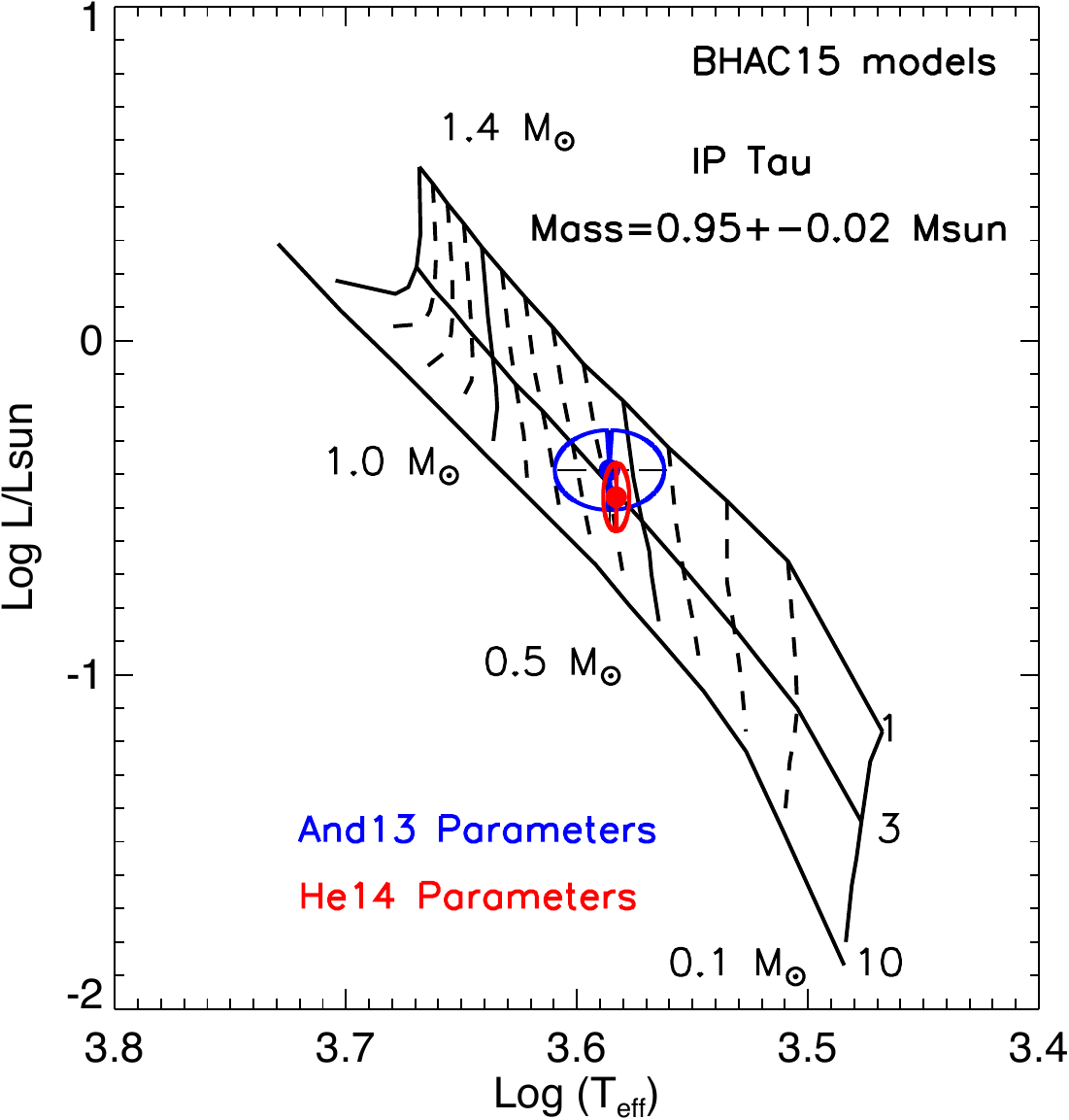}
\caption{The HRD shows IP Tau plotted at the \teff\ and $L/L_{sun}$ 
given by \citet[][And13]{Andrews+etal_2013} ({\it blue}) 
and \citet{Herczeg+Hillenbrand_2014} ({\it red}) and uses their
uncertainties to calculate the uncertainty ellipses. The uncertainty 
ellipse is used as a unit for measuring the distance between 
IP Tau's stellar parameters 
and the BHAC15 evolutionary track corresponding to its absolute 
dynamical mass, $ 0.95 \pm 0.02$ \msun. }
\label{fig:6}
\end{figure}

\begin{table}
\caption{Results: Comparison with Tracks  in L1495 and L1688}
\begin{tabular}{lr|rrr|rrr}
\hline
\hline
Name    &Mass (\msun) & And13&SpTy   & Age (MY)& He14&SpTy& Age (MY)\\
&&\multicolumn{6}{c}{Stars in L1495}\\
\hline
CY Tau    &$0.31\pm0.02$& $< 1\sigma $&Ok&  2 &$< 1\sigma$&Ok& 2 \\
FP Tau    &$0.37\pm0.02$& $\sim 2\sigma$&C& 1 &$<1\sigma$&Ok& 3 \\
CX Tau    &$0.37\pm0.02$& $< 1\sigma $&Ok&  1 &$\sim 1\sigma$&Ok& 2 \\
IP Tau    &$0.95\pm0.02$& $\sim 2\sigma$&C& 3 &$\sim 4\sigma$& \textbf{C} & 3 \\
FM Tau    &$0.36\pm0.01$& $\sim 2\sigma$&H& 3 &$\sim 5\sigma$&C&2 \\
GM Aur    &$1.14\pm0.02$& $\sim 2\sigma$&H& 3 &$\sim 10\sigma$&C&3    \\
DL Tau    &$1.05\pm0.02$& $\sim 2 \sigma$&C&1 &$\sim  4\sigma$&C& 3  \\
V807 Tau Ba&$0.86\pm0.03$& $\sim 5\sigma$ &C&$<1$&$\sim 5\sigma$&C&$<1$ \\
V807 Tau Bb&$0.73\pm0.03$& $\sim 4\sigma$ &C&1 &$\sim 4\sigma$&C& 1 \\
CI Tau    &$0.92\pm0.02$& $\sim 1\sigma$ &Ok&3&$\sim 1\sigma$&C& 1  \\
\hline
&&\multicolumn{6}{c}{Stars in L1688}  \\
Name    &Mass (\msun) & R10a&SpTy   & Age (MY)\\
\hline
GSS 26   &$1.51\pm0.02$& $ \sim 5\sigma $&C & 1/2&&& \\
GSS 39    &$0.47\pm0.01$& $< 1\sigma$    &Ok & 1 &&& \\
YLW 16C    &$1.80\pm0.10$&$>\sim 10\sigma$&C &1/2&&& \\
ROXs 25   &$1.10\pm0.07$& $<1\sigma$     &Ok &1/2&&& \\
YLW 58    &$0.09\pm0.01$& $<1\sigma$     &Ok &1  &&& \\
\hline
\end{tabular}
\label{tab:5}
\end{table}

The spacing of the evolutionary tracks in Fig.~\ref{fig:4} is 0.1 \msun.
Since this is greater than the precision of the measured
masses we can investigate whether the measured mass is consistent
with an evolutionary track of corresponding  mass.
In fact, Fig.~\ref{fig:4} shows that the comparison is often limited
by uncertainties in {\teff}.  We make a rough assessment 
of the level of agreement between the absolute mass and the 
evolutionary track (here BHAC15) as follows and illustrated in
Fig.~\ref{fig:6} for IP Tau (absolute mass = $0.95\pm0.02~\msun$).
We draw an error ellipse centered at the 
$(\log(\teff), \log(L/\lsun))$ for each star using the stellar parameters 
and their uncertainties quoted in Table \ref{tab:3}.  We estimate
the level of agreement by the distance in the HRD plane 
between the error ellipse  and the evolutionary track 
corresponding to the measured absolute mass. The mass 
uncertainty  is smaller than the filled circle used 
to plot the position of IP Tau in Fig.~\ref{fig:6} and all the stars
plotted in Figs.~\ref{fig:4}-\ref{fig:5}.  We emphasize that the masses in Table \ref{tab:3}
are measures of a fundamental property of the stars
completely independent of their $\teff$ and $L$.  As such
if we find, as we do, that the positions on the HRD
of most of the stars lie close to the evolutionary track
corresponding to their measured masses, but some do not,
the fault for the discrepant stars lies not with the models 
but probably mostly with their separately determined 
$\teff$s. Their luminosities are less likely at fault
because the evolutionary tracks are nearly 
vertical at young ages.  This approach assumes 
that the tracks have no uncertainty and we return to 
this point in \S 5.1. Table \ref{tab:5} summarizes 
the comparisons and lists the ages indicated. 

Fig.~\ref{fig:4} shows that masses measured for CY Tau, CX Tau, 
and CI Tau are consistent to about $1 \sigma$ with the ($L$,{\teff}) values
given by And13 and the BHAC15 models at ages 1-3 MY. The relatively
large discrepancy for  V807 Tau Ba and Bb, is probably attributable to
the limited data available to estimate their ($L$,{\teff})
\citep{Schaefer+etal_2012}.  We estimate also in Table \ref{tab:5}
the sense of the discrepancy with respect to the
spectral type for the star.  C indicates that the listed spectral type is too
cool and H that it is too hot.  In most cases the reported spectral type
is too cool, that is, the measured mass would be better represented by
a hotter {\teff}. In the case of IP Tau the discrepancy in $\teff$~could
be attributable to the possibility that it is actually a binary. IP Tau's continuum
image (Fig.~\ref{cont:taurus}) clearly shows an inner cavity.  We discuss this possibility further
 in \S 5.3.   Fig.~\ref{fig:4}  shows that the comparison with
He14's ($L$,{\teff}) values is similar. The measured masses of CY Tau,
FP Tau, CX Tau, and CI Tau are consistent with the tracks. The stars' ages also
lie in the range 1-3 MY.  That the differences are much
greater for FM Tau, GM Aur, and DL Tau, seems attributable
to the very small uncertainties He14 assigned to their spectral types.

\begin{figure}
\figurenum{7}
\centering
\includegraphics[width=7cm]{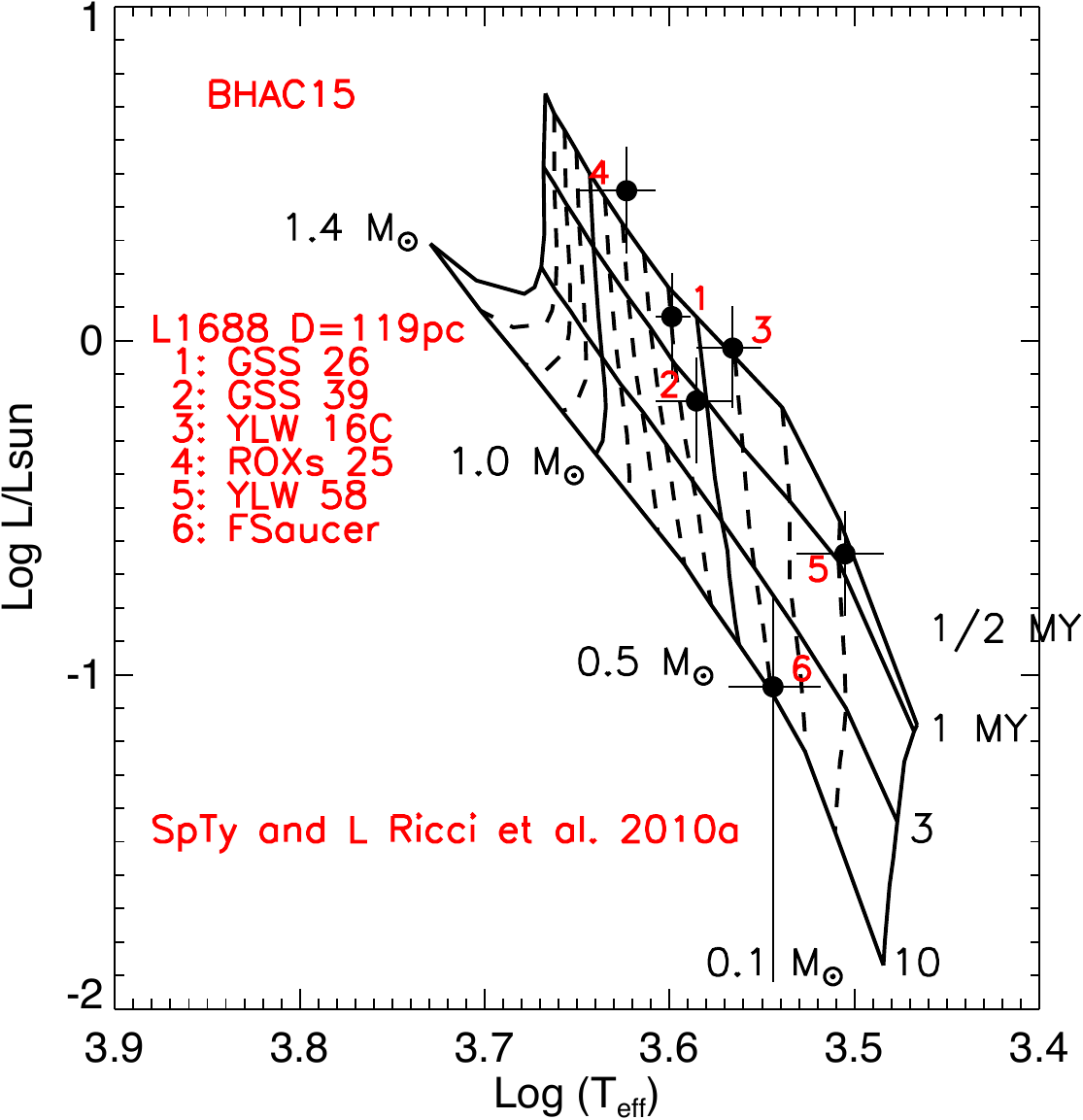}
\includegraphics[width=7cm]{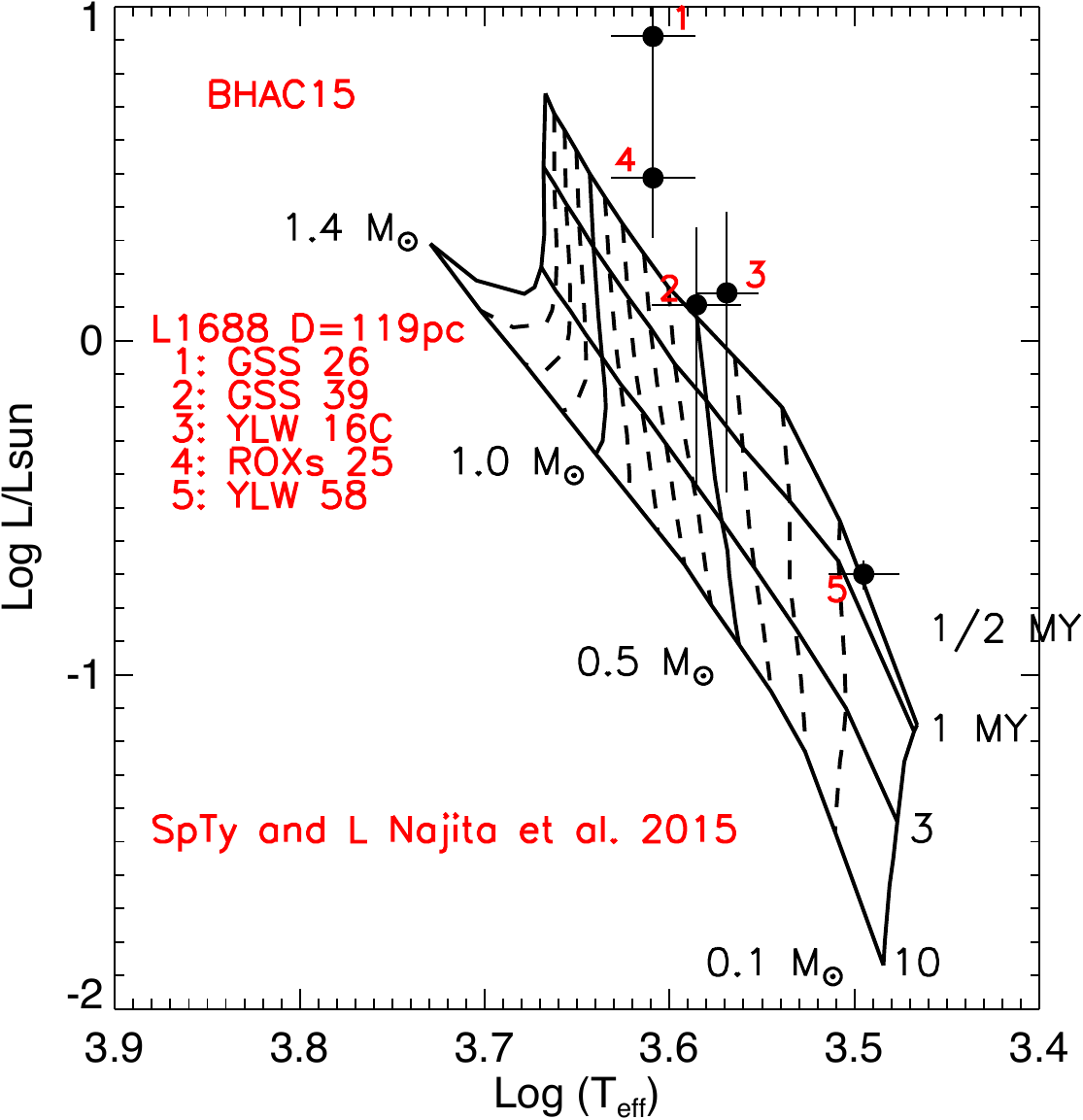}
\caption{An HRD for single PMS stars in the L1688 of the Ophiuchus SFR
with absolute masses measured dynamically in our ALMA Cycle 2 observations.
Left: the luminosities and effective temperatures are from  Ricci
et al. (2010a) for all stars except that the stellar parameters for the
Flying Saucer are from Grosso et al. (2003) and Pontoppidan et al. (2007),
see text. Right: the luminosities and spectral types
are from Najita et al. (2015) (See \S 4.3).}
\label{fig:7}
\end{figure}

\subsection{Masses for ALMA Cycle 2 - Ophiuchus Stars with Known Distances}

Table \ref{tab:3} indicates that the precision of the mass measurements is better
than $7\%$ for all but one of the stars; the exception is YLW 58
measured to $11\%$~ precision. The measured masses
of the stars in L1688 are therefore good estimates of their absolute
values and we plot their ($L$,{\teff}) values on conventional HRDs in
Fig.~\ref{fig:7}.

Because the spectral types reported by R10a and \citet{Najita+etal_2015} are identical or
nearly so, the plotted positions of the stars along the \teff\ axis are nearly the
same in both panels of Fig.~\ref{fig:7}.  Their differing positions in the two figures
are attributable to different luminosities given by R10a and \citet{Najita+etal_2015}.
The evolutionary tracks at the masses and ages of interest here, are nearly vertical.
The luminosity differences therefore correspond to differences in ages.
Comparing Figs.~\ref{fig:4} and \ref{fig:5}  with Fig.~\ref{fig:7}
it is clear that the stars in L1688
are younger than those in the Taurus SFR with ages less than 1 MY.

Table \ref{tab:5}  assesses the positions of the stars in the L1688 with respect to the
HRD in Fig.~\ref{fig:7}. We make the
comparisons with the $L$,{\teff} values given by R10a because the luminosities given by
\citet{Najita+etal_2015} indicates ages $<0.5$ MY; models for stars this young 
are uncertain and are not given by BHAC15.

\subsection{Masses for ALMA Cycle 0 Stars with Average Association
Distance}
\label{sub:taurus:aver}

\begin{figure}
\figurenum{8}
\centering
\includegraphics[width=7cm]{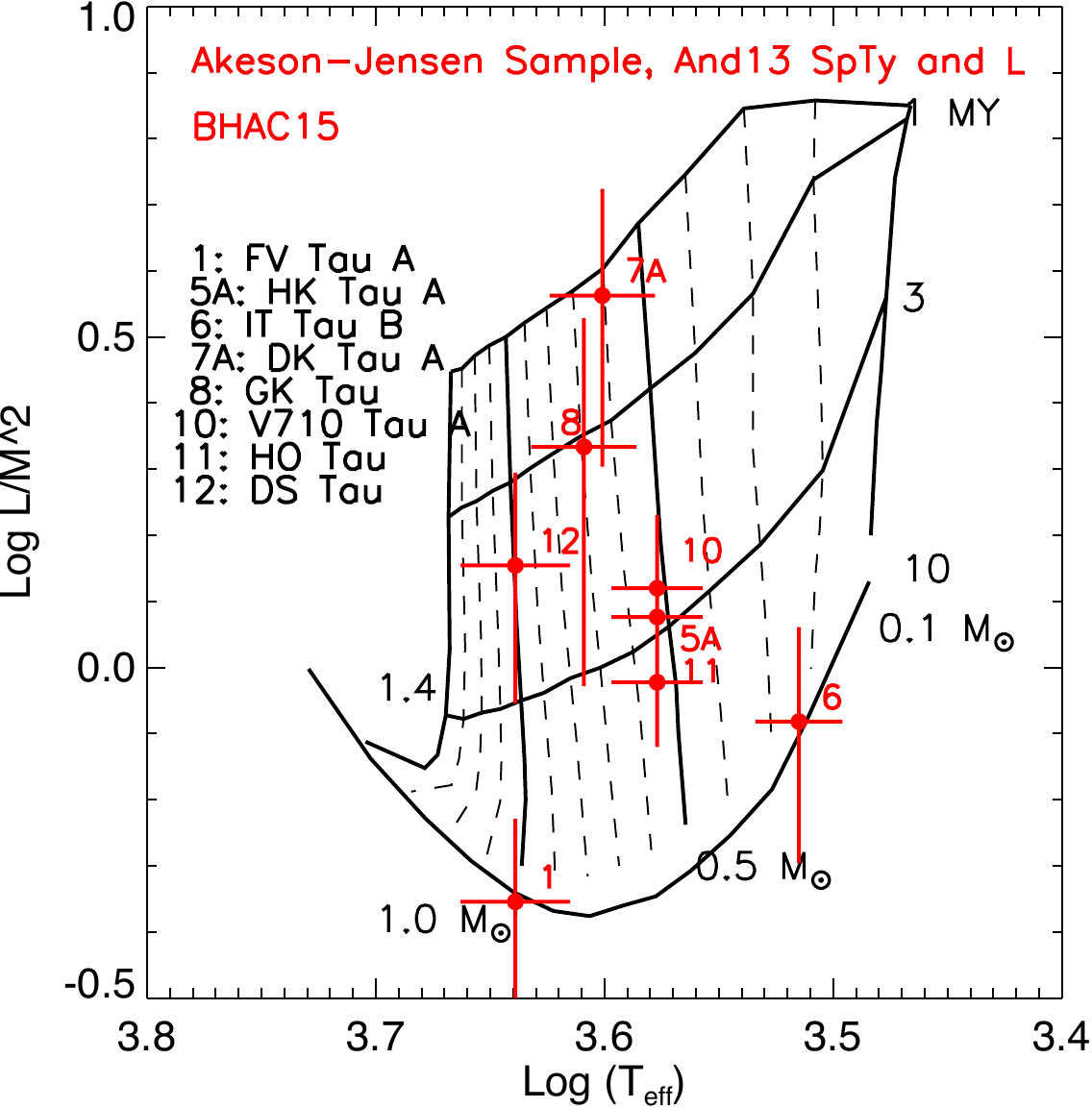}
\includegraphics[width=7cm]{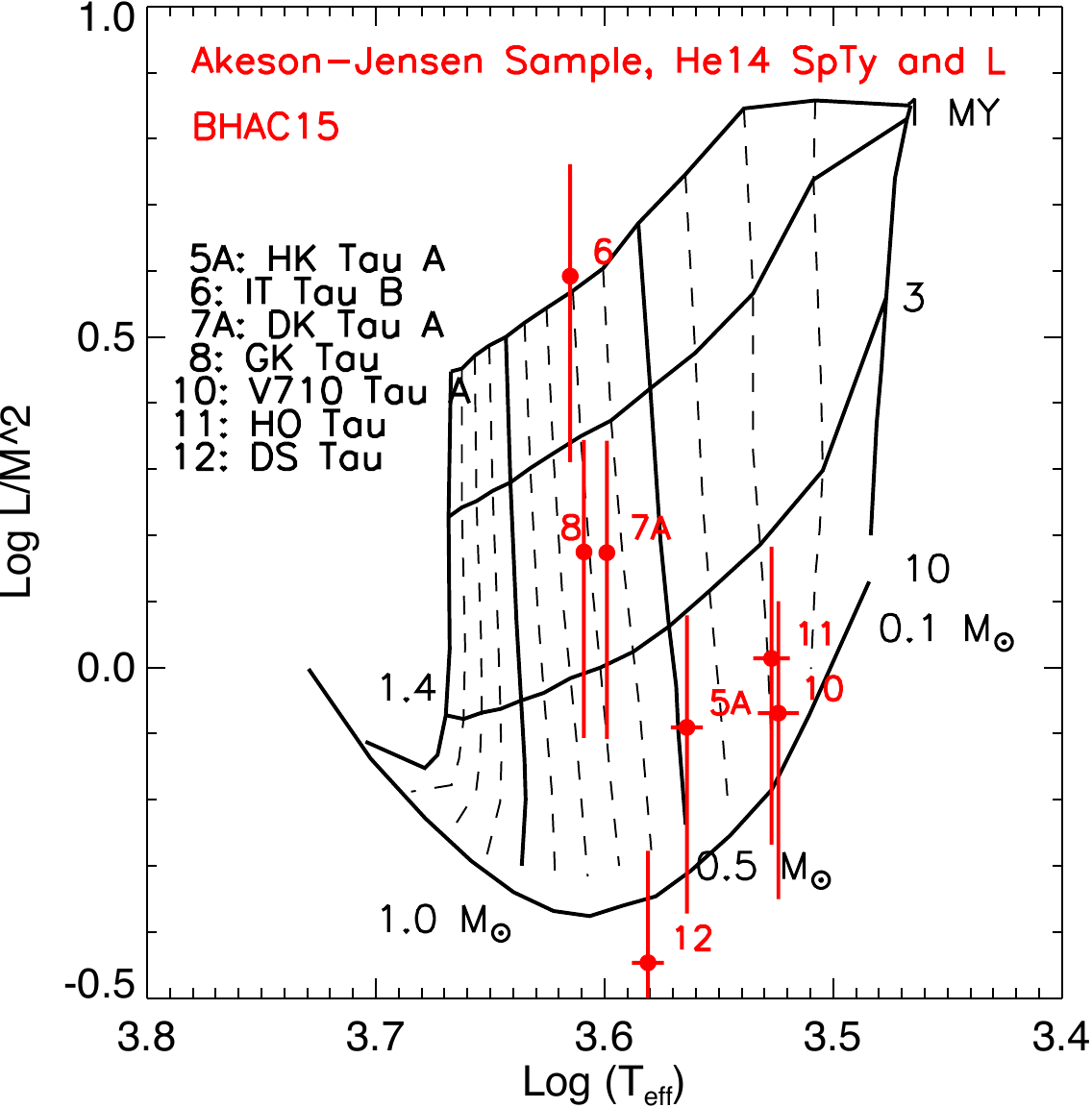}
\caption{A modified HRD (see \S \ref{sub:analysis} and \ref{sub:hrd}) 
for components of
PMS binaries distributed over the Taurus SFR with masses 
measured dynamically using
Akeson \& Jensen's archived ALMA Cycle 0 observations. 
Left: luminosities and effective
temperatures given by Andrews et al. (2013, And13), 
right: luminosities and effective
temperatures given by Herczeg and Hillenbrand (2014, He14).}
\label{fig:8}
\end{figure}

All the masses in Table \ref{tab:4} are measured to an internal precision
better than $25 \%$ and 5 are measured to better than $10 \%$.
Of the 9 binaries in Table \ref{tab:4}, we were able to measure
the masses of both components only for HK Tau and DK Tau.
Our derived masses for HK Tau A and B improve on the
initial determination from \citet{Jensen+Akeson_2014}'s. The
measured value of the inclination of B's disk, $i_{line} = 84\pm2^\circ$,
is in excellent agreement with \citet{McCabe+etal_2011} value
determined from high resolution IR images.  We also derive
a new value for the inclination  of A's disk,
$i_{A, line}=51\pm 2^\circ$, improving on the original determination
of \citet{Jensen+Akeson_2014}, $43\pm5 ^\circ$. We discuss this further 
in Appendix \ref{app:hktau}. 
The new values do not affect Jensen\&Akeson's finding that the
A and B disks are not co-planar, although the 
misalignment can be somewhat smaller.  For DK Tau A, the difference
in inclinations measured in the continuum and line emission is $45^\circ$.
DK Tau A has a small disk (col. 9) and a solution with 
$i_{line}\sim70^\circ$~ is 
also possible.  This would however imply an unacceptably low  mass 
of $\sim 0.15\msun$, which would be completely inconsistent with its late  
K spectral type (Table \ref{tab:2}).  Higher spatial and spectral observations 
are needed for DK Tau A.

Because most of the stars in the Akeson-Jensen sample do not have precise
distances, we  plot  them on modified HRDs in which the variable along the
abscissa is the distance independent parameter $L/M^2$.
As in the conventional HRD, a star's vertical position in the modified HRD
measures its age with respect to model isochrones.  The measured
masses in Table \ref{tab:4} are evaluated at 140 pc (except for GK Tau and HO Tau).
Their actual values may differ by $\pm \sim 14\%$ reflecting the dependance
of the measured mass on distance and the known spread
of distances, $\pm \sim 20$pc of members of the Taurus SFR.

Fig.~\ref{fig:8} uses the stellar parameters of And13 and He14.
Of the 14 stars in Table \ref{tab:4},
six (FX Tau A, HBC 411 B, CIDA-9 A, HK Tau B, DK Tau B, HN Tau A)
do not appear in Fig.~\ref{fig:8} (left) because their $L/M^2$~ values
are too small, which, taken at face  value would imply unreasonably great ages
$>>10$ MY. The same 6 stars and also a 7th, FV Tau A, do not appear
in Fig.~\ref{fig:8} (right). In fact, in Fig.~\ref{fig:8} (left), FV Tau A falls on the 10 MY
isochrone, an age that is suspect given the younger ages of the other stars.
We consider the 7 stars with small values of $L/M^2$ together
and identify them in bold in Table \ref{tab:4}.  We discuss these stars
further in \S 5.3.

\begin{table}
\caption{ALMA Cycle 0 Results: Comparison with Tracks and Age Estimates}
\begin{tabular}{lr|rrr|rrr}
\hline
\hline
Name    &Mass (\msun) & And13&SpTy   & Age (MY)& He14&SpTy& Age (MY)\\
\hline
HK Tau A   &$0.58\pm0.05$ & $<1 \sigma$    &Ok & 3 &$>3\sigma$     &C  & 2 \\
IT Tau B   &$0.50\pm0.08$ & $\sim 3\sigma$ &C  &10 &$>3\sigma$     &H  & $<0.5$ \\
GK  Tau    &$0.79\pm0.07$ & $ 1\sigma$     &Ok & 1 &$\sim 3\sigma$ &C  & 2 \\
V710 Tau A &$0.66\pm0.06$ & $ 1\sigma$     &Ok & 3 &$\sim 3\sigma$ &C  & 8    \\
HO Tau     &$0.37\pm0.03$ & $\sim 2\sigma$ &H  & 5 &$\sim 2\sigma$ &C  & 7  \\
DS Tau     &$0.73\pm0.02$ & $\sim 2\sigma$ &H  & 2 &$\sim 2\sigma$ &C  & $>10$ \\
\hline
\end{tabular}
\label{tab:6}
\end{table}

Table \ref{tab:6} provides an assessment of the parameters of the stars plotted
in Fig.~\ref{fig:8} as measured with respect to the BHAC15 evolutionary models
in the same format as Table \ref{tab:5}. We omit DK Tau A because of the difference in its
inclination as measured in the continuum and lines.
The number of stars whose spectral type is too cool with respect to the
evolutionary track corresponding to its mass is somewhat greater than
the number that is too hot, similar to the situation  in L1495 and L1688
described in Table \ref{tab:5}. Also, there is a greater dispersion in ages among 
the stars in Table \ref{tab:6}.  Both effects could be the result of the technical 
difficulties of measuring spectral types  and luminosities in binaries. 
The measured masses of two stars,  GK Tau and HO Tau, are absolute 
because their distances are provided by association with  L1495 and 
L1529, respectively. They lie within $\sim 1 $ and $\sim 2 \sigma$ 
of the tracks indicated by their mass and their ages  are 
$\sim 1$ and $\sim 5$ MY, respectively. We made
no entry for FV Tau A because its $L/M^2$ is anomalously small.

\section{Discussion}

\begin{table}
\caption{Assessment of $M_{*}$, $L$, \teff\ and Models}
\begin{tabular}{ll|ll}
\hline
\hline
\multicolumn{4}{c}{Dynamical masses are given in parentheses}\\
\multicolumn{2}{c}{Known Distance} &\multicolumn{2}{c}{Average Distance}\\
Agreement $\le 1 \sigma$ & $\le 2\sigma$&Agreement $\le 1\sigma$ &$\le 2 \sigma$\\
\hline
YLW 58 (0.09) & FM Tau (0.36) &  DM Tau (0.53)    & GO Tau (0.48)\\
CY Tau (0.31) & FP Tau (0.37) &  V710 Tau A (0.66)& DS Tau (0.73)\\
CX Tau (0.37) & HO Tau (0.37) &  LkCa15 (1.01)    & IQ Tau (0.79)\\
GSS 39 (0.47) & GM Aur (1.14) &  HK Tau A (0.58)  &              \\
GK Tau (0.79) & IP Tau (0.95) &                   &              \\
CI Tau (0.92) &               &                   &              \\
DL Tau (1.05) &               &                   &              \\
ROXs 25 (1.10)&               &                   &              \\
\hline
\end{tabular}
\label{tab:7}
\end{table}

\subsection{Comparison  of Measured Masses and Evolutionary
tracks on the HRD}
\label{sub:hrd}

Tables \ref{tab:5} and \ref{tab:6} assess the measured stellar masses and
effective temperatures with respect to the evolutionary 
tracks and present estimates of the stellar ages. This 
information is summarized in Table \ref{tab:7} for stars 
with and without known distance.  For the stars in Taurus, 
Table \ref{tab:7} extracts assessments using And13's parameters only.
The assessments using He14's parameters are similar
but with greater discrepancies owing to their very small 
uncertainties on \teff . For the stars in Ophiuchus, 
we made  assessments with respect to Ricci et al. (2010a)'s 
parameters.  Table \ref{tab:7} includes 4 stars DM Tau, LkCa15, GO Tau
and IQ Tau whose parameters are reported in  
\citet{Guilloteau+etal_2014}.  Their masses are measured
to an internal precision $<5\%$ but, because the distances
to the groups in which they lie are unknown the masses
are evaluated at 140pc.  We estimated their separation 
in the ($L/M^2,\teff$) from the evolutionary track
corresponding  to their mass in the same manner as in 
Tables \ref{tab:5} and \ref{tab:6} \citep[e.g. Fig.~7 in][]{Guilloteau+etal_2014}.
The evolutionary tracks used in \citet{Guilloteau+etal_2014}
are essentially F16 non-magnetic tracks which in the mass range of
these stars is very similar to those of BHAC15 (see \S 4.1
and Fig.~\ref{fig:5}). 

 The stars with known distance are listed in Cols. 1 and 2
of Table \ref{tab:7}.  These masses are model independent in that they
depend only on the interferometric observations and the cluster distances.  
Thus, the masses in Cols. 1 and 2 are absolute, that is, an 
intrinsic property of the stars.  All have mass precisions $<11\%$ 
and most $<6 \%$. Cols. 1 and 2 list the stars which lie
within $1\sigma$~and $2 \sigma$~, respectively, of the evolutionary
tracks corresponding to their mass. The measured mass is indicated
in parentheses.  For the stars in Col. 1, it is truly significant
that the PMS evolutionary tracks and the measured absolute
masses and {\teff}s are consistent to $1 \sigma$ in the range 0.09
to 1.1 ~\msun~ and at ages $<3 $ MY. 

The stars in cols. 3 and 4  have mass measured to an internal 
precision $10\%$ or better. When accurate and sufficiently precise 
{\it GAIA} parallaxes for these stars become available, their 
masses will become absolute. These stars will then enable further 
tests of the evolutionary models. Columns
3 and 4 include stars with $L/M^2$ large enough (ages sufficiently young) 
to appear in Fig.~\ref{fig:8}. We discuss in \S 5.3 the 
stars with $L/M^2$ too small to appear in Fig.~\ref{fig:8}.

Where do the problems lie when there is a discrepancy of the stellar 
parameters with respect to the evolutionary tracks? Do they  lie 
with the measured mass, luminosity, spectral type, and hence 
effective temperature, and/or the evolutionary models? That the
BHAC15 and F16's non-magnetic models agree well gives us confidence 
that they  are a good starting point for the evaluation of our results.

Discrepancy in the HRD position of a star can be attributed either to its
measured mass or spectral type and hence \teff .  The luminosity does
not enter because a  young star contracts at almost constant \teff .
Uncertainty in luminosity produces an age uncertainty.  We are confident
of the mass measurements of the stars in cols. 1 and 3 because their disks
are positioned such that their inclinations are not small and are sufficently
extensive that we were able to measure the inclinations and rotation
accurately. This suggests that the  departure of a star's
\teff~from the model value is responsible for the discrepancy.
These departures include the aleatory parameters that can vary from star
to star according to its circumstances determined by chance.  These
include the accretion luminosity, stellar activity, starspots and
their time variation.

\subsection{Ages in Taurus and Ophiuchus SFRs}
\label{sub:ages}
Most of the stars in Taurus considered here have ages in the range 1 to 
3 MY.  Although the luminosity uncertainties for the
very obscured stars in L1688 can be very much greater than for the 
less obscured stars in Taurus, all the stars in Fig.~\ref{fig:7}
 except the Flying Saucer indicate an age younger than 1 MY.  Because 
the uncertainty of the Flying Saucer's luminosity is especially 
large and its spectral type uncertain, we regard its present position 
on the HRD as preliminary. That the stars in L1688 are younger than 
those in L1495 supports the finding that the Ophiuchus SFR 
is younger than the Taurus SFR  \citep{Kenyon+etal_2008,Wilking+etal_2008}.

These estimated ages are based on the BHAC15 models that do not include
the effects of internal magnetic fields.
F16 and \citet{MacDonald+Mullan_2017} showed that the pressure of 
magnetic fields included in the convective regions of PMS stars 
slows their contraction.  Thus, given a star's luminosity and effective 
temperature, the star will appear older when magnetic effects are 
considered.  The effect on estimated  age is mass dependent and
a full analysis will have to await a larger sample of measured masses
than presently available.  

\subsection{Stars with Small $L/M^2$~Values}
\label{sub:low-lum}

The  $L/M^2$ ~values of 7 stars identified in bold in Table \ref{tab:4} 
are so small that they imply  ages unrealistically large for members of 
the Taurus SFR.  They can be smaller than expected either because their 
masses are large or because their luminosities are small for their mass, 
or both. Two characteristics of this sample stand out. Their masses are 
all greater than the masses of those that do appear in Fig.~\ref{fig:8};
all are greater than 1 \msun~ while the largest mass of the plotted stars 
is $0.79\pm 0.07$~\msun.  Binary 
separations may also play a role but a lesser one.  The components of 
two binaries with the smallest separations, FV Tau and FX Tau, $<1''$, 
have small $L/M^2$ ~values while three of the stars that AJ14 found are
singles, GK Tau, HO Tau, and DS Tau, appear in Fig.~\ref{fig:8}.
However, other stars in both groups are in binaries that have separations 
$>2''$ indicating that more is involved than binarity.

It is not possible that FV Tau A (2.3 \msun), HBC411 B (2.07 \msun), and 
FX Tau A (1.7 \msun) are single stars because their luminosities are too 
small and {\teff}s are too cool for stars of such masses. It seems 
likely that the stars with small $L/M^2$ values are actually 
unresolved binaries or higher order multiples themselves.
With the evolutionary models as 
a guide we find that it is possible to account for the ($L$,{\teff}) 
of the stars with low $L/M^2$ values as unresolved multiples of 
lower mass stars. The details of the possible composite systems 
depend on whether the And13 or He14 parameters are used. 
Observations that can identify the hypothetical components 
by either radial velocity measurements or interferometric imaging
will elucidate the nature of these stars.

Further evidence for a higher multiplicity as the main cause of discrepancy
between spectral types and dynamical masses is provided by IP Tau, MHO 2,
and YLW 16c.  We drew attention (\S 4.1) to the cavity detected at the
center of IP Tau's continuum image.  Using its measured mass and luminosity
(either the And13 or He14 values) we find that its $L/M^2$~and $\teff$~ 
values would place it on or below the 10 MY isochrones on modified HRDs 
such as those in Fig.~\ref{fig:6}.  As for the other stars with small $L/M^2$~values, 
this supports the possibility that IP Tau is an unresolved binary.
The finding of a third component in the GG Tau Aa, Ab binary 
\citep{DiFolco+etal_2014}, namely that Ab is itself a close binary Ab1 and
Ab2, is an example of components yet to be found. 
 MHO 2 is surrounded by a striking circumbinary ring of dust emission
(Fig. \ref{cont:taurus}), from which an accurate orientation and inclination can be derived. 
CO J=2-1 emission is also detected, but strong confusion with the 
molecular cloud limits our ability to measure the dynamical
mass 
\ifdefined\PREPRINT
 (see Fig.\ref{fig:mho_2}). 
\else
  \textbf{(see Figure Set \ref{fig:gss39}.8).} 
\fi 
Yet, the presence of emission at sufficiently 
high velocities rule out very low mass: allowed values are in the range 
0.5 -- 0.8 \msun, while the spectral type is M2.5 \citep{Briceno+etal_1998}.
The measured mass of YLW16C, $1.80\pm 0.10$ \msun ~draws attention because it 
is the highest of those we measured in L1688 and its $\teff$ is too low
for its mass. Its $L/M^2$~ and $\teff$ would also place it below a 10 MY
isochrone in the modified HRD.  It seems likely that YLW16C is an unresolved
binary too.

\subsection{Spectral lines from Disks} 
\label{sub:lines}

A very serious obstacle for mass determination by the rotation of the host 
star's circumstellar disk is contamination of the disk line emission by 
molecular
clouds along the line of sight. To overcome this, we tested the detectability  
of the hyperfine-split CN 2-1 transition radiated by circumstellar disks.  We 
found that the CN lines were good tracers of disk emission of targets in the 
Taurus SFR and were free of molecular absorption
\citep{Guilloteau+etal_2013}. We  then used the Plateau de Bure interferometer 
to measure the masses of  11 stars in the Taurus SFR \citep{Guilloteau+etal_2014}.
However, \citet{Reboussin+etal_2015} showed that CN line emission from disks 
was more difficult to detect in the $\rho$ Oph region than in the Taurus SFR. 
Smaller disks, perhaps because of the much denser environment of L1688, 
different chemistry because of the younger ages were suggested as possible 
causes for this difference.

L1688 is significantly different from L1495 in at least two respects.  The 
number and number density of stars in L1688 is greater than in L1495
\citep{Wilking+etal_2008,McClure+etal_2010,Kenyon+etal_2008}.  Using 
their counts of stars and maps of the two regions, the stellar density is 
45 stars pc$^{-3}$ and 0.27 stars pc$^{-3}$ in L1688 and L1495 respectively.
The greater extinction to the stars in L1688 than in L1495 indicates that 
the density of interstellar material also is greater (McClure et al. 2010; 
And13; He14).  The different environments to which the disks are
exposed  may affect the abundances of CN in ways not yet explored.

Anticipating that the stars we planned to observed in L1688 would 
suffer strong line of sight contamination in CO J=2-1, our strategy 
with ALMA was to observe simultaneously in the CN 2-1 transitions 
and the CO 2-1 and  H$_2$CO lines.  Most disks in our sample were 
found to be smaller than those in the \citet{Guilloteau+etal_2013} 
sample. However, this is probably  a selection bias: the targets 
of our ALMA study have {\teff}s cooler than the stars in our 
pilot studies \citep{Guilloteau+etal_2013,Guilloteau+etal_2014}.
Indeed, there is no difference in disk size in our new sample between the
L1495 and L1688 regions. Nevertheless, CN is detectable in most 
sources in the L1688 region, but the N=2-1 line emission is 
heavily contaminated by apparent absorption from the molecular cloud, 
except for the Flying Saucer which lies in the outskirts
of the region. The apparent absorption is partly due to filtering of
the extended emission by the interferometer, but true absorption 
against the strong continuum of some disks may also occur. The 
intensity ratio of the hyperfine components indicates
that the CN line from the molecular cloud has moderate optical thickness.
A similar contamination is seen towards MHO 1-2 in the Taurus SFR. In 
other Taurus sources, CN is barely detectable because of the limited 
brightness sensitivity of the observations, a result of observing 
with longer baselines than initially requested in the proposal.

Contamination in H$_2$CO is much less significant than in CN, but still 
detectable in L1688. In practice, the precision of our mass measurements 
relies mostly on the higher signal to noise obtained with the CO J=2-1 
transition, but the other lines nevertheless play a key role in allowing 
a more accurate measurement of the systemic velocity.

\section{Summary}

1)  Our ALMA Cycle 2 program has yielded absolute dynamical masses
for 5 stars in L1495 and 6 in L1688.  7 of these are at masses $<0.6$~\msun~,
and of these 6 are at precision $<5\%$.

2) We find good  agreement with the measured parameters of 8
stars with the BHAC15 and F16 evolutionary models over the mass 
range $\sim 0.09$ to 1.10 \msun.  It seems reasonable to attribute 
the lesser agreement of 5 stars to the aleatory properties of 
PMS stars that can affect the \teff ~measurement.

3) Positions on the HRD with respect to the isochrones of the BHAC15 models
indicate the stars we observed in three Lynds clouds in Taurus have age 
1 to 3 MY and those in L1688 in Ophiuchus with reliable
luminosities are younger than 1 MY.

4) We also measured masses of 14 stars in the ALMA Cycle 0 archival data for
Akeson and Jensen's (2014, AJ14) study of disks associated with binaries in
Taurus. Most of the measured masses are greater than those in our
Cycle 2 program. We confirm AJ14's  measurement of the masses of
HK Tau A and B.

5) The masses measured for 7 targets in the AJ14 sample are sufficiently large
as to be inconsistent with their {\teff}s and luminosities.  The
most plausible explanation is that these components are actually binaries
or higher order multiples. Similar considerations suggest that IP Tau
and YLW 16c in our ALMA sample may also be unresolved binaries.

6) We detected strong contamination of disk emission in CN lines by the 
molecular cloud in L1688, and towards  MHO 1-2 in L1495. H$_2$CO is 
less affected, but fainter. Dynamical mass measurements in dense regions 
will require a combination of spectral lines with different opacities 
to overcome the contamination on one hand, and the sensitivity limitation 
on the other hand. 

\bigskip

\facility{ALMA}.

\acknowledgments
We thank the referee for a thorough reading of our paper and comments that
improved the presentation.
We thank R. Akeson and E. Jensen for their interest in having us
re-analyze their Cycle 0 data. We are grateful to I. Baraffe, G. Feiden, 
G. Herczeg, J. Najita and L. Ricci for their quick and thorough replies 
to our questions and especially to M. McClure
for a thorough assessment of the luminosity uncertainties in L1688. MS thanks
J. Toraskar for help with the HRD plots.

This paper makes use of the following ALMA data: ADS/JAO.ALMA\#2011.0.00150.S, \\
ADS/JAO.ALMA\#2013.2.00163.S and ADS/JAO.ALMA\#2013.2.00426.S. ALMA is a partnership of ESO
(representing its member states), NSF (USA) and NINS (Japan), together with NRC (Canada)
and NSC and ASIAA (Taiwan), in cooperation with the Republic of Chile. The Joint ALMA
Observatory is operated by ESO, AUI/NRAO and NAOJ.
This research made use of the SIMBAD database, operated at CDS, Strasbourg, France.
This research was supported by the "Programme National de Physique Stellaire" and the
"Action Sp\'ecifique ALMA" of INSU/CNRS (France).


\appendix

\section{Method of analysis}
We use the {\it DiskFit} tool \citep{Pietu+etal_2007} to derive the 
disk properties by fitting truncated power law disk models. The disk 
model is that of a flared disk with power laws for the temperature 
($T(r) = T_0(r/R_0)^{-q}$), surface density ($\Sigma(r) = 
\Sigma_0(r/R_0)^{-p}$) and scale height ($H(r) = H_0(r/R_0)^{-h}$) 
radial distributions, and sharp inner ($R_\mathrm{in}$) and outer 
($R_\mathrm{out}$) radii. The emission from the disk is computed using 
a ray-tracing method. For spectral lines, we assume that the velocity 
field is Keplerian, $v(r) = V_0 (r/R_0)^{-0.5}$ and use a constant 
local linewidth $\delta V$. Besides the above disk parameters, 4 
geometric parameters are required to derive the disk emission: the disk 
position ($x_0,y_0$), orientation PA and inclination $i$, plus the 
systemic velocity $V_\mathrm{sys}$ for spectral lines.

The $\chi^2$ difference between the predicted model visibilities (using 
the $uv$ coverage from the observations) and the observed ones is 
minimized using a modified Levenberg-Marquardt method, and the 
errorbars are derived from the covariance matrix. In practice, $R_{in}$ 
is set to a small value (0.1\,au), and the scale height is kept fixed 
in the process.

In this analysis, we are only interested in recovering the rotation 
velocity  $V_0$ ~at $R_0$ ~from which we derive the stellar mass 
$M_* = R_0 V_0^2 /G$, where $G$ is the gravitational constant. The validity of 
the errorbars was further assessed by running Monte Carlo Markov Chain 
(MCMC) in several cases and comparing their results to the simpler 
minimization method. The agreement is in general excellent because most 
parameters are only loosely coupled. Apart from the strong coupling 
between $V_0$ and sin($i$), since only the projected velocity is 
measured, the MCMC only revealed a weak coupling between 
$V_\mathrm{sys}$ and $V_0$.

This weak coupling is not insignificant because contamination by the 
foreground (or background) cloud emission prevents  using a fraction of 
the available velocities, especially in the CO J=2-1 line. Contaminated 
velocities are ignored in our minimimization process. However, in some 
cases (e.g. GSS 39), only one wing of the overall velocity spread of 
the disk can be seen, leaving a substantial uncertainty for 
$V_\mathrm{sys}$. We mitigated these problems by using the velocities 
derived from the CN or H$_2$CO lines, if bright enough (e.g. in the 
Flying Saucer or GSS 39).  We also ensured that the results did not 
depend critically on the masked velocity range.

The reliability of our measurements is confirmed in several ways. 
First, repeated observations (e.g. CX Tau observed in two ALMA 
projects) yield consistent results. Second, when more than one spectral 
line is useable, the results also agree (e.g. CO J=2-1 and CO J=3-2 in 
CX Tau, or CN and H$_2$CO in the Flying Saucer). Most importantly, 
consistent inclinations and orientations are derived from the continuum 
and the spectral line data.  DK Tau A is an exception attributable
to its small disk; see \S 4.3 for discussion. 
This is significant because these two types 
of observations are limited by different problems: the continuum data 
is affected by residual phase and amplitude calibration errors, while 
the line data is affected by bandpass calibration errors. Furthermore, 
the disk sizes are quite different in line and continuum, so that any 
bias due to an inappropriate disk model should have a different impact. 
We also expect dust to be settled towards the disk mid-plane, as found 
for example in the Flying Saucer by \citet{Guilloteau+etal_2016} and HL 
Tau by \citep{Pinte+etal_2016}. Molecules, however, sample different 
heights above the disk mid-plane. The agreement between the 
inclinations derived in different ways show that these structural 
differences do not dominate the current errors on the inclination.

The masses of a few sources had already been derived before. 
\citet{Pietu+etal_2014} reported a mass of $0.77\pm0.07$ for DS Tau (at 
140 pc)  from $^{13}$CO J=2-1 observations with the IRAM array, in 
perfect agreement with our new result. However, for CY Tau 
\citet{Guilloteau+etal_2014} cited a most likely mass of 0.48 \msun 
(after correction for the 131 pc distance), but the measurement was 
hampered by ambiguities in the inclination determination because of 
insufficient angular resolution. The much higher angular resolution 
used gives an inclination of $30^\circ$ and yields a lower mass.

\section{HK Tau}
\label{app:hktau}

HK Tau was analyzed by \citet{Jensen+Akeson_2014}. Overall,
we find similar results, but with improved uncertainties. 
Our analysis differs from theirs in 
several points. \citet{Jensen+Akeson_2014} adjusted an 8 parameter disk 
model, comprising PA and $i$, 3 parameters characterizing the CO 
surface density (using a viscous disk model shape), 2 parameters for 
the temperature, and the last one is the stellar mass. The disk is in 
hydrostatic equilibrium, with a constant CO abundance. They assumed 
that both stars have the same systemic velocity, and fix the positions 
from the centroid of the velocity integrated CO emission. The 
orientation and inclination of HK Tau B were taken from optical images. 
A model of the HK Tau B disk is first subtracted from the data before a 
model of HK Tau A is fit to the residual.
\citet{Jensen+Akeson_2014} only used velocities between 0.3–-5.4 
and 7.9–-11.3 km\,s$^{-1}$, ignoring the faint wings at higher 
velocities.

It is possible that the centroid of the CO emission derived by 
\citet{Jensen+Akeson_2014} is biased because of contamination at the 
surrounding molecular cloud velocity. Furthermore, in an optically 
thick disk (as expected for CO), the disk flaring produces an 
asymmetric emission because of the radial temperature gradient 
\citep{Dartois+etal_2003}, so that the centroid of emission is not on 
the disk center, but offset along the projected minor axis. Combined
with the assumption on velocities, this may have biased the derived
inclination and stellar masses.

To evaluate the possible biases, we use a similar, but more complete 
approach of separate disk fitting. We fitted one disk at a time,
and checked that the results did not depend whether the fit
used the residual from the other star disk fit, or the whole
original data set. We improved on AJ14 analysis by running MCMC chains using either the 
J=2-1 or the J=3-2 line, or both. Furthermore, our model uses a more 
comprehensive set of parameters, in particular leaving the positions 
and systemic velocity as free parameters.  We actually find different 
systemic velocities for both stars. Our derived inclination and 
orientations for HK Tau B is quite consistent with the optical results, 
and the two independent determinations (CO J=3-2 and CO J=2-1) of the 
orientation and inclination of HK Tau A yield consistent results (the 
continuum at 345 GHz is not sufficiently resolved for this purpose).

Our derived inclination of $51\pm2^\circ$ is slightly higher
than found by \citet{Jensen+Akeson_2014}. Interestingly, we found that
the derived inclination was dependent on the sharpness of the
surface density profile. The above value is for steep profiles
(the power law exponent of the surface density being $p=7$, 
essentially a sharp-edge disk), while for exponents $p<4$, there
is another solution for the inclination, $i\approx 65^\circ$. The derived
stellar mass remains unaffected, however. Furthermore, this
does not affect the main conclusion of \citet{Jensen+Akeson_2014} on 
the mis-alignment, which only drops down to $52$ or $58^\circ$ 
in the most extreme case instead of 60 or 68$^\circ$, 
the two values depending on which side of the HK Tau A disk is closest to us.

The disk around HK Tau A is smaller than that of HK Tau B, as would be
expected if caused by tidal truncation because of the stellar mass difference.
Interpreting the disk sizes as due to tidal truncation then implies a semi-major
axis of about 300-400 au for the orbit, close to the projected separation of 
300 au. However, the measured velocity difference is only 0.45 km\,s$^{-1}$, 
inconsistent with a circular orbit. A significant orbital eccentricity would 
be required to explain the disk sizes by tidal truncation.


\figsetstart
\figsetnum{\ref{fig:gss39}}
\figsettitle{Images and integrated spectra of the detected sources.}

    \figsetgrpstart
    \figsetgrpnum{\ref{fig:gss39}.1}
    \figsetgrptitle{Molecules towards GSS 39}
\figsetplot{GSS_39-co21-eps-converted-to.pdf}
\figsetplot{GSS_39-h2co-eps-converted-to.pdf}
\figsetplot{GSS_39-cn21-eps-converted-to.pdf}
    \figsetgrpnote{CO J=2-1 (top left), H$_2$CO (top right) and CN emission (bottom).
In the maps, red (resp. blue) contours indicate red-shifted (blue-shifted) emission, black
contours emission near the systemic velocity. The black ellipse is the location
of the disk outerradius, and the green ellipse the region used to derive
the integrated spectra. The CN lines have hyperfine structure.
}

    \figsetgrpstart
    \figsetgrpnum{\ref{fig:gss39}.2}
    \figsetgrptitle{Molecules towards the Flying Saucer}
\figsetplot{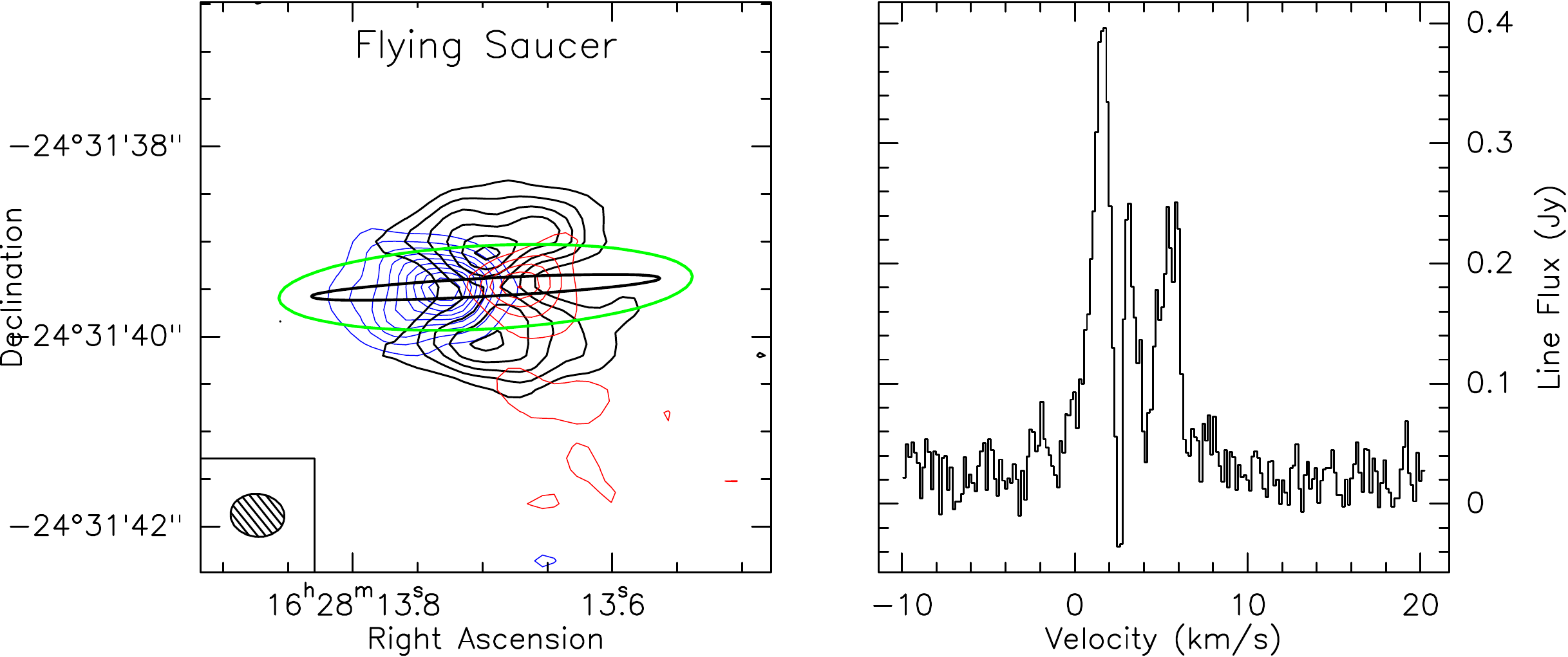}
\figsetplot{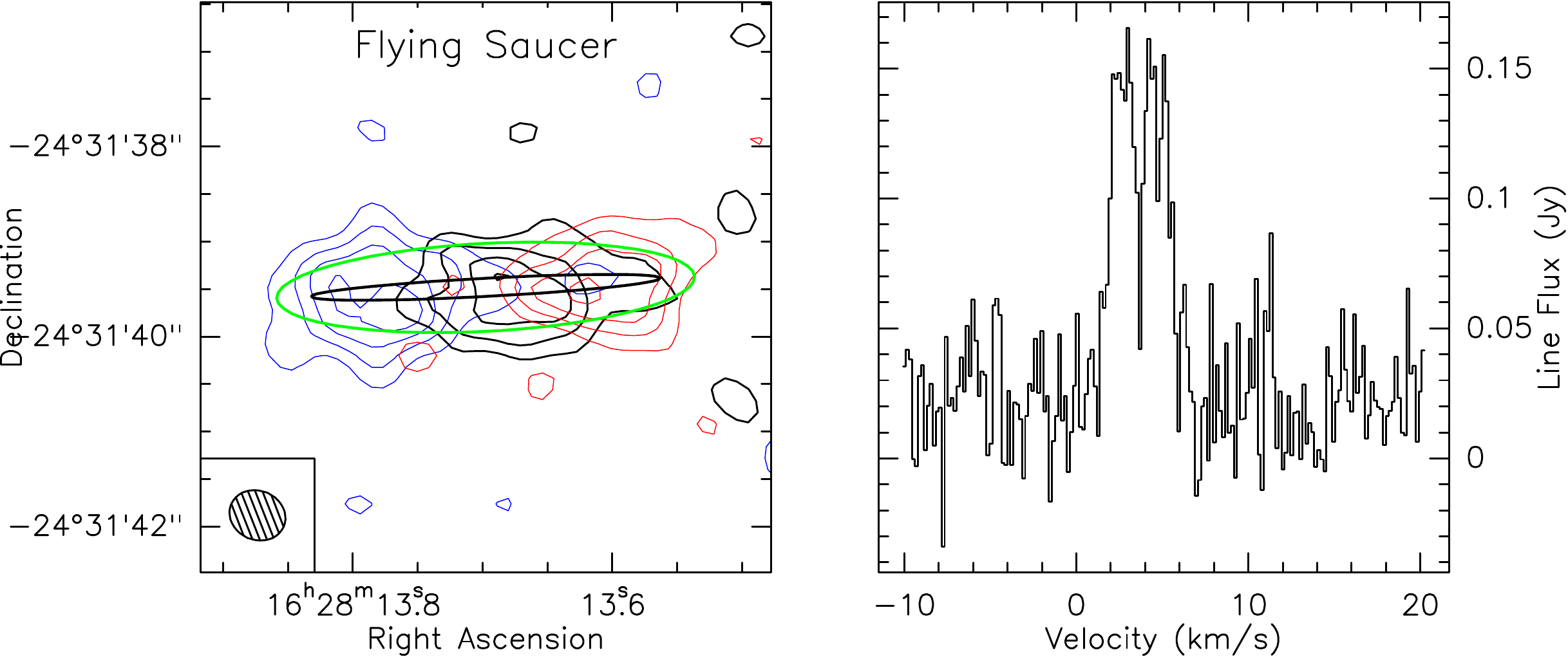}
\figsetplot{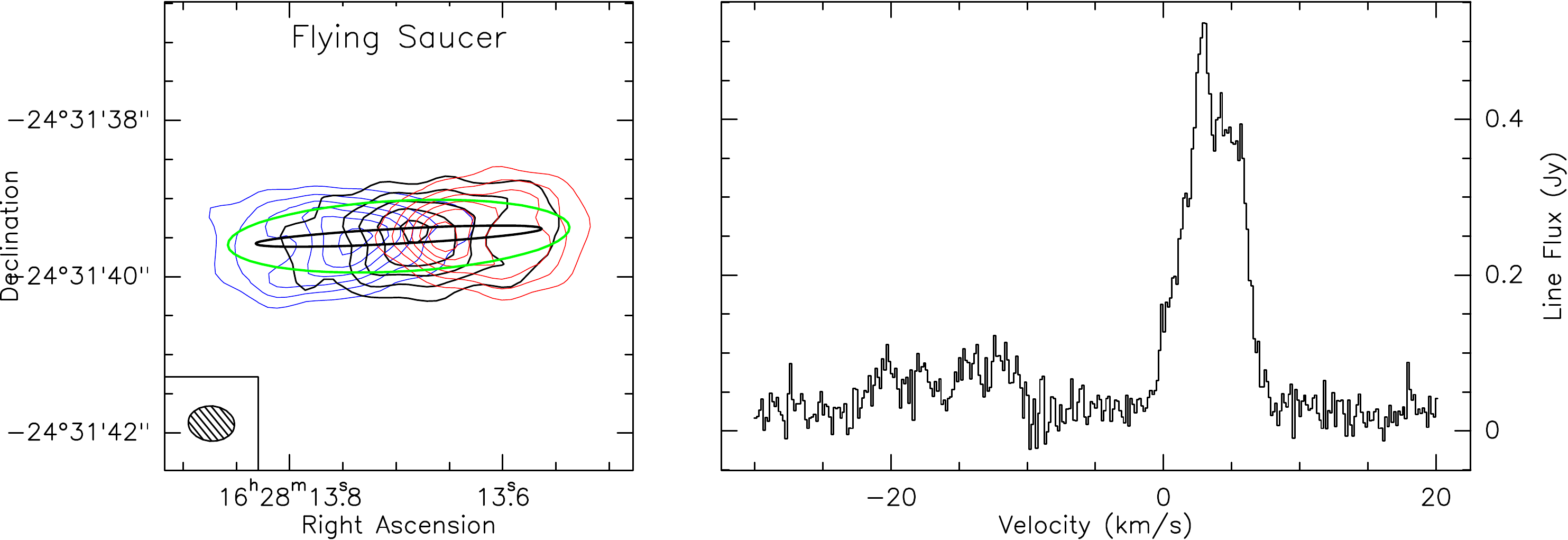}
    \figsetgrpnote{CO J=2-1 (top left), H$_2$CO (top right) and CN emission (bottom).}
    \figsetgrpend

    \figsetgrpstart
    \figsetgrpnum{1.3}
    \figsetgrptitle{Molecules towards YLW 58}
\figsetplot{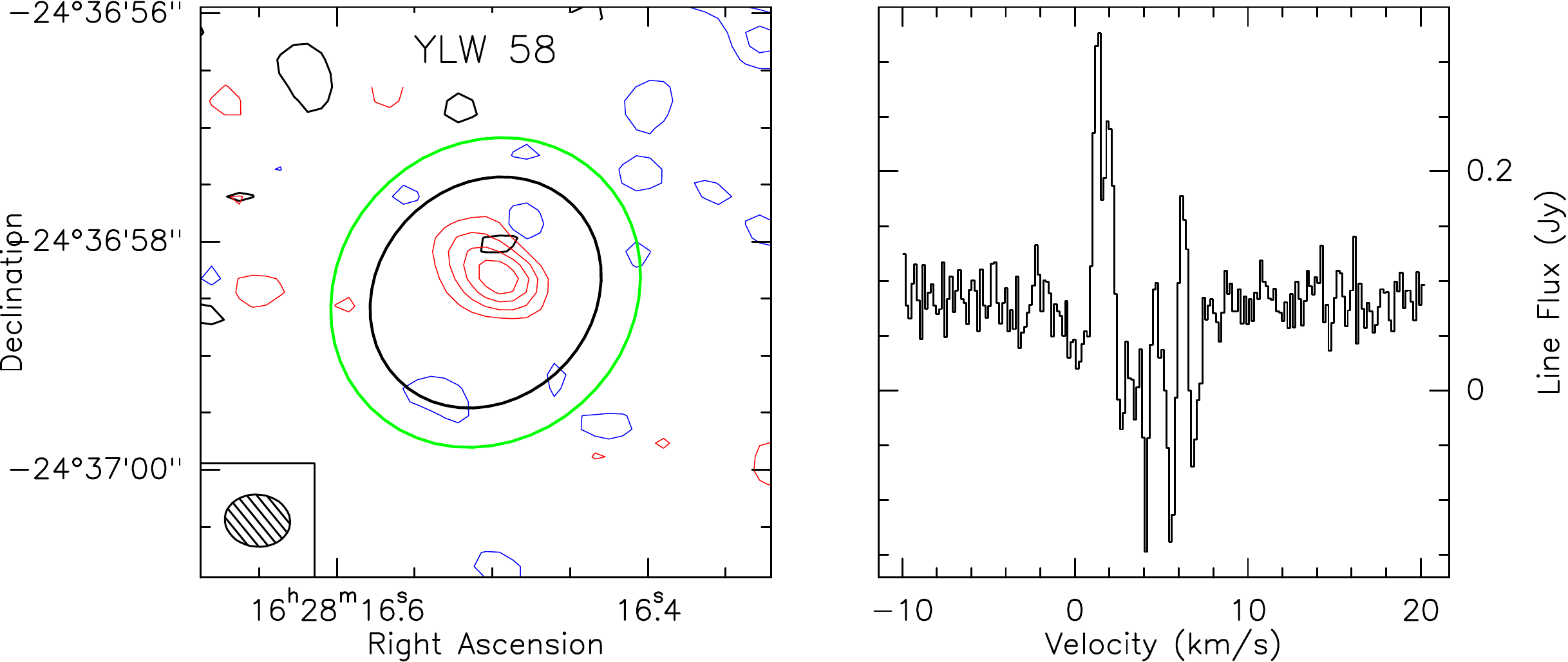}
\figsetplot{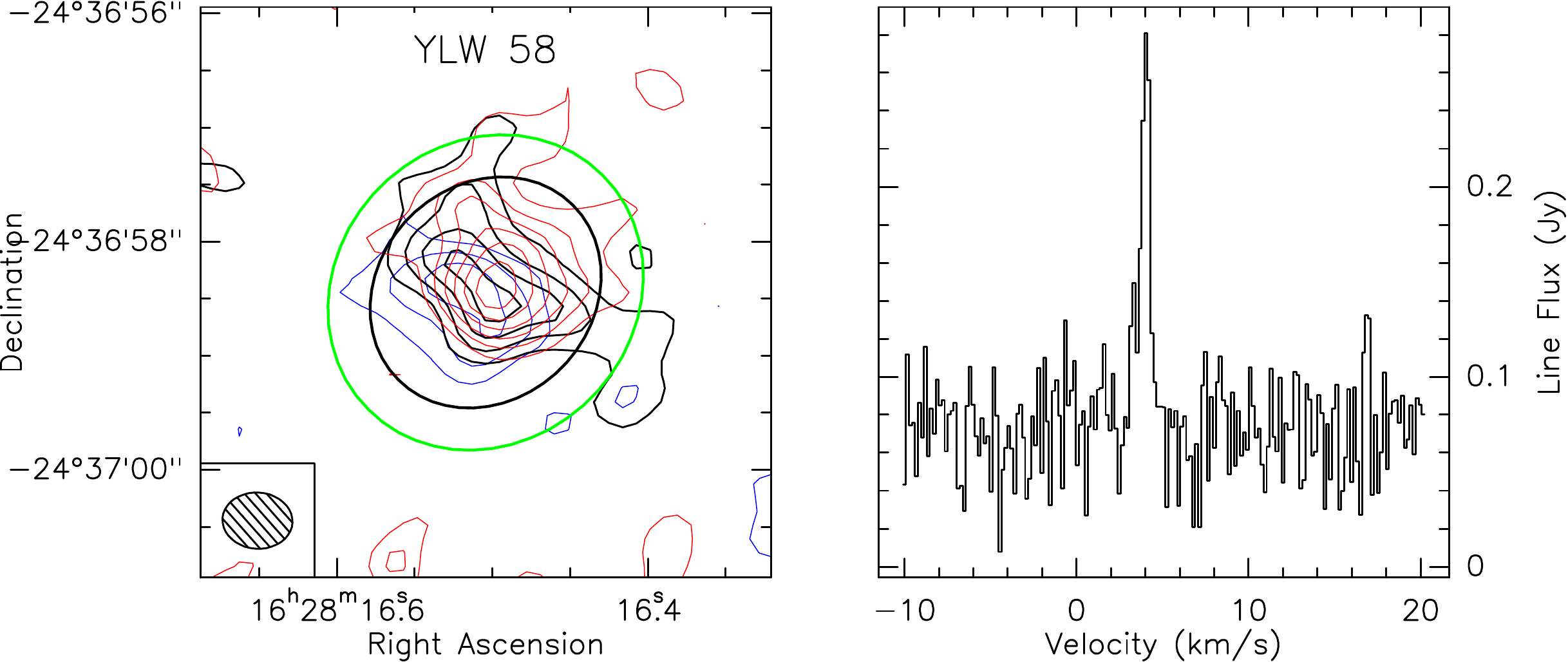}
    \figsetgrpnote{CO J=2-1 (left) and H$_2$CO (right).}
    \figsetgrpend

    \figsetgrpstart
    \figsetgrpnum{1.4}
    \figsetgrptitle{Molecules towards CY Tau}
\figsetplot{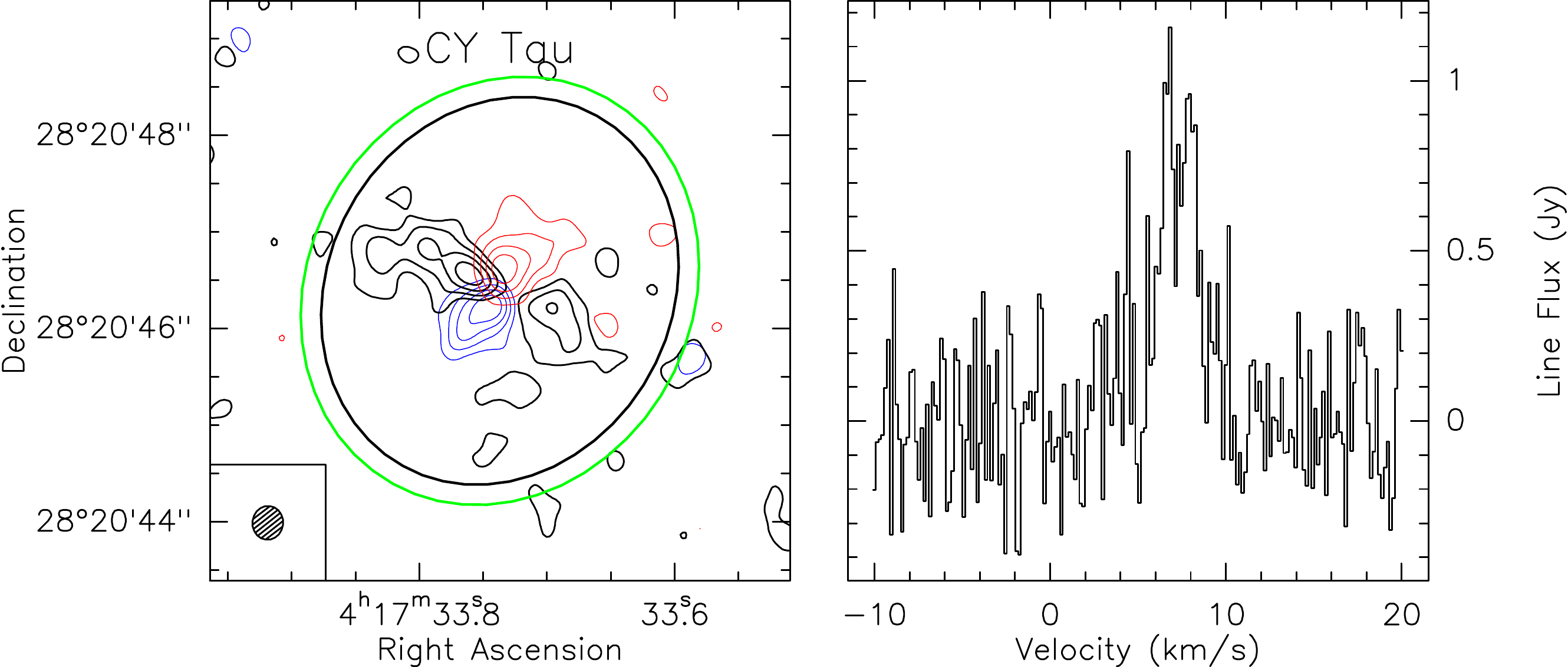}
\figsetplot{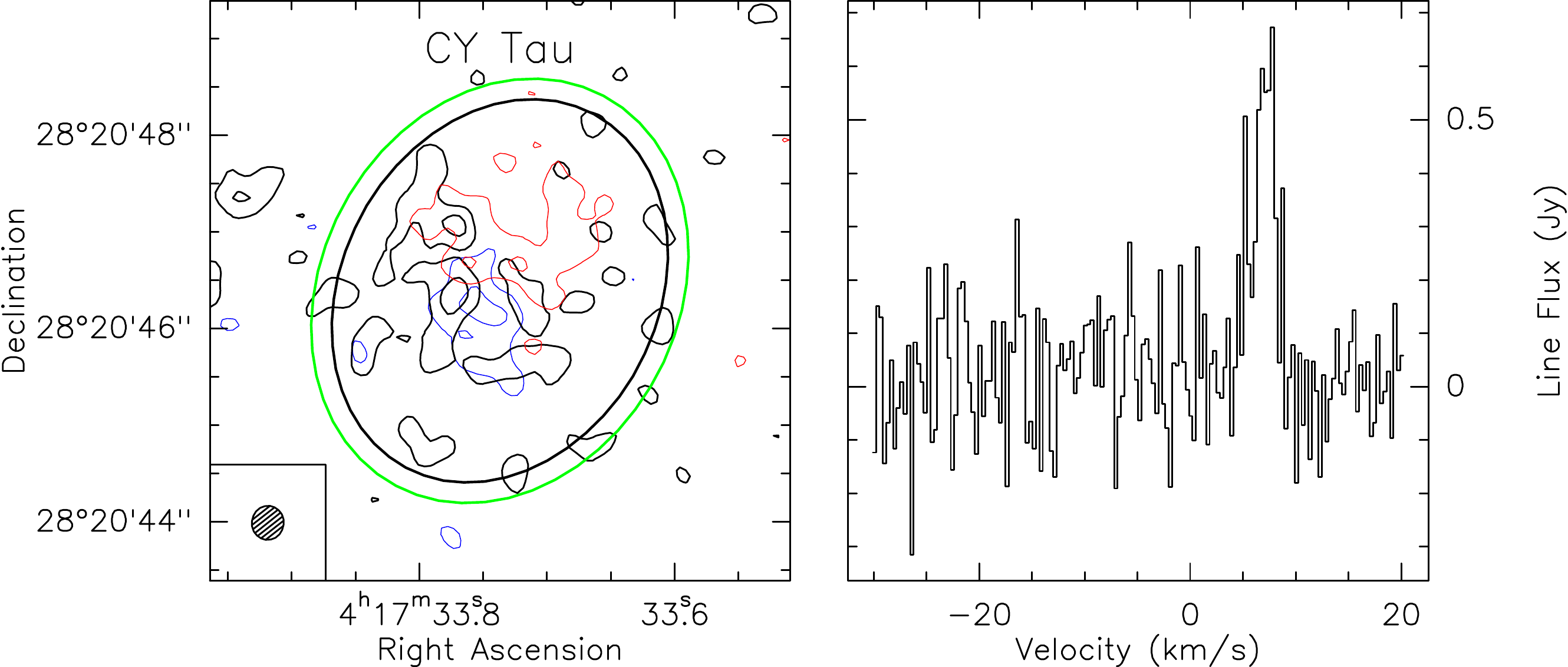}
    \figsetgrpnote{CO J=2-1 (left) and H$_2$CO (right).}
    \figsetgrpend

    \figsetgrpstart
    \figsetgrpnum{1.5}
    \figsetgrptitle{CO J=2-1 towards GSS 26}
\figsetplot{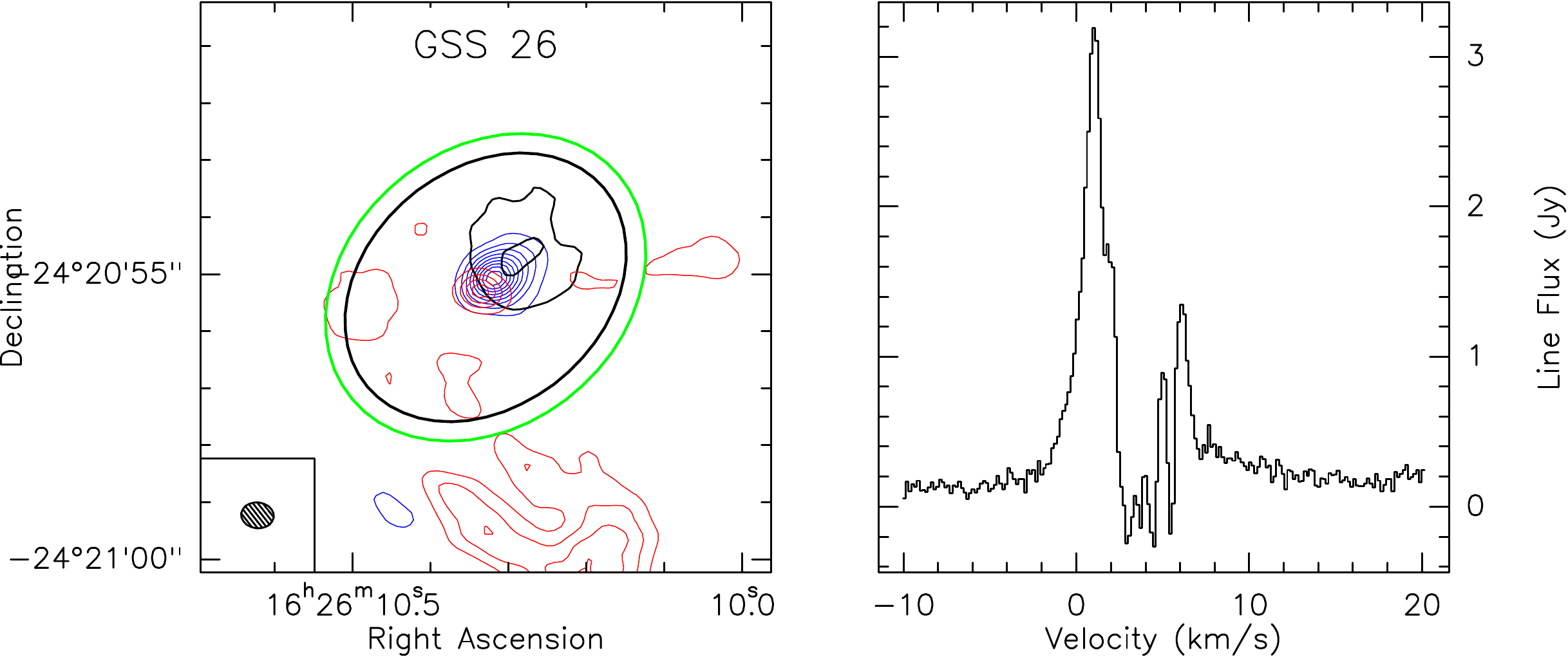}
    \figsetgrpend

    \figsetgrpstart
    \figsetgrpnum{1.6}
    \figsetgrptitle{CO J=2-1 towards CX Tau}
\figsetplot{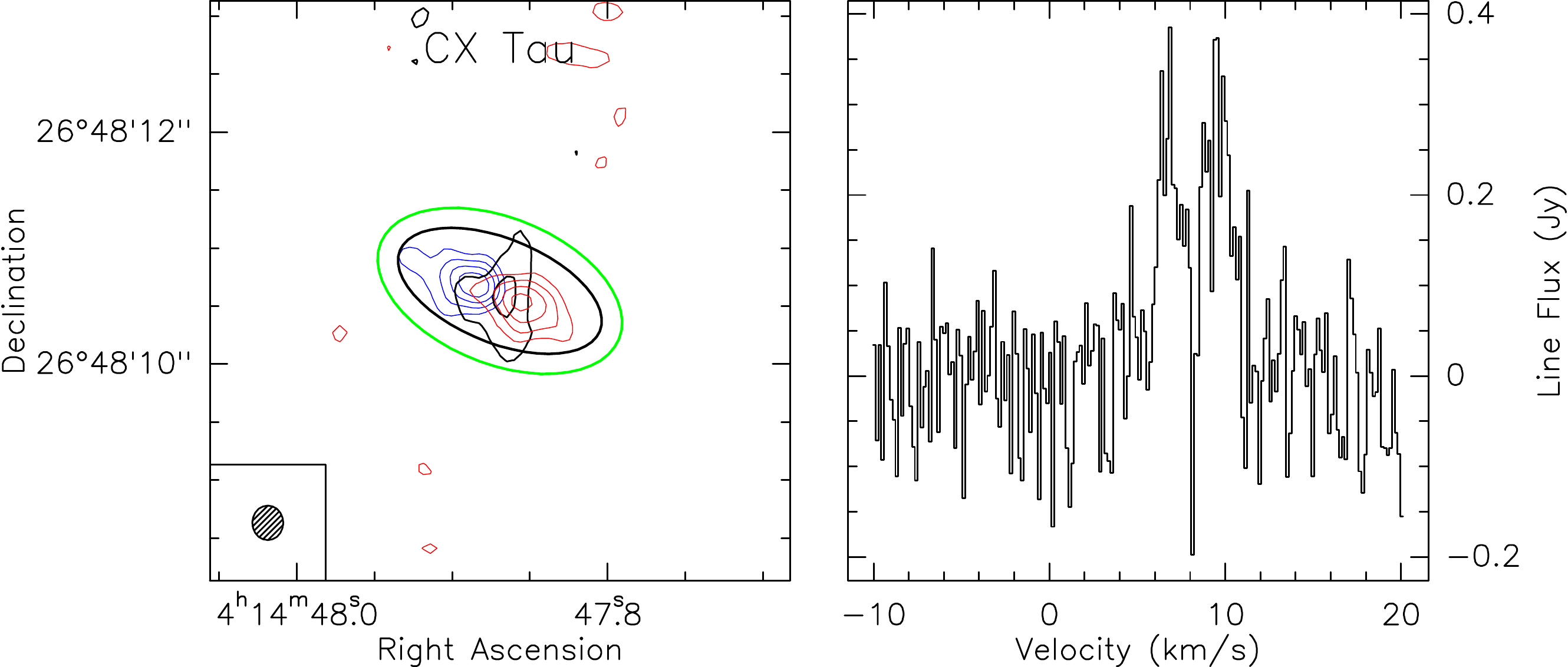}
    \figsetgrpend

    \figsetgrpstart
    \figsetgrpnum{1.6}
    \figsetgrptitle{CO J=2-1 towards FP Tau}
\figsetplot{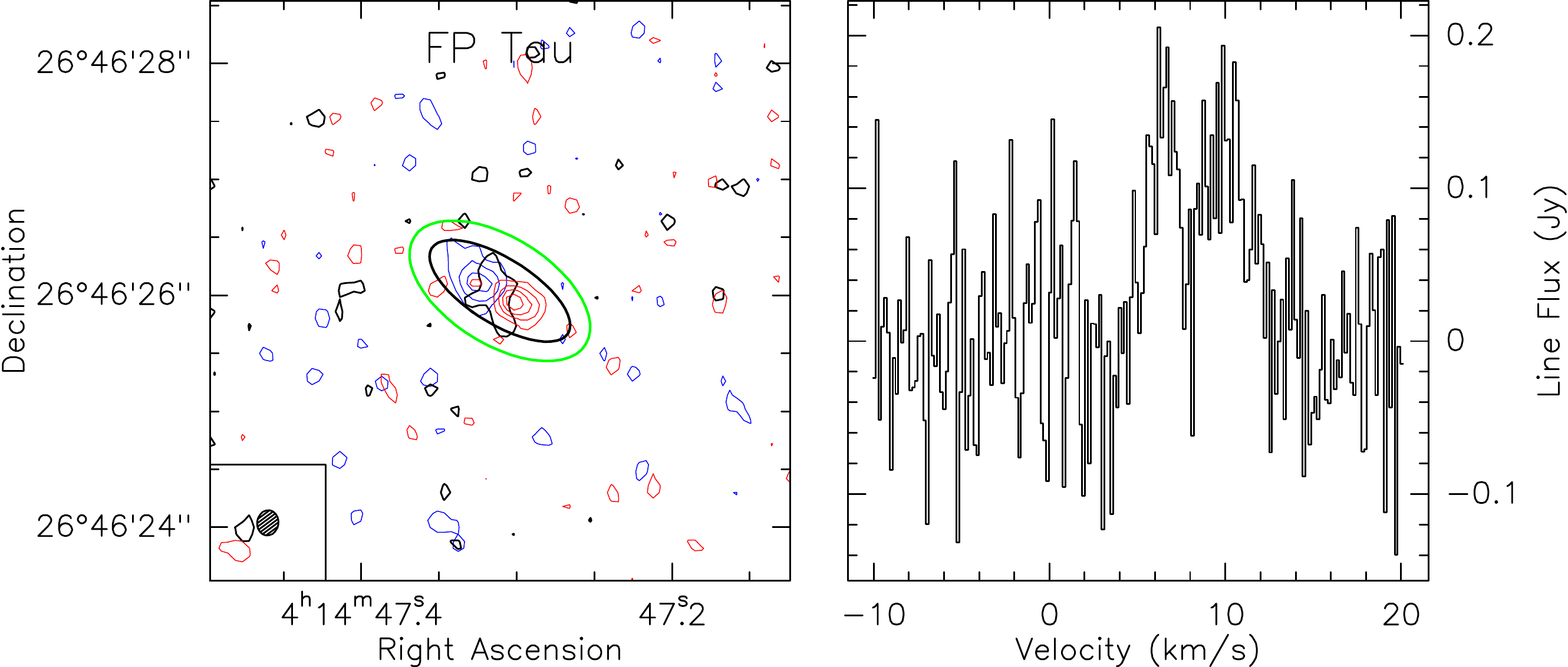}
    \figsetgrpend

    \figsetgrpstart
    \figsetgrpnum{1.7}
    \figsetgrptitle{CO J=2-1 towards IP Tau}
\figsetplot{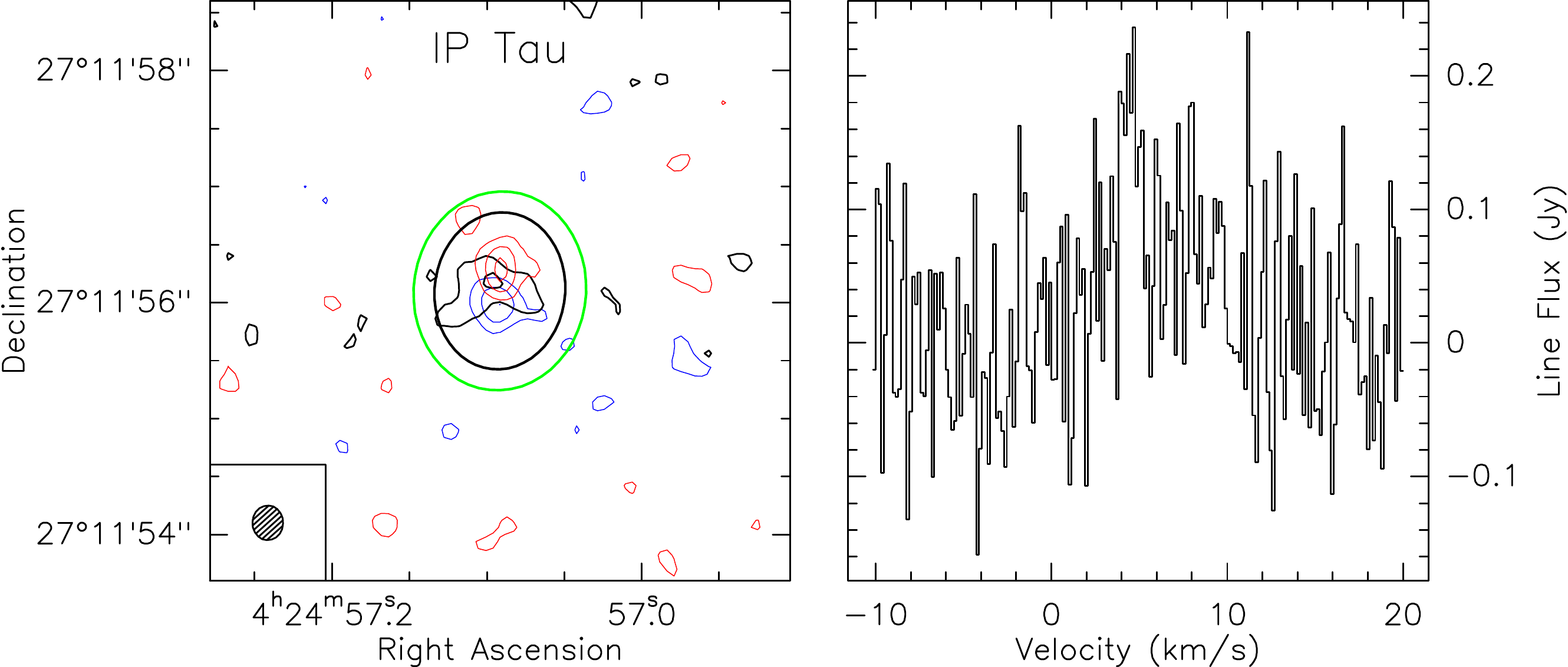}
    \figsetgrpend

    \figsetgrpstart
    \figsetgrpnum{1.8}
    \figsetgrptitle{CO J=2-1 towards MHO 2}
\figsetplot{mho-2-eps-converted-to.pdf} 
    \figsetgrpend

    \figsetgrpstart
    \figsetgrpnum{1.9}
    \figsetgrptitle{CO J=2-1 towards FM Tau}
\figsetplot{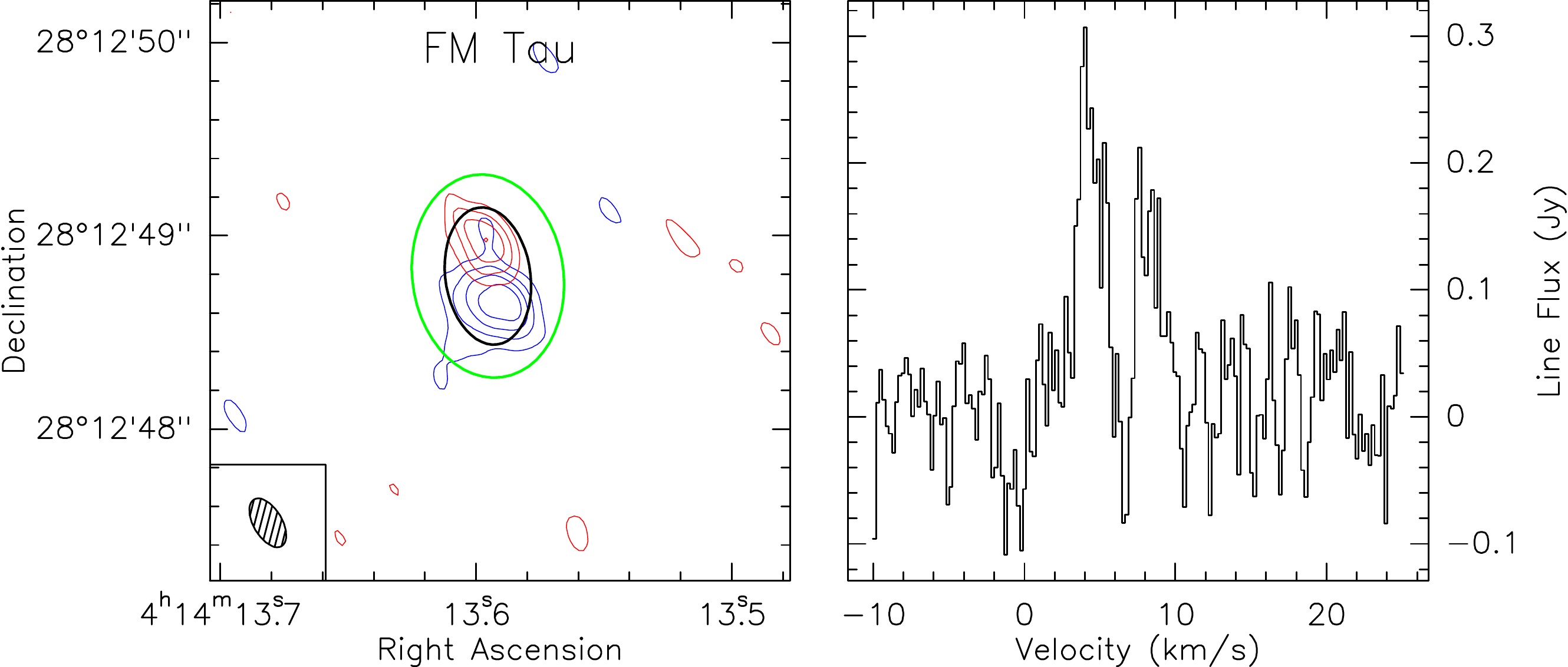} 
    \figsetgrpend
\figsetend

\ifdefined\PREPRINT
\else
  \COFigSet{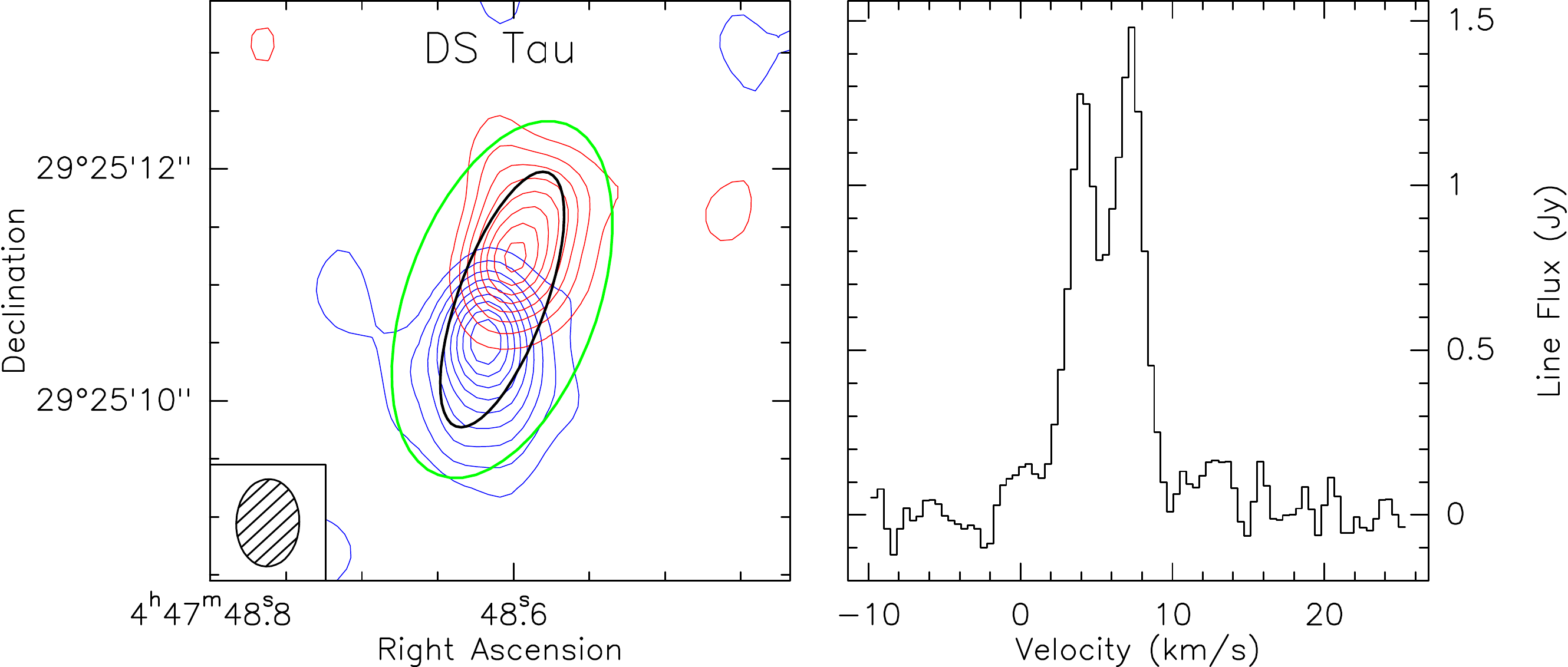}{DS Tau}{3-2}
\fi
\figsetstart
\figsetnum{\ref{fig:aj}}
\figsettitle{CO J=3-2 from Akeson \& Jensen source sample.}
\COSet{ds_tau}{DS Tau}{3-2}{1}
\COSet{fx_tau}{FX Tau}{3-2}{2}
\COSet{hn_tau}{HN Tau}{3-2}{3}
\COSet{ho_tau}{HO Tau}{3-2}{4}
\COSet{i05022}{IRAS05022}{3-2}{5}
\COSet{dk_tau-a}{DK Tau A}{3-2}{6}
\COSet{dk_tau-b}{DK Tau B}{3-2}{7}
\COSet{v710_tau}{V710 Tau}{3-2}{8}
\COSet{gk_tau}{GK Tau}{3-2}{9}
\COSet{hbc_411}{HBC 411}{3-2}{10}
\COSet{it_tau}{IT Tau}{3-2}{1}
\COSet{hk_tau-a}{HK Tau A}{3-2}{11}
\COSet{hk_tau-b}{HK Tau B}{3-2}{12}
\COSet{fv_tau}{FV Tau}{3-2}{13}
\figsetend

\ifdefined\PREPRINT
%
\begin{figure}
\includegraphics[height=3.5cm]{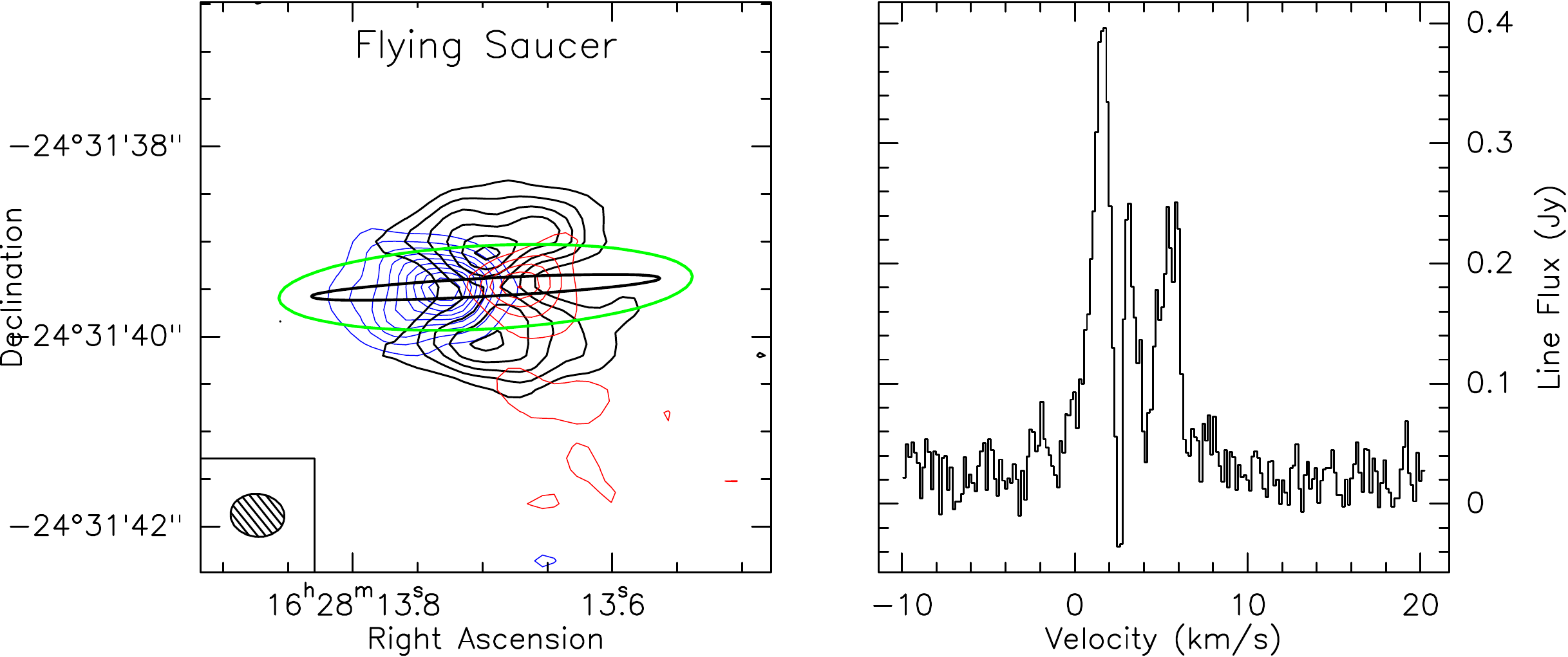}\hspace{0.5cm}
\includegraphics[height=3.5cm]{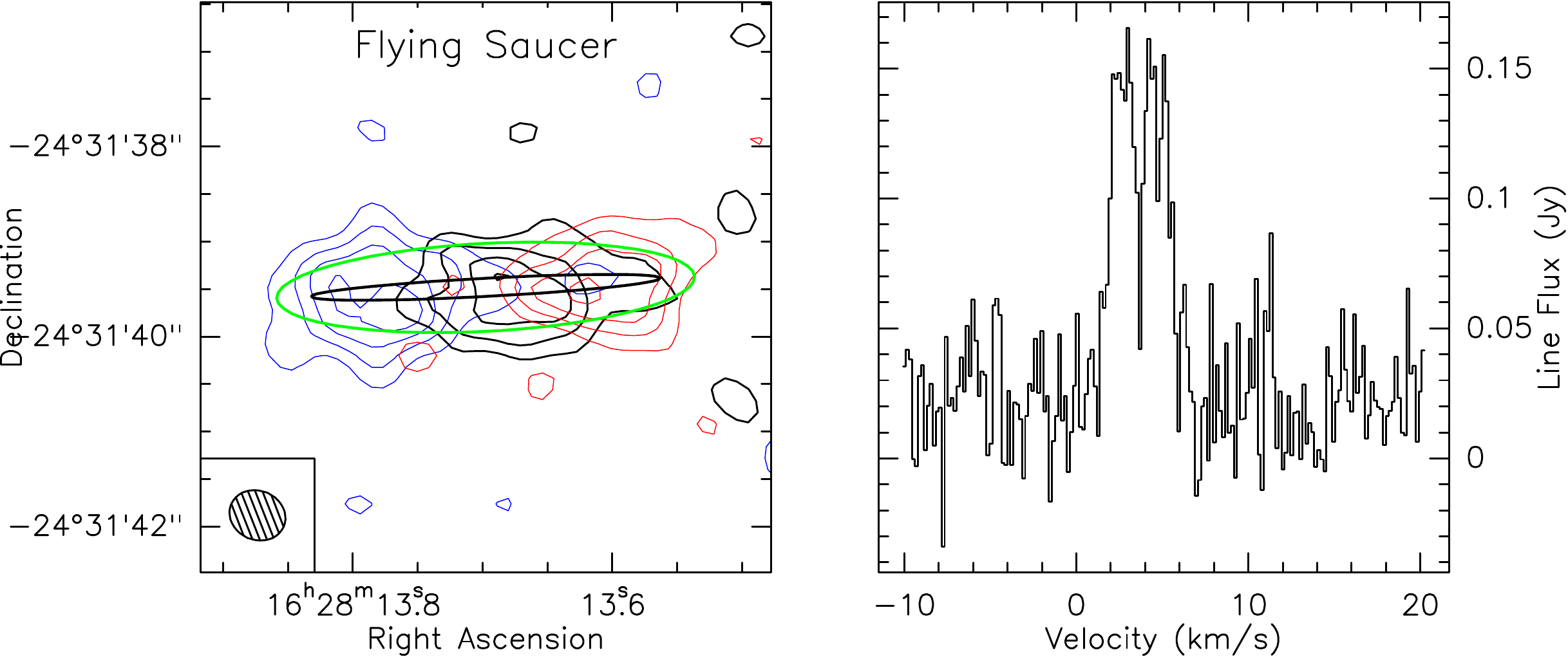}
\includegraphics[height=3.5cm]{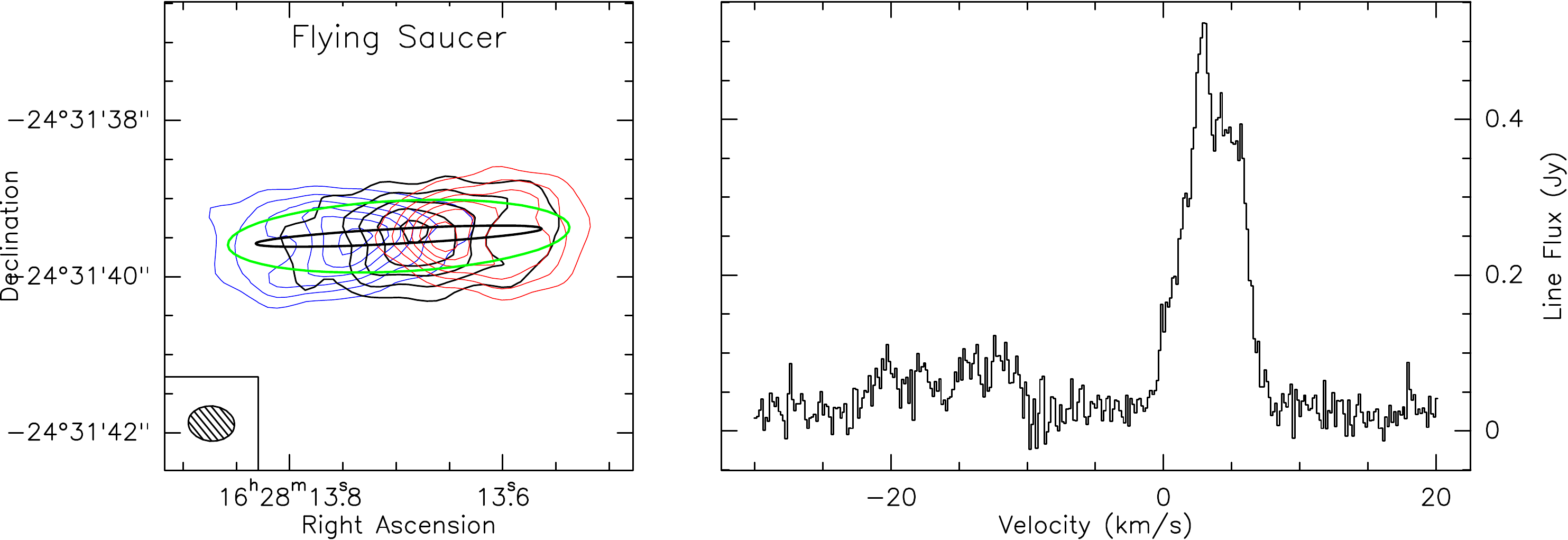}
\caption{As Fig.\ref{fig:gss39} for the Flying Saucer: CO (top left), H$_2$CO (top right), CN (bottom).}
\label{fig:flying}
\end{figure}

\begin{figure}
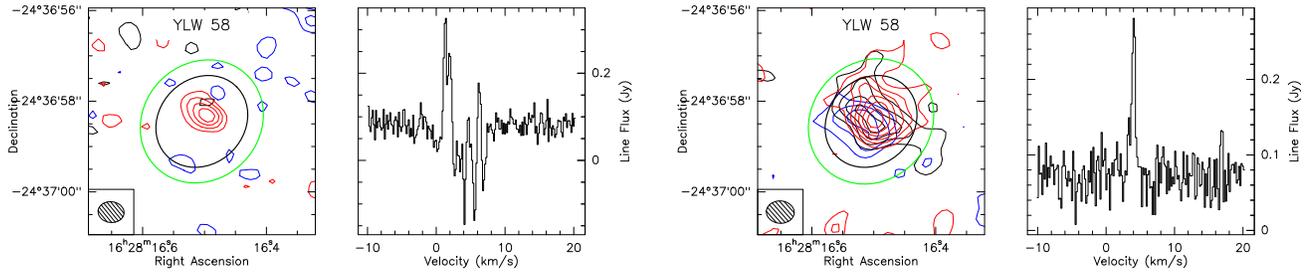

\includegraphics[height=3.5cm]{YLW_58-co21-eps-converted-to.pdf}\hspace{0.5cm}
\includegraphics[height=3.5cm]{YLW_58-h2co-eps-converted-to.pdf}
\caption{Molecules towards YLW 58: CO (left), H$_2$CO (right).}
\label{fig:ylw58}
\end{figure}

\begin{figure}
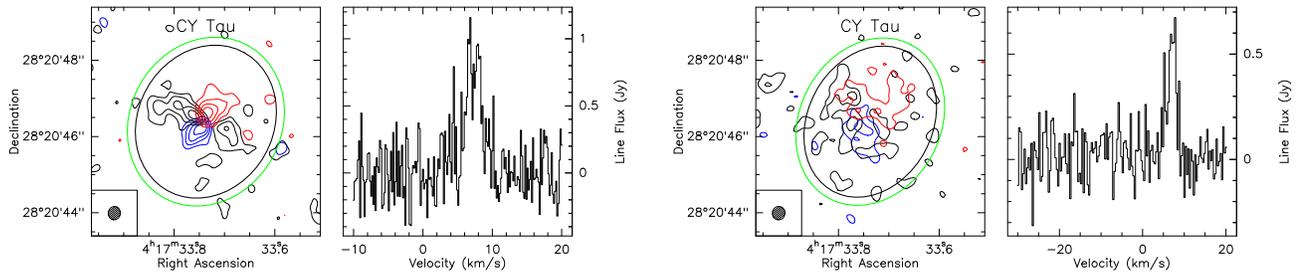

\includegraphics[height=3.5cm]{cy_tau-co21-eps-converted-to.pdf}\hspace{0.5cm}
\includegraphics[height=3.5cm]{cy_tau-cn21-eps-converted-to.pdf}
\caption{Molecules towards CY Tau: CO (left), CN (right).}
\label{fig:cytau}
\end{figure}

\COFigure{gss_26}{GSS 26}{2-1}
\COFigure{cx_tau}{CX Tau}{2-1}
\COFigure{fp_tau}{FP Tau}{2-1}
\COFigure{ip_tau}{IP Tau}{2-1}
\COFigure{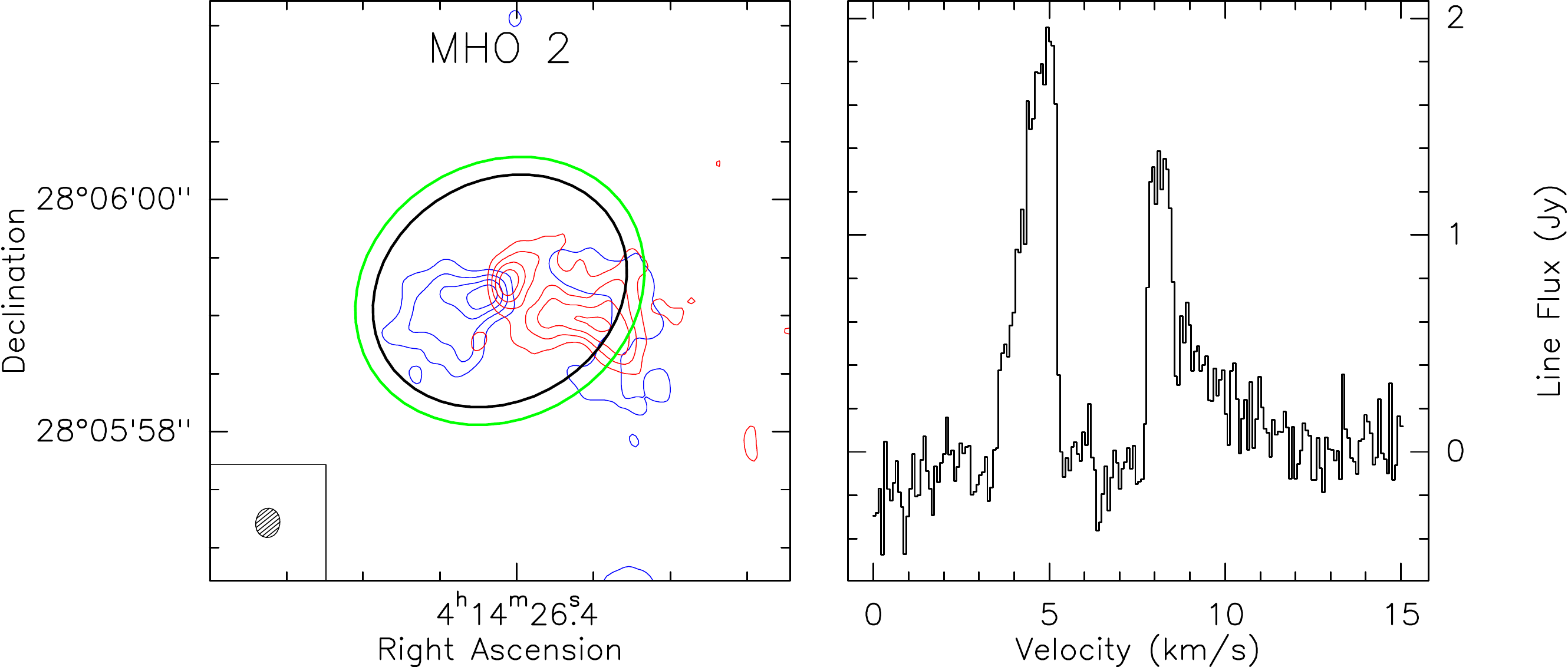}{MHO 2}{2-1}
\COFigure{fm_tau}{FM Tau}{3-2}

\begin{figure*}
\COpanel{ds_tau}{DS Tau}{3-2}
\COpanel{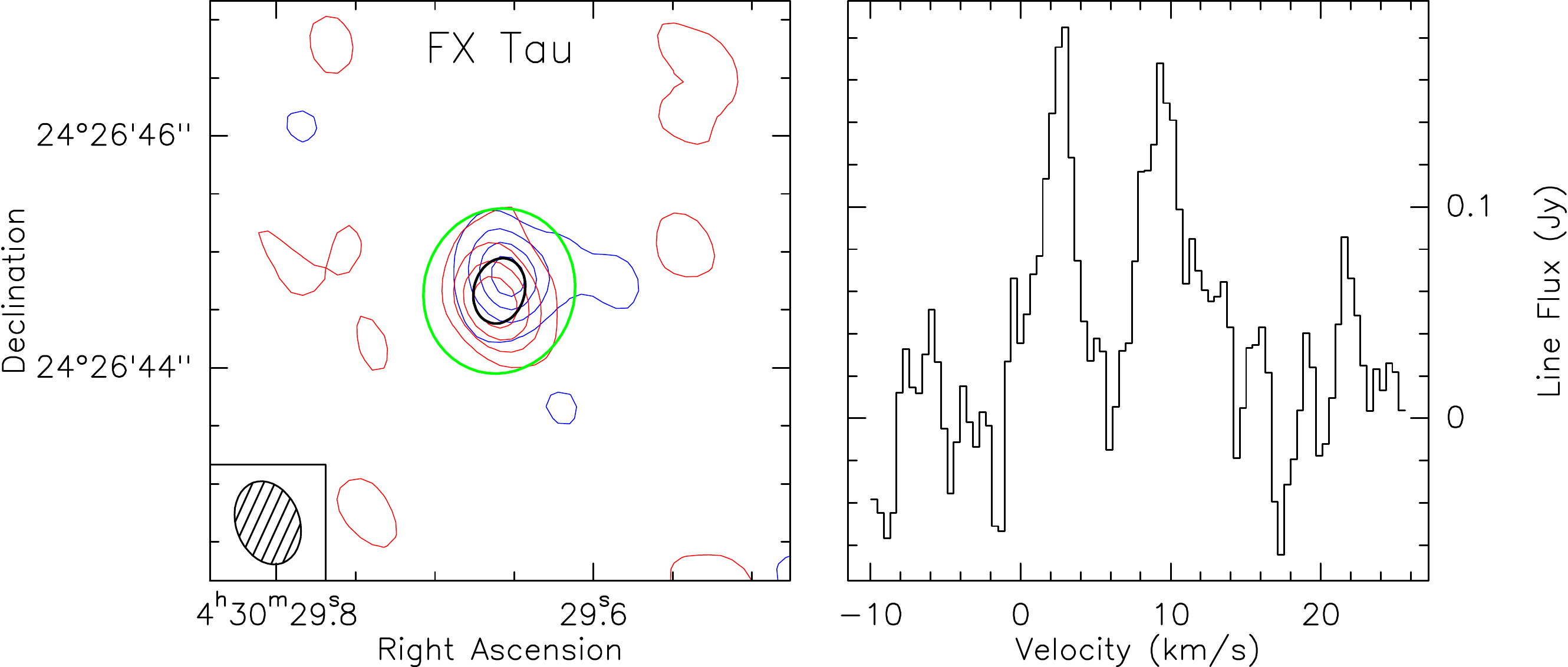}{FX Tau}{3-2}
\COpanel{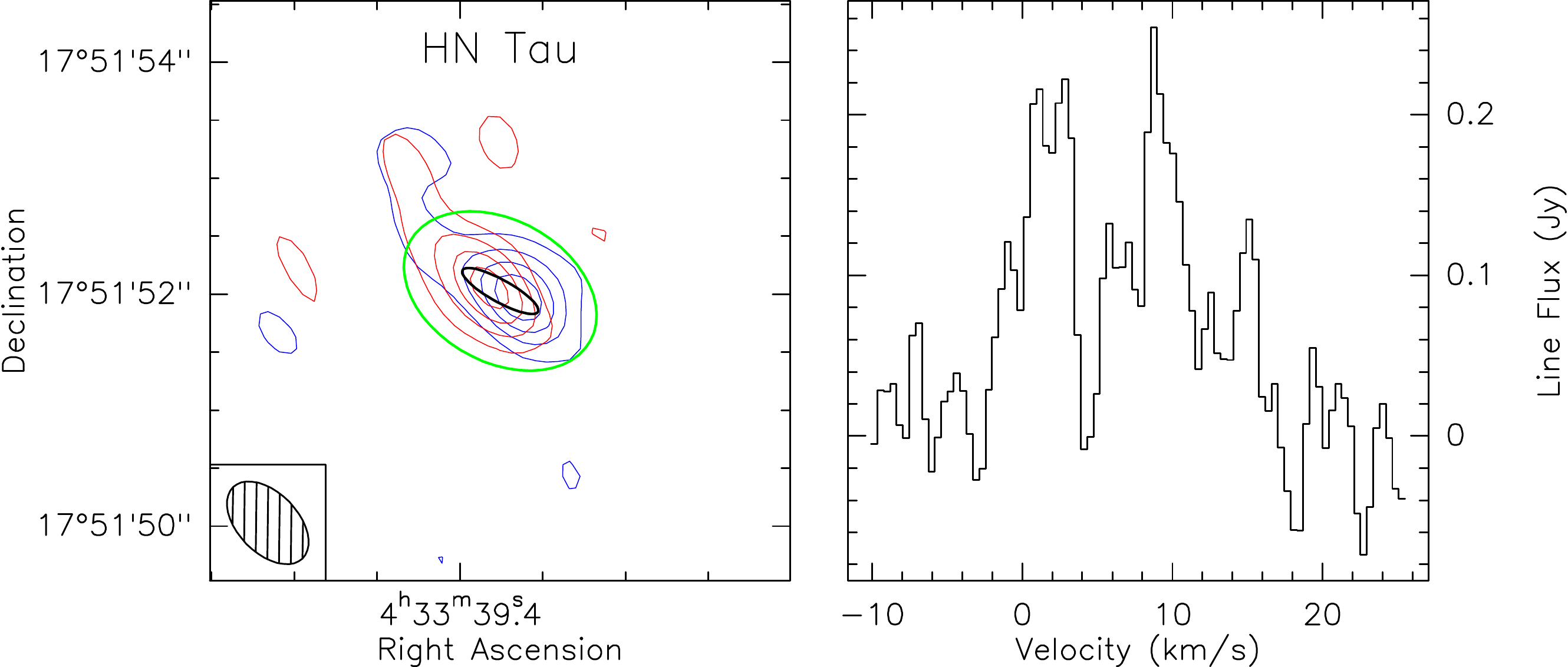}{HN Tau}{3-2}
\COpanel{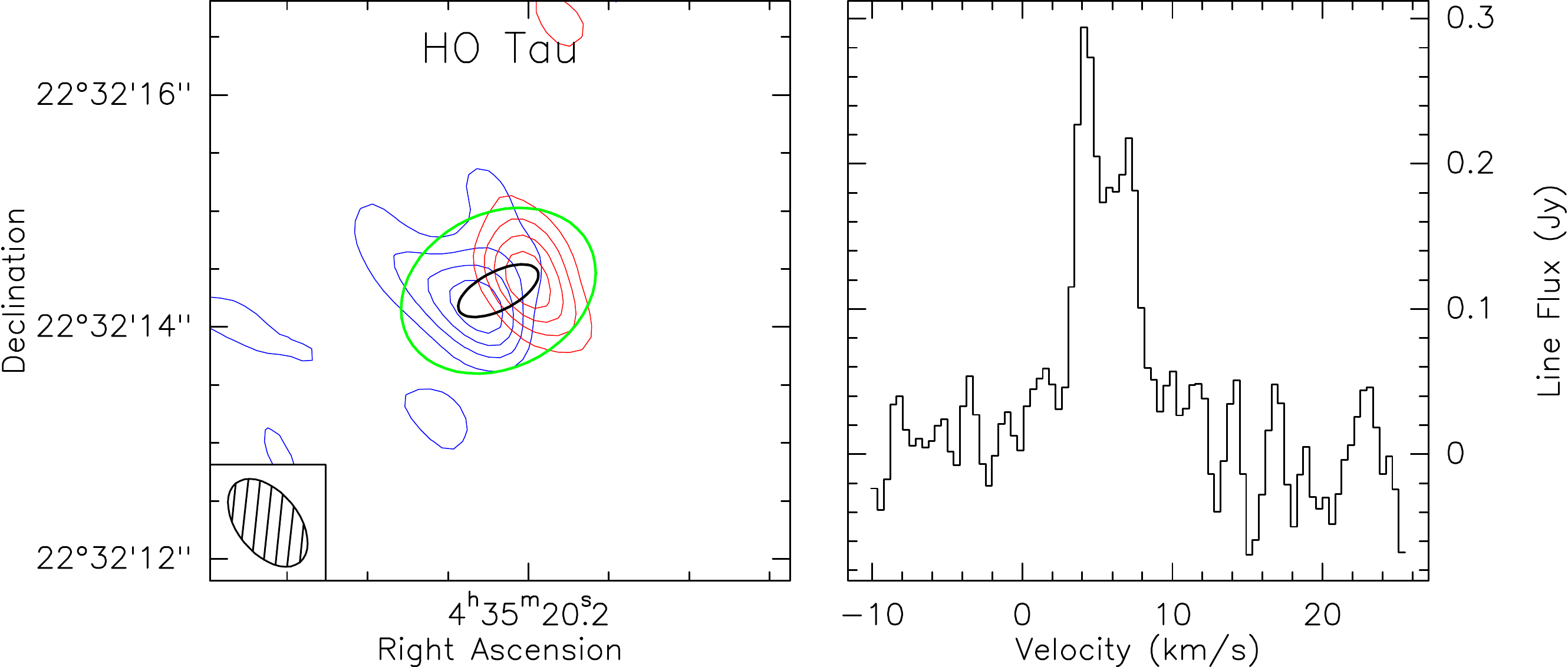}{HO Tau}{3-2}
\COpanel{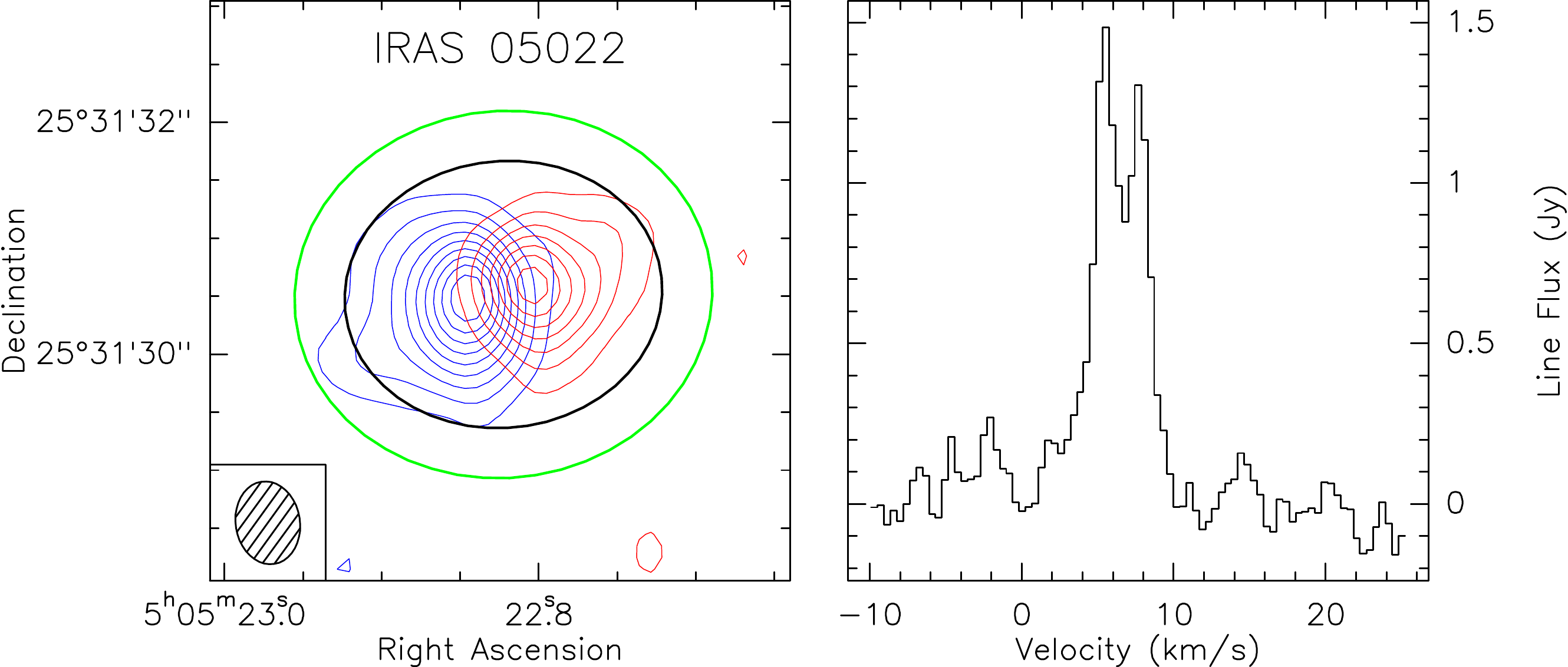}{IRAS05022}{3-2}
\COpanel{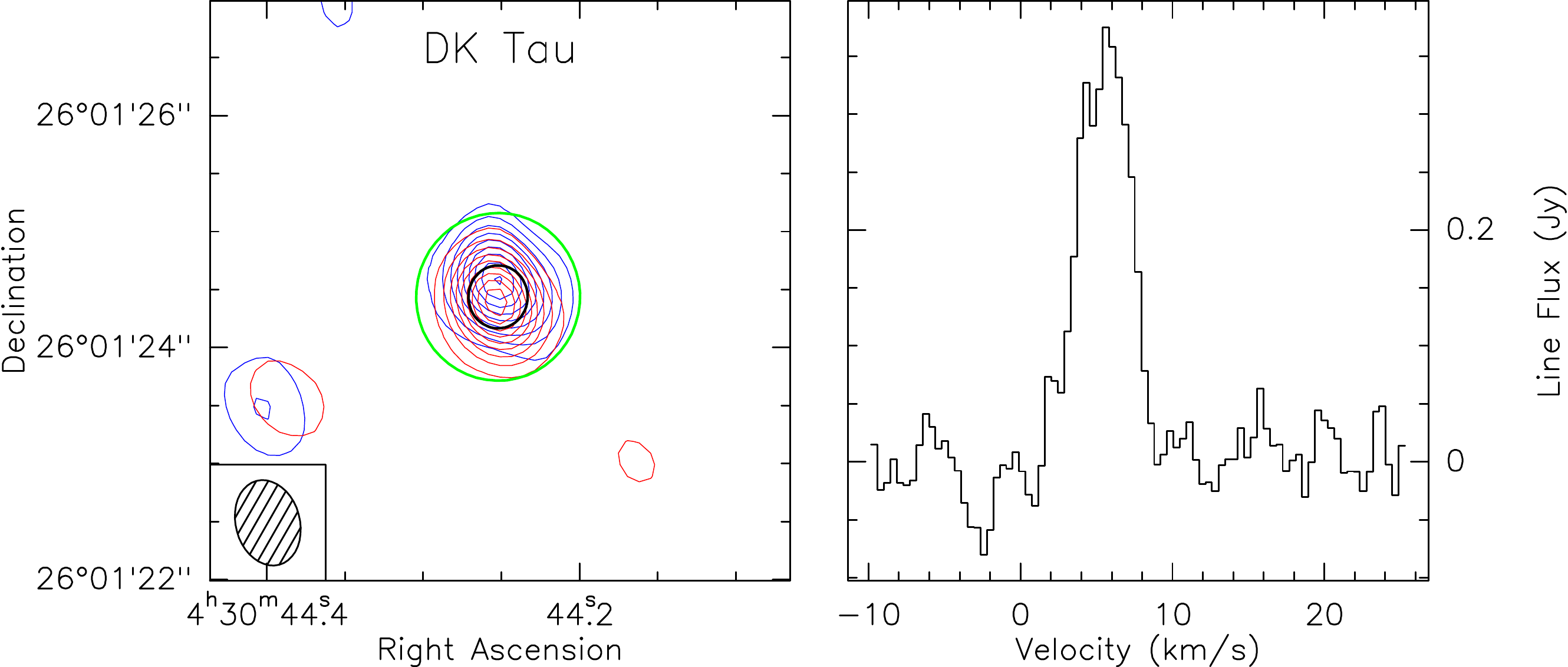}{DK Tau A}{3-2}
\COpanel{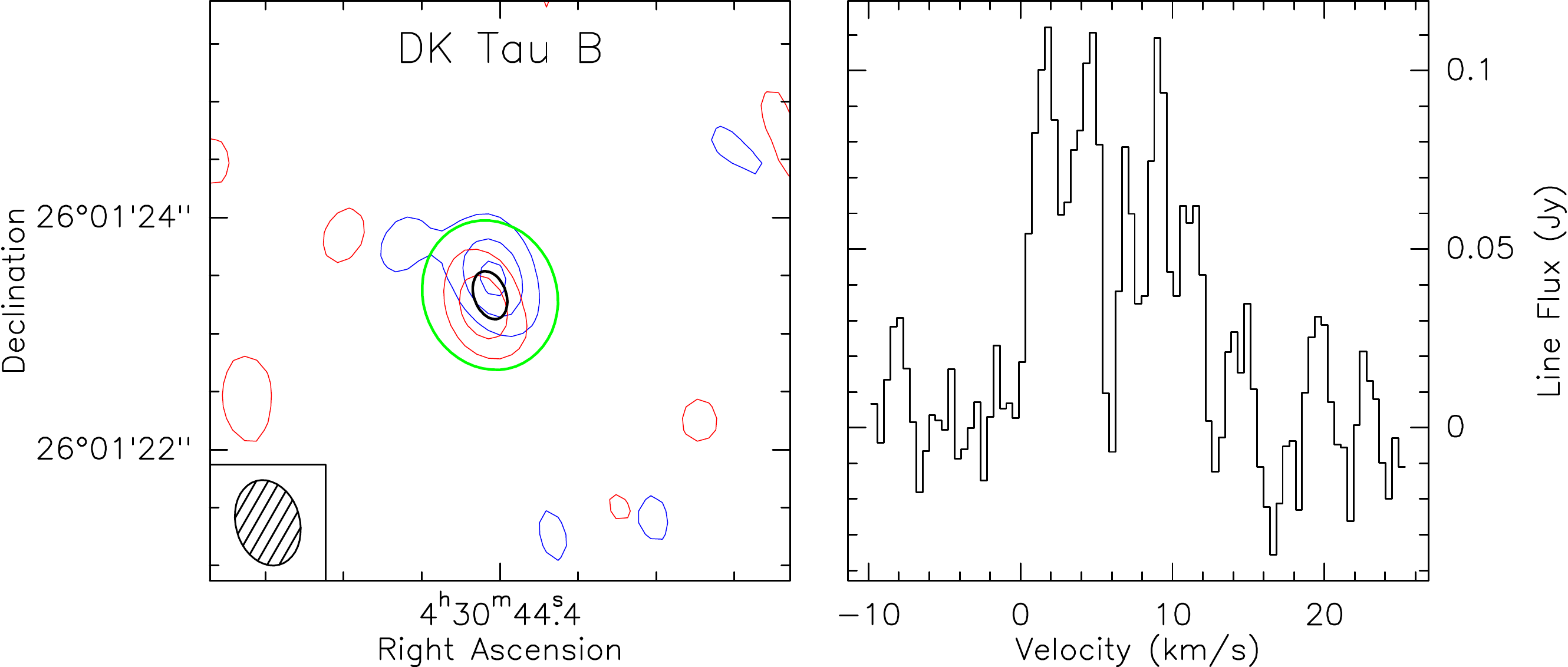}{DK Tau B}{3-2}
\COpanel{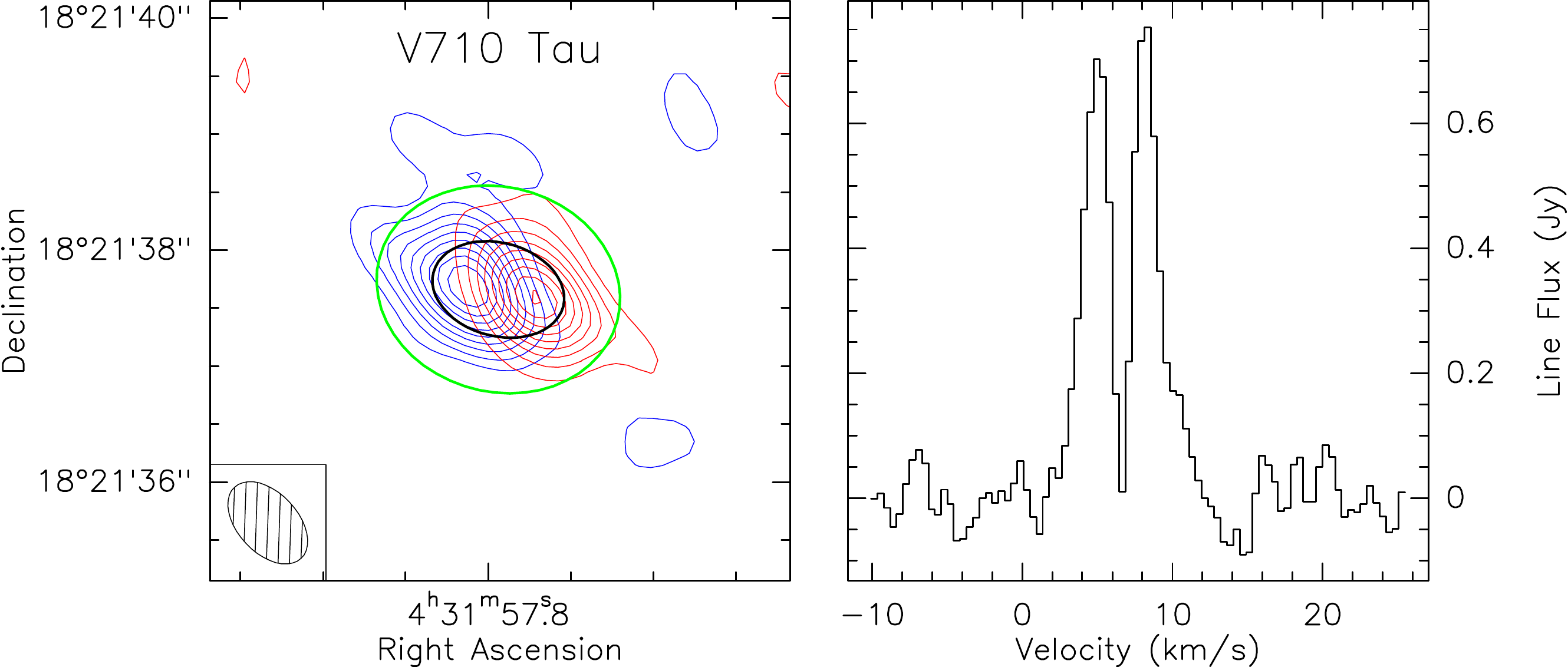}{V710 Tau}{3-2}
\COpanel{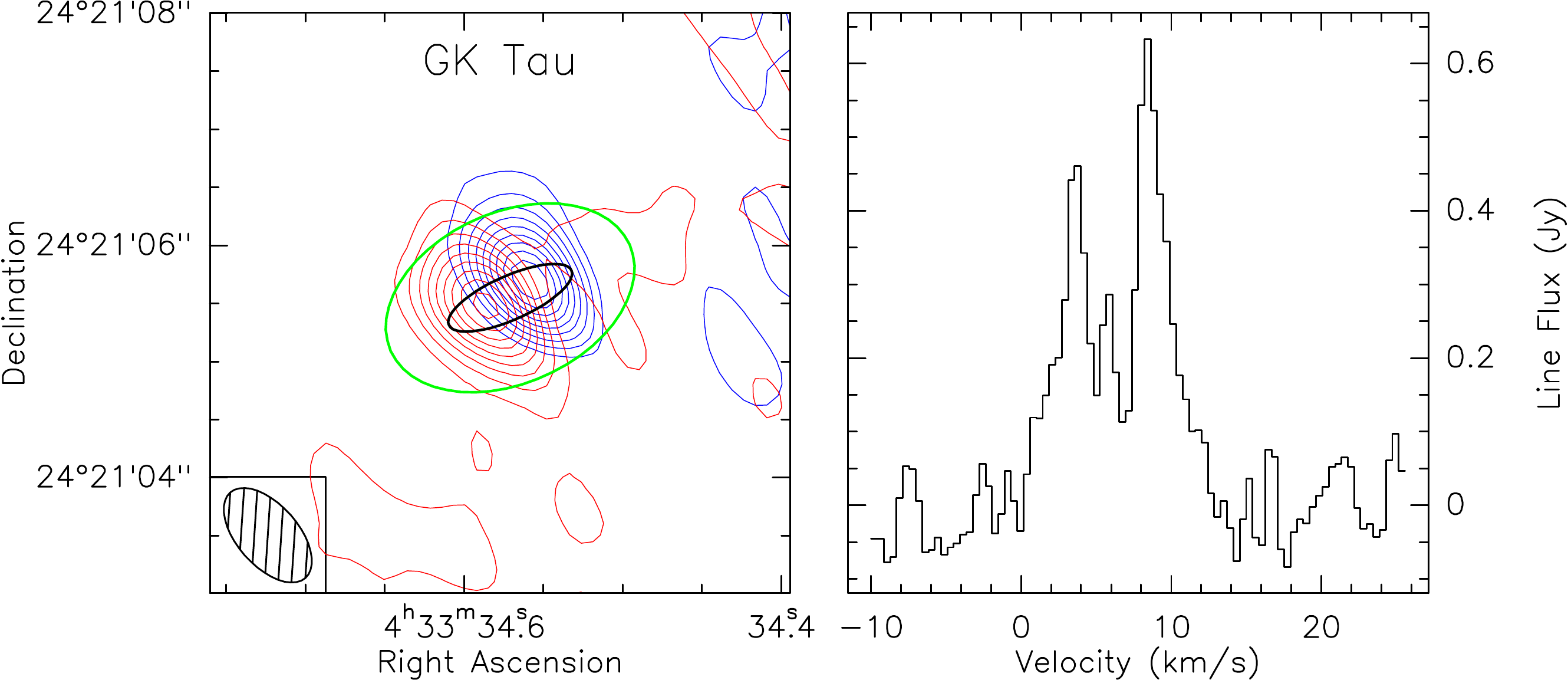}{GK Tau}{3-2}
\COpanel{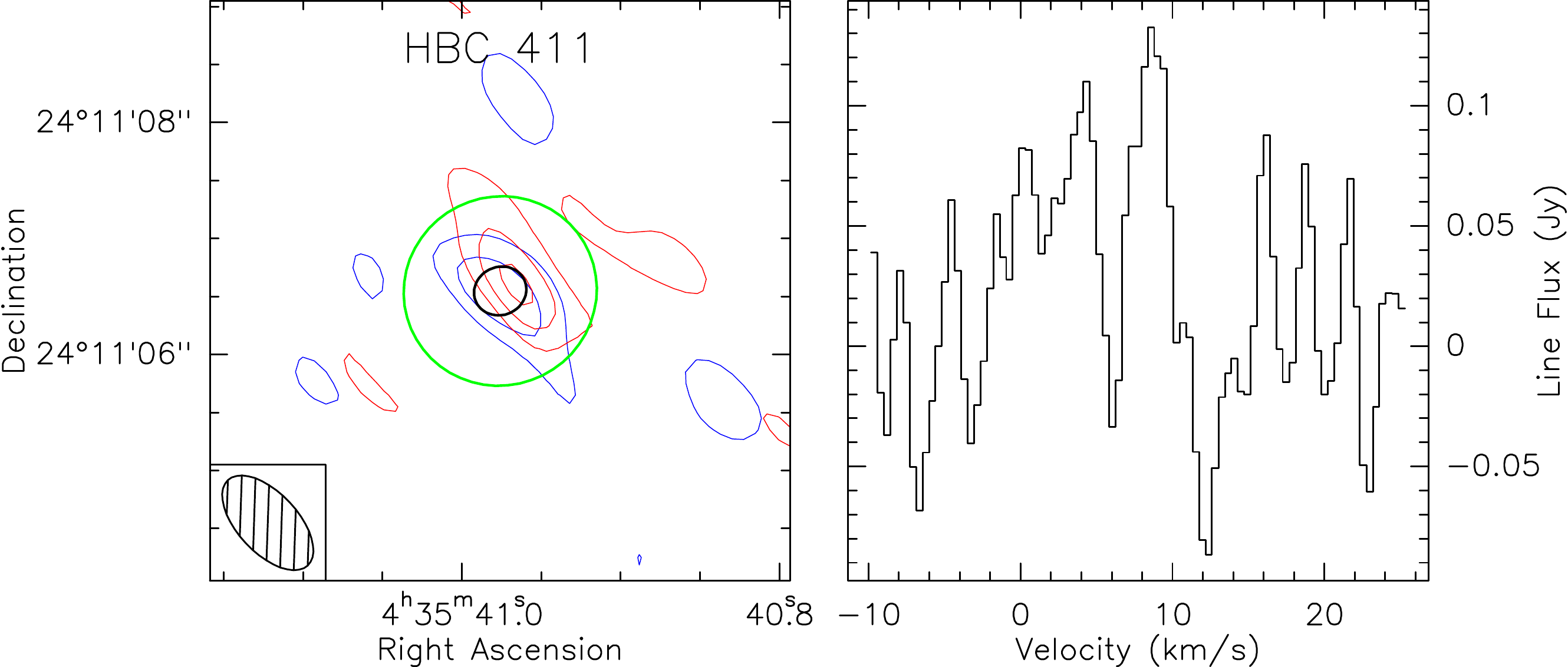}{HBC 411}{3-2}
\COpanel{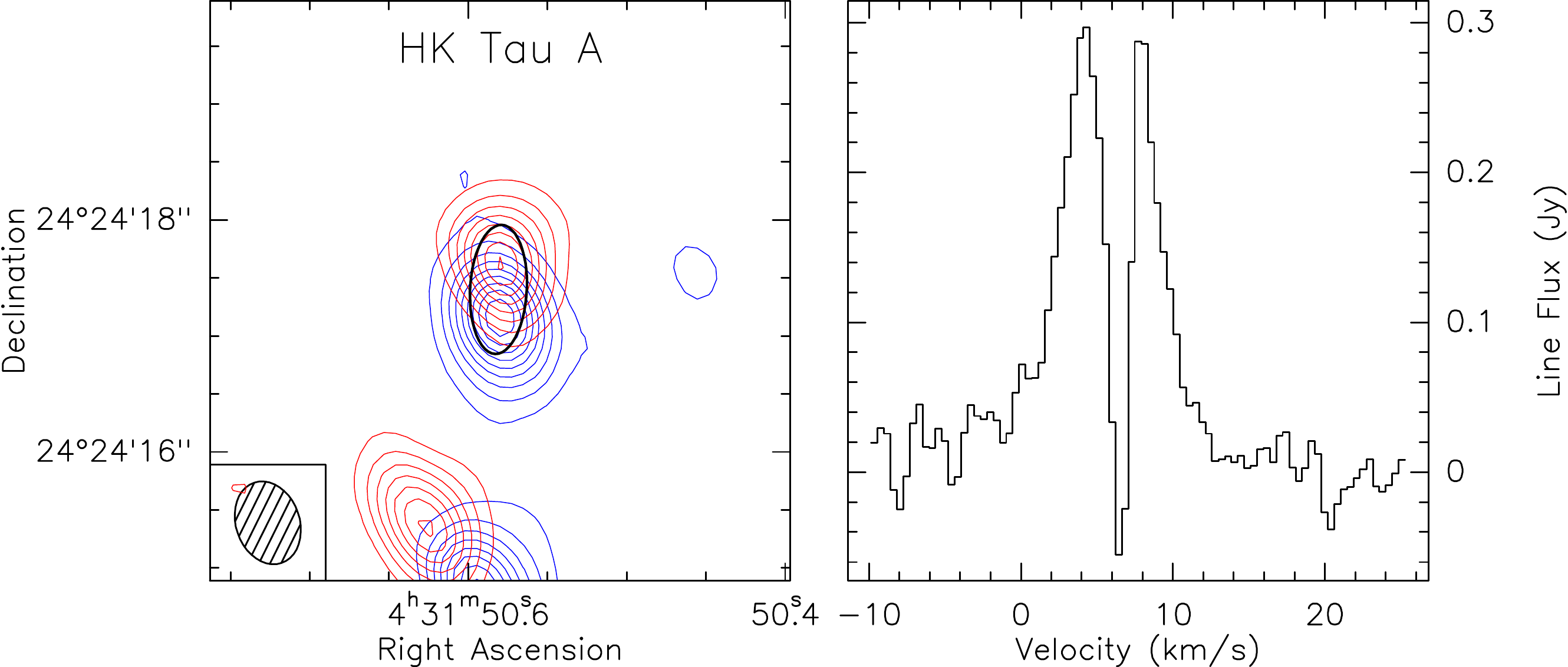}{HK Tau A}{3-2}
\COpanel{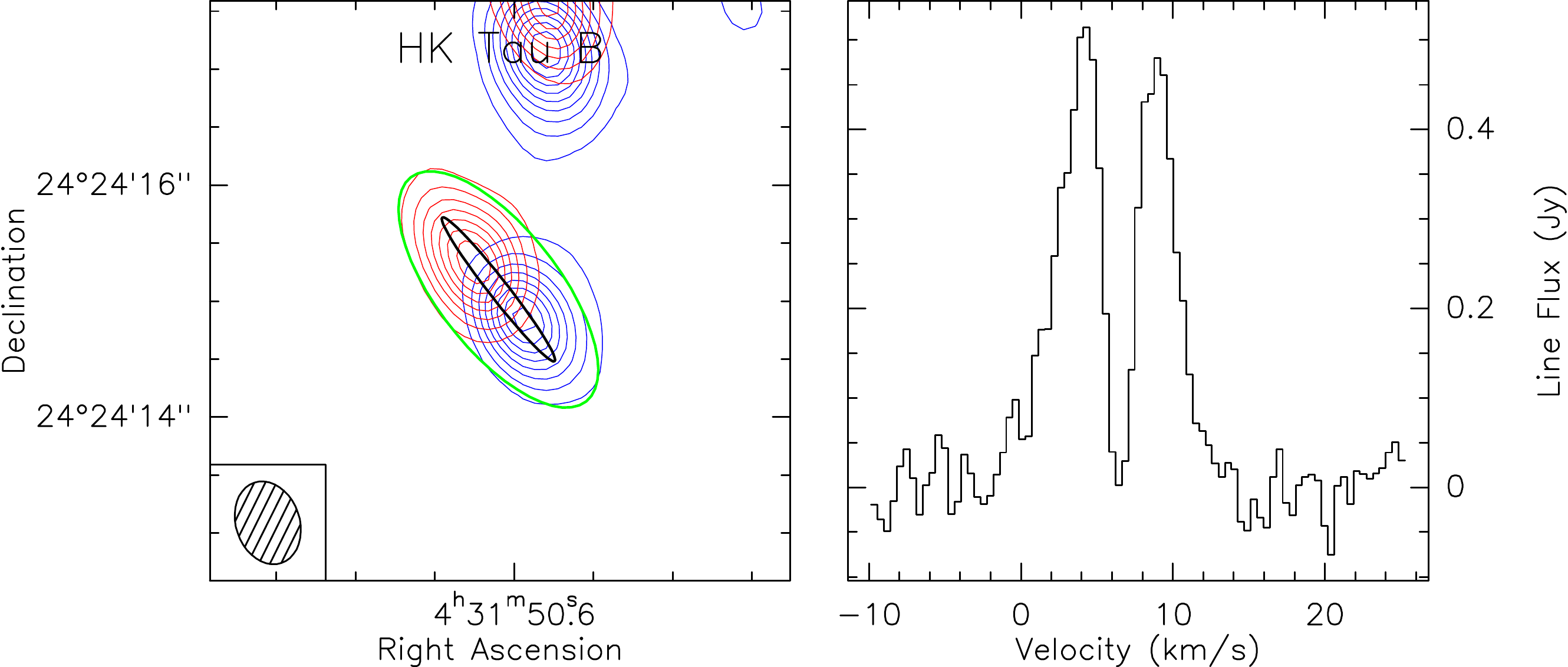}{HK Tau B}{3-2}
\COpanel{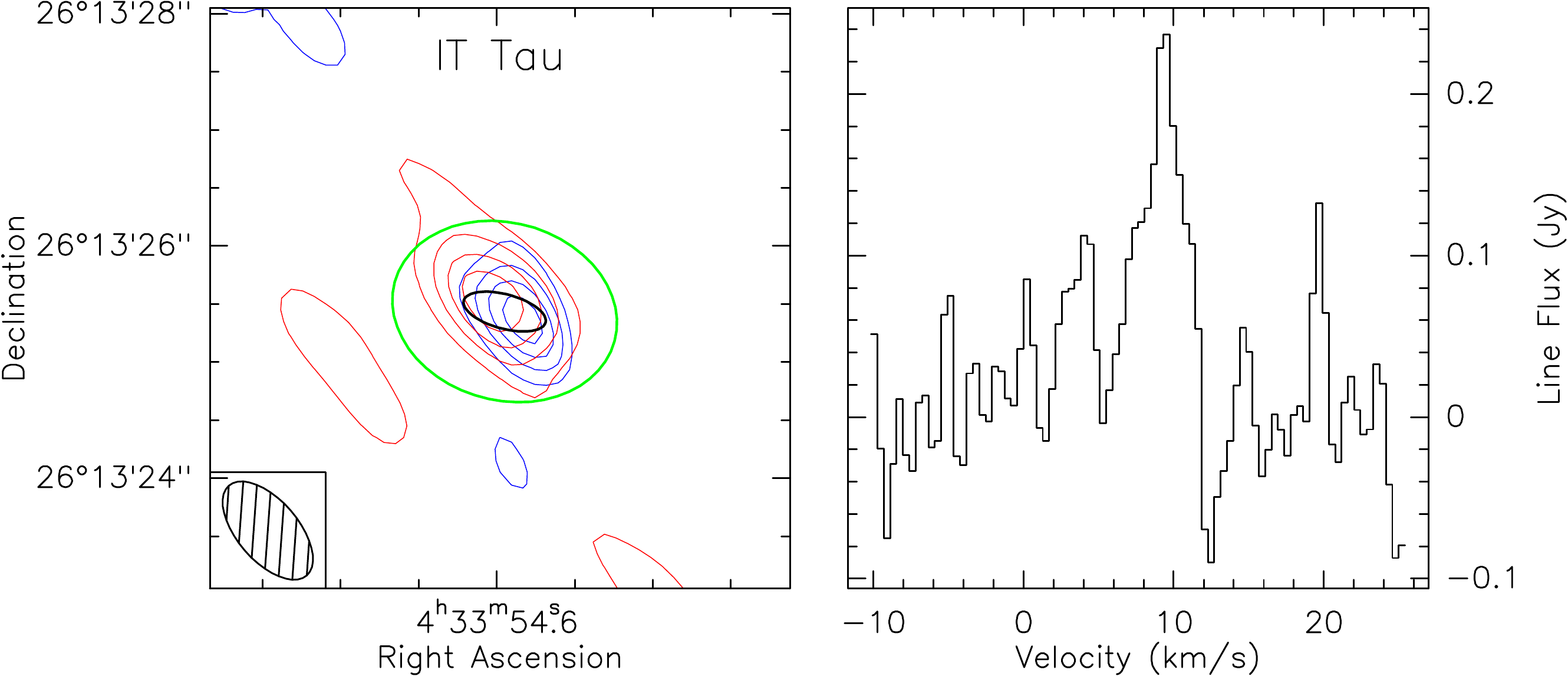}{IT Tau}{3-2}
\hspace{1.9cm}
\COpanel{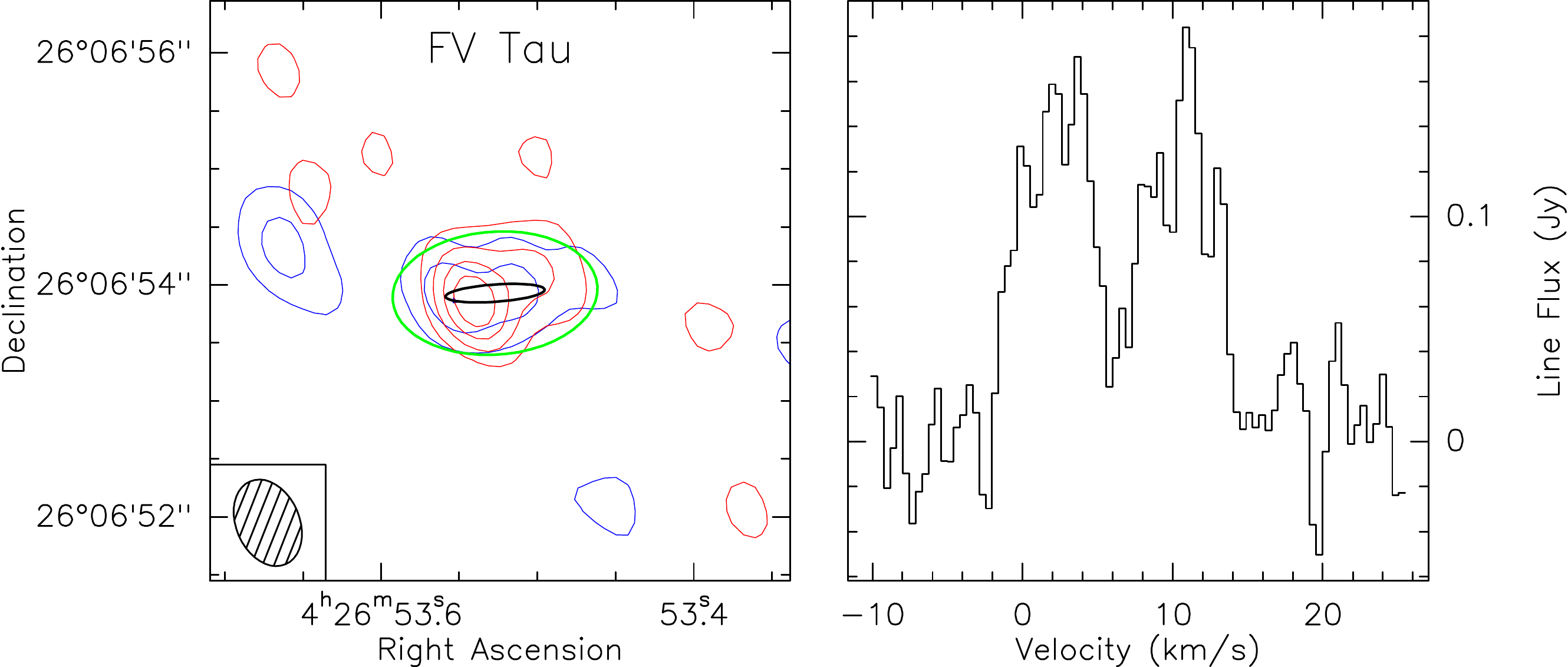}{FV Tau}{3-2}
\caption{CO J=3-2 from Akeson \& Jensen 2014 sample}
\label{fig:aj}
\end{figure*}
\fi
%


\begin{figure}
\includegraphics[height=16.0cm]{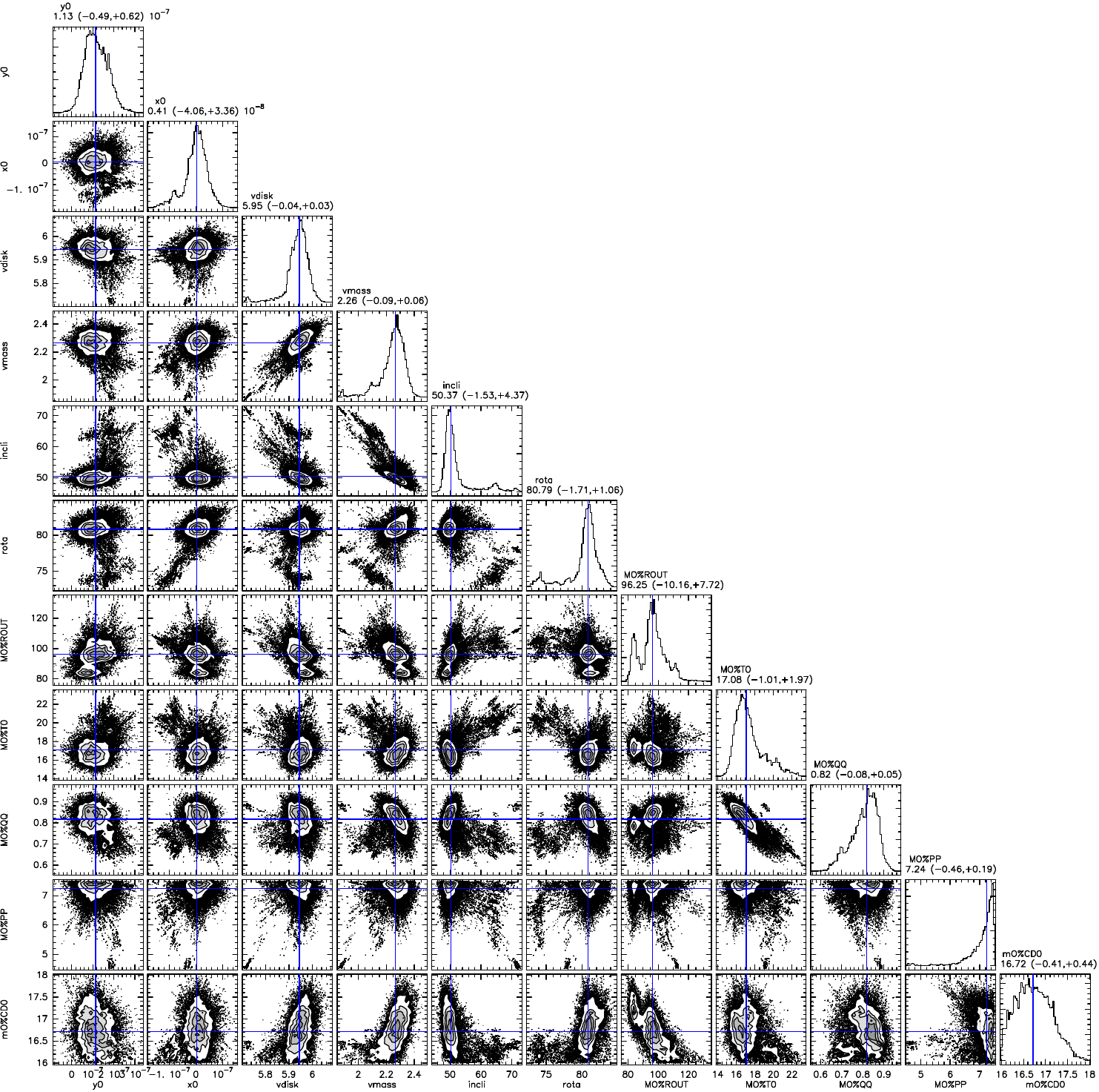}
\caption{Correlation plot for parameters of the HK Tau A disk resulting from
an MCMC analysis.}
\label{fig:co-mcmc}
\end{figure}

\end{document}